# Efficient, Divergence-Free, High Order MHD on 3D Spherical Meshes with Optimal Geodesic Meshing

By

Dinshaw S. Balsara[1,2], Vladimir Florinski[3], Sudip Garain[2], Sethupathy Subramanian[2], Katharine F. Gurski[4]

(dbalsara@nd.edu, sgarain@nd.edu, ssubrama@nd.edu) [1]ACMS Department and [2]Physics Department, University of Notre Dame
(vaf0001@uah.edu) [3]Space Physics, University of Alabama, Huntsville,
(kgurski@howard.edu) [4]Department of Mathematics, Howard University

**Abstract**

There is a great need in several areas of astrophysics and space-physics to carry out high order of accuracy, divergence-free MHD simulations on spherical meshes. This requires us to pay careful attention to the interplay between mesh quality and numerical algorithms. Methods have been designed that fundamentally integrate high order isoparametric mappings with the other high accuracy algorithms that are needed for divergence-free MHD simulations on geodesic meshes. The goal of this paper is to document such algorithms that are implemented in the geodesic mesh version of the *RIEMANN* code. The fluid variables are reconstructed using a special kind of WENO-AO algorithm that integrates the mesh geometry into the reconstruction process from the ground-up. A novel divergence-free reconstruction strategy for the magnetic field that performs efficiently at all orders, even on isoparametrically mapped meshes, is then presented. The MHD equations are evolved in space and time using a novel ADER predictor algorithm that is efficiently adapted to the isoparametrically mapped geometry. The application of one-dimensional and multidimensional Riemann solvers at suitable locations on the mesh then provides the corrector step. The corrector step for the magnetic field uses a Yee-type staggering of magnetic fields. This results in a scheme with divergence-free update for the magnetic field. The use of ADER enables a one-step update which only requires one messaging operation per complete timestep. This is very beneficial for parallel processing. Several accuracy tests are presented as are stringent test problems. PetaScale performance is also demonstrated on the largest available supercomputers.

**I) Introduction**

Several problems in astrophysics, space-physics, metrology and engineering entail carrying out fluid dynamical, magnetohydrodynamical or electromagnetic simulations on spherical meshes. Logically Cartesian meshing, based on $(r,\theta,\phi)$ geometry, provides an imperfect solution to the problem of meshing the sphere. Such meshes have two very prominent deficiencies. First, the timestep is diminished by the smaller zones that are closer to the poles of the mesh. Second, the presence of a coordinate singularity at the poles results in substantial build-up of error at the poles. The overarching goal of this paper is to describe an efficient, divergence-free MHD scheme at high orders that operates on geodesic meshes with the same level of sophistication as the high order divergence-free MHD schemes that have been developed for structured, logically Cartesian, meshes.

Such a scheme would have to map the more complicated geometries that arise when the sphere is mapped to a geodesic mesh – i.e. a mesh that overcomes the deficiencies that are inherent in logically Cartesian $(r,\theta,\phi)$ meshes. A geodesic mesh starts with one of the Platonic solids (usually a cube or an icosahedron) and uses it to find the optimal subdivision of the sphere. Recursive bisection with geodesic curves are then used to further refine the meshing of the spherical surface. The mesh is then extruded in the radial direction. The zones of such a mesh have curved boundaries and one has then to overcome the challenge of defining optimal quadrature and cubature on curved surfaces and volumes. This is achieved by using isoparametric meshing of the resulting zones using ideas developed in Zienkiewicz & Taylor (2000). Since isoparametric mapping is well-described in the above-cited reference, we describe it only briefly in this paper. Moreover, our discussion is restricted to nuances that arise on spherical geodesic meshes.

A good scheme for numerical MHD should also have to have excellent conservation and shock-handling capabilities for the conserved fluid variables. This necessitates a highly accurate treatment of the fluid fluxes, while not losing accuracy in the presence of the curved meshes that are inherent in mapping the sphere. This is achieved by a recent extension of the WENO (Weighted Essentially Non Oscillatory) algorithm that adapts to any curvilinear mesh (Balsara *et al*. 2018). For background on WENO schemes, an incomplete list would consist of Jiang & Shu (1996), Balsara & Shu (2000), Hu & Shu (1999), Dumbser & Käser (2007), Herrick *et al*. (2006), Castro *et al.* (2011), Zhu & Qiu (2016), Balsara, Garain & Shu (2016) Cravero & Semplice (2016), Dumbser *et al.* (2017) and Balsara *et al.* (2019). Magnetic fields can assume very large values in astrophysics and space physics with the result that the thermal energy is a small fraction of the total energy. This makes the preservation of positive pressures problematic for some problems where a conservative formulation is used. WENO methods can also be ruggedized to have a positivity-preserving property for MHD flows (Balsara 2012b) making the very suitable for astrophysical applications. Please also see a recent review on higher order schemes by Balsara (2017). The variant of the WENO scheme that we use here is the WENO-AO algorithm; i.e. a WENO scheme that has Adaptive Order. Consequently, the WENO-AO scheme can drop order of accuracy in shock-dominated regions where the stability from a lower order scheme is desirable. However, the WENO-AO accuracy can also achieve its full higher order accuracy when locally smooth flow makes it profitable to retain higher order accuracy. Since the finite volume WENO-AO scheme is fully described in (Balsara *et al.* 2018), we do not describe the details here. However, the finite volume WENO-AO scheme is also a building block for the reconstruction of the magnetic field, which we describe next. Therefore, we provide some essential insights on the WENO-AO algorithm just to give the paper a modicum of completeness.

A good scheme for MHD should also use a Yee-type collocation of face-centered magnetic fields and edge-centered electric fields while not losing any accuracy in the constraint-preserving reconstruction of the magnetic field. This is achieved by extending the higher order constraint-preserving reconstruction methods described in Balsara (2001, 2004, 2009) and Balsara & Dumbser (2015a) and Xu *et al.* (2016). Such a constraint-preserving reconstruction strategy for vector fields that have an involution-constraint has never been presented for isoparametrically mapped meshes. We present it for the first time in this paper.

There is also a substantial interest in carrying out highly parallel MHD simulations. As a result, it is beneficial to minimize the number of messaging operations per timestep in the MHD algorithm. For achieving that goal it is extremely beneficial to use ADER (Arbitrary DERivatives in space and time) methods. While these methods were initially based on extensions of the generalized Riemann problem (Titarev & Toro 2002, 2005, Toro & Titarev 2002) modern ADER schemes are based on a different predictor-corrector style of formulation (Dumbser et al. 2008, 2013, Balsara *et al.* 2009, 2013, Boscheri & Dumbser, 2013, 2016, 2017). Because the predictor step provides a high order space-time reconstruction of the solution within a zone, the MHD corrector step (which consists of applying one-dimensional Riemann solvers at facial quadrature points and multidimensional Riemann solvers at the edges of the mesh) is reduced to a single step operation. This economical simplification permits a single step time-update which only requires one messaging operation per time-step. For this reason we describe an ADER scheme in considerable detail that works well with isoparametrically mapped meshes. Our ADER formulation has the additional novelty that it uses serendipity elements, thereby introducing further efficiencies in the ADER algorithm. The ADER method provides the predictor step. For fluid variables, the corrector step simply consists of the invocation of one-dimensional Riemann solvers at a suitable number of high order quadrature locations in the faces of the mesh. This provides us with a higher order numerical flux that is properly upwinded for the fluid variables. The corrector step for the magnetic fields is more intricate and is described next.

Yee (1966) type meshes have found great appeal in computational electrodynamics (Taflove & Brodwin 1975a, 1975b) because Faraday's law and Ampere's law both have associated involution constraints. Because the MHD equations also evolve the magnetic field according to Faraday's law, these ideas were subsequently imported from computational electrodynamics into the MHD literature by Brecht *et al.* (1981). The ideas have been subsequently developed by Evans & Hawley (1989), DeVore (1991), Dai & Woodward (1998), Ryu *et al.* (1998), Balsara & Spicer (1999) and Londrillo & Del Zanna (2004). A Yee-type mesh requires that edge-averaged electric field should be used for the update of the facially-averaged magnetic field components. However, it was recognized quite early that even on a structured mesh, there will be four states abutting each edge. As a result, a one-dimensional Riemann solver cannot provide the multidimensional upwinding that is needed at such an edge. Efficient, implementable approximate multidimensional Riemann solvers were first developed in Balsara (2010, 2012a, 2014). Such Riemann solvers provide a natural strategy for obtaining multidimensionally upwinded electric fields at the edges of the mesh. In Balsara (2014) such multidimensional Riemann solvers were named MuSIC Riemann solvers, where the acronym stands for **Mu**ltidimensional **S**elf-similar solver that is based on strongly-**I**nteracting states that are **C**onsistent with the hyperbolic system. MuSIC Riemann solvers were extended to unstructured meshes (Balsara, Dumbser & Abgrall 2014, Balsara & Dumbser 2015b). By now it is also well-understood that one can introduce sub-structure (i.e. intermediate waves in multiple directions) into such Riemann solvers, thereby reducing their dissipation (Balsara 2014, Balsara *et al*. 2016, Balsara & Nkonga 2017). On a geodesic mesh, the zones that come together at an edge of the mesh are not mutually orthogonal. In fact, the zones can have substantial deviation from a Cartesian mesh with the result that the mesh geometry must be incorporated into the multidimensional Riemann solver; and MuSIC Riemann solvers provide a natural way of doing that.

Section II very briefly describes geodesic meshes and isoparametric mapping and its benefits. Section III describes some essential background on the finite volume WENO-AO reconstruction used here while pointing the reader to other papers which describe the algorithm in more detail. Section IV describes the constraint-preserving reconstruction of magnetic vector fields on isoparametrically mapped meshes. Section V describes the ADER algorithm on isoparametrically mapped meshes. Section VI describes the implementation of the entire scheme in pointwise fashion. Section VII shows accuracy tests. Section VIII shows other stringent results. Section IX draws conclusions. The Appendices that are mentioned in this paper have been provided as electronic supplementary material.

**II) Geodesic Meshes and Isoparametric Mapping and its Advantages**

To support a high accuracy MHD calculation on a sphere, the sphere should be mapped with a very high quality mesh. In Sub-section II.a we document such a geodesic mesh that is based on spherical icosahedra. Volumetric, area and linear quadratures at sufficiently high order of accuracy are required on such a curved mesh in order to support the high order accurate numerical methods that we document here. While it is not the goal of this paper to document such geometric mapping technologies, Sub-section II.b gives a sufficiently detailed flavor of isoparametric mapping for spherical meshes while pointing the reader to further literature. Sub-section II.c provides a broad brushstroke sketch of a divergence-free MHD algorithm on a geodesic spherical mesh. This is meant to prepare the reader for subsequent sections.

**II.a) Geodesic Meshes Based on Spherical Icosahedra**

As mentioned in the Introduction, a logically Cartesian mesh in $(r,\theta,\phi)$ geometry provides an imperfect solution to the problem of meshing the sphere and carrying out computations on it. The two prominent deficiencies are a diminishment of the timestep due to the smaller zones at the poles and a loss of accuracy at the poles. As a result, we wish to have a strategy for the uniform meshing of the sphere. Cubed sphere meshes have been tried, but they result in significant distortion of the mesh at the vertices of the cube, resulting in substantial build-up of error at those locations. Even so, higher order MHD schemes, albeit without a divergence-free aspect in the magnetic field, have been tried on cubed sphere meshes (Ivan *et al*. 2013, 2015) and meshes with Voronoi tessellation (Florinski *et al*. 2013). For hydrodynamical simulations on cubed sphere meshes, see Woodward *et al*. (2015), and for MHD simulations, see Koldoba *et al*. (2002). Yin-Yang meshes have their own deficiencies with conservation at the interface where the two meshes dovetail with each other. For a use case of Yin-Yang meshes, see Jiang *et al*. (2012). There is, therefore, much room for improvement in meshing the sphere and carrying out higher order divergence-free MHD simulations in such an environment.

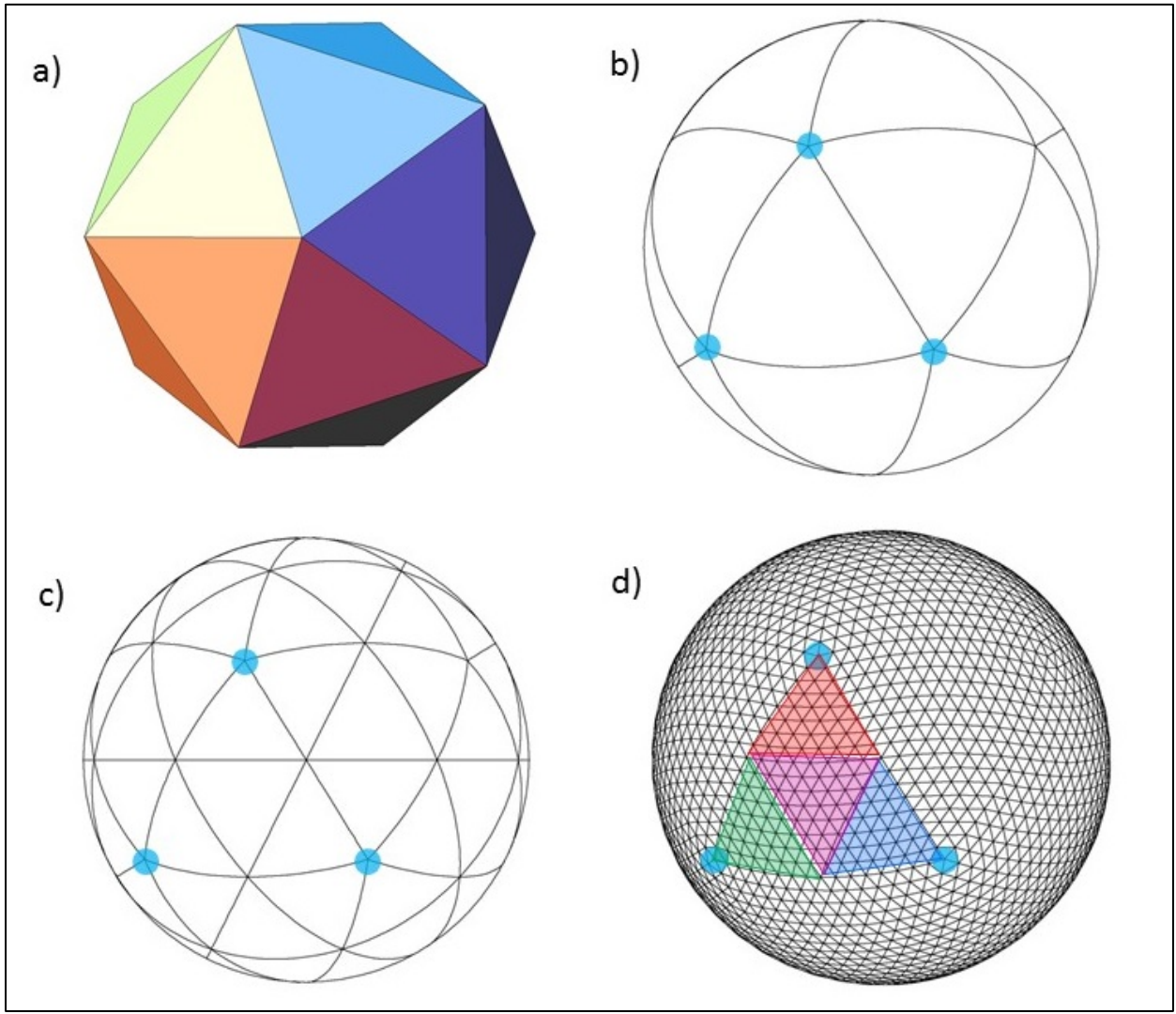

An optimal strategy for meshing the sphere in the most isotropic way possible was provided by Euclid and is shown in Fig. 1. Fig. 1 shows how one starts with an icosahedron in Fig. 1a and uses it to obtain a spherical icosahedron in Fig. 1b. The first subdivision of the spherical icosahedron is shown in Fig. 1c. Continuing this process three more times yields Fig. 1d. Fig. 1b is referred to as a level 0 sectorial division and produces 20 great triangles whose sides subtend an angle of $(\pi/2) - \tan^{-1}(1/2) \cong 63.4^o$ at the center of the sphere. Fig. 1c is a level 1 sectorial division and produces 80 great triangles whose sides subtend an angle of $33.9^o$ (on average) at the center of the sphere. The sides of the level 4 zones in Fig. 1d subtend an angle of $4.33^o$ at the center of the sphere. The level 1 sectors are still shown in Fig. 1d and can be used to form the three-dimensional computational patches that are efficiently processed on each processor. Different computational patches are distributed to different processors resulting in efficient parallel computation. For the purposes of this paper, the angular resolution of a geodesic mesh will refer to the mean central angle subtended by an edge. We clarify that the central angle is the angle with its vertex at the center of the sphere and its end points located on the surface of the sphere. In Table 2 of Florinski et al. (2018) we give the central angle subtended by the average edge for each level of subdivision of the sphere. In Section 3 of Florinski *et al*. (2018) we also provide quantitative details that convincingly suggest that the spherical icosahedral strategy presented here is the optimal method for meshing the sphere. Since this paper has an algorithms-based focus, we do not provide such detail here.

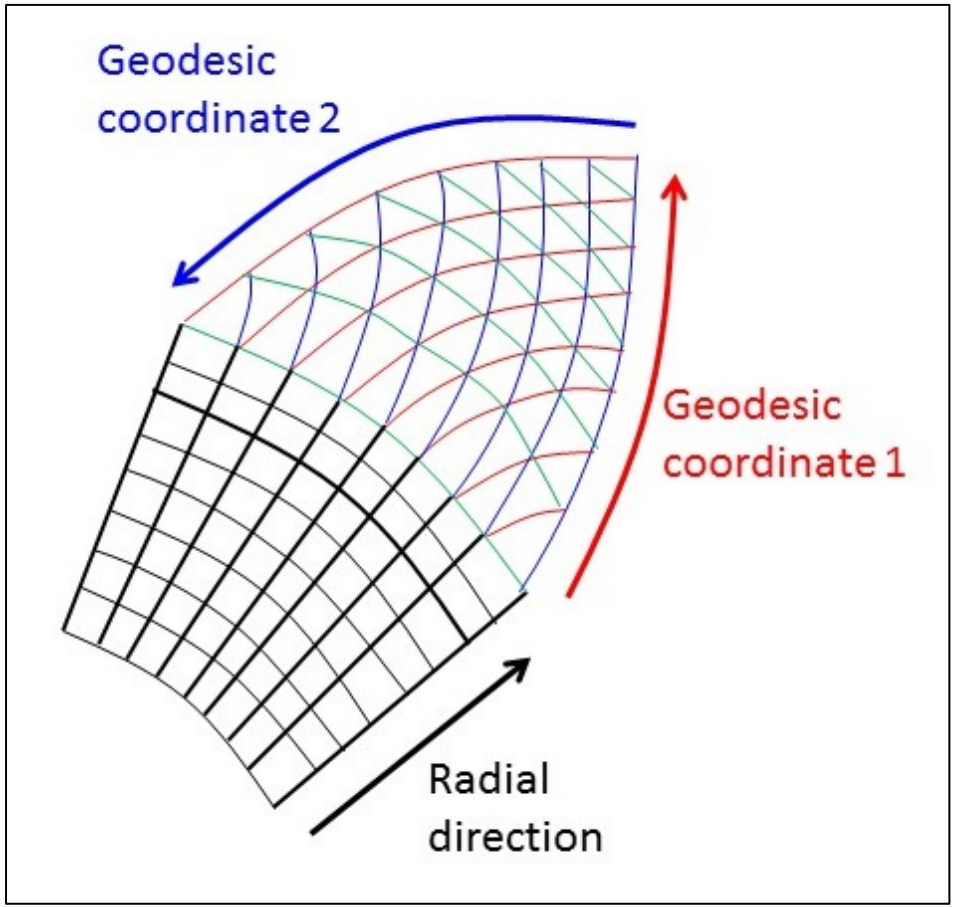

*Fig. 2 shows how a sectorial mesh on the surface of the sphere is extruded in the radial direction in order to form a three dimensional geodesic mesh. We see 64 Delaunay triangulations within the surface of the sphere (shown in red, blue and green) and 8 sub-divisions in the radial direction (shown in black). Any two great circles in the surface of the sphere (see red and blue geodesics) form a coordinate system which can be used to label the triangles. (Count zones; notice square numbers.) Zones on a geodesic mesh can, therefore, be accessed almost as simply as zones on a structured mesh. This facilitates optimal processing speeds. It also facilitates efficient parallelization. Data packing and unpacking is quite simple owing to the close analogy with structured meshes.*

Once the sphere has been optimally mapped, it is possible to extrude that mapping in the radial direction in order to obtain a three-dimensional mesh. Fig. 2 shows how a sectorial mesh on the surface of the sphere is extruded in the radial direction in order to form a three dimensional geodesic mesh. We see 64 Delaunay triangulations within the surface of the sphere (shown in red, blue and green) and 8 sub-divisions in the radial direction (shown in black). Any two great circles in the surface of the sphere (see red and blue geodesics) form a coordinate system which can be used to label the triangles. (Count zones; notice square numbers.) Zones on a geodesic mesh can, therefore, be accessed almost as simply as zones on a structured mesh. This facilitates optimal processing speeds. It also facilitates efficient parallelization. Data packing and unpacking is quite simple owing to the close analogy with structured meshes. For all of these reasons, we wish to develop a high order, divergence-free MHD algorithm on such a geometrically optimal mesh.

**II.b) Isoparametric Mapping of Geodesic Meshes and its Advantages**

From Fig. 2 it is easy to see that the zones of a geodesic mesh of the type we are considering here are logically closest to a three dimensional triangular prism. The technical name for such a three dimensional shape is a frustum. Because the spherical icosahedron-based geodesic mesh strives to keep each zone as close to the shape of an equilateral triangular prism, we choose an equilateral triangular prism as our reference element. Any high order fluid or MHD algorithm has to carry out certain high order accuracy quadrature operations within a zone, or at the faces and edges of a zone. The only way to achieve this on curved elements, while accurately preserving the curvature of the element, and the curvature of its faces and edges, is to resort to isoparametric mapping. Isoparametric mappings have been discussed in detail by Zienkiewicz & Taylor (2000); please see Chapter 8 of their text. As a result, we just provide some intuitive results, without necessarily going into detail here. Our primary goal in this section is to illustrate some classes of isoparametric mappings that are most suitable for our needs.

Isoparametric mappings have been documented in the literature at various orders and the order of the mapping can (optionally) be different from the order of accuracy of the scheme. When the order of accuracy of the mapping is lower than the order of accuracy of the scheme, the mapping is referred to as sub-parametric. When the order of accuracy of the mapping is equal to the order of accuracy of the scheme, the mapping is referred to as isoparametric. Sub-parametric mappings are not preferred, because it is thought that the accuracy of the geometric representation of the mesh should keep up with the order of accuracy of the scheme. For frustums, which will always be curved on a spherical mesh, the first mapping that nominally retains the curvature of the zone is the third order mapping which uses quadratic functions to map from the reference element to the frustum. As a result, we base our higher order schemes on third and fourth order accurate mappings. For second order schemes, the linear isoparametric mapping only provides frustums with straight line sides, and flat rather than spherical faces. However, the straight edges don't seem to impact the order of accuracy of a second order scheme. For third order accurate schemes, we will use quadratic polynomial-based isoparametric mappings, which are indeed third order accurate. The fourth order mapping uses cubic functions to map from the reference element to the frustum. For fourth order schemes, we will use cubic polynomial-based isoparametric mappings, which are indeed fourth order accurate. Figs. 3a, 3b and 3c show isoparametric mappings at second, third and fourth order respectively from the reference equilateral triangular prism to the spherical frustum. Nodes at vertices are shown in black; nodes within the edges of triangles are shown in red; and nodes at the centroids of triangular faces are shown in blue. Notice from Fig. 3 that the variation in the radial direction is always linear for a frustum; as a result, we can use fewer nodes in the radial direction and posit only a linear variation in that direction.

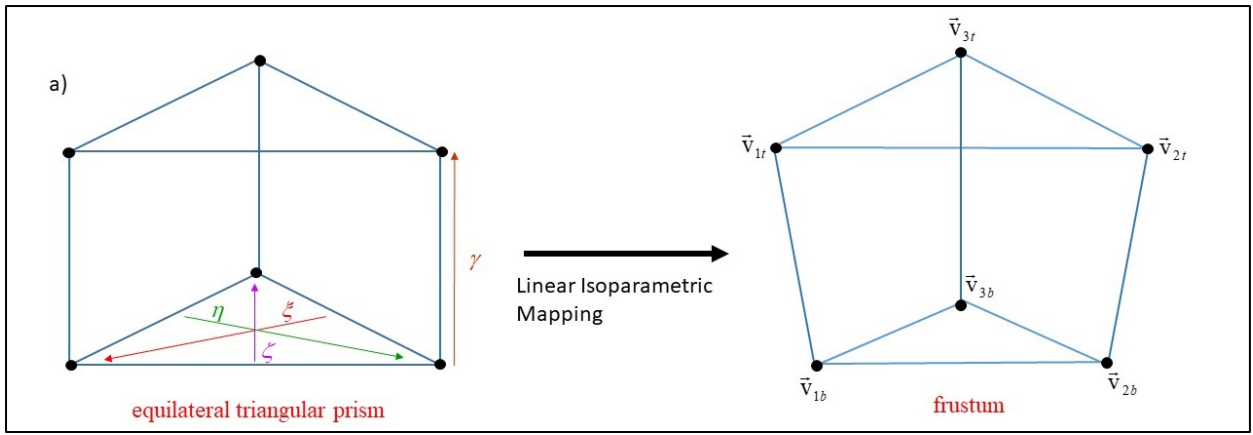

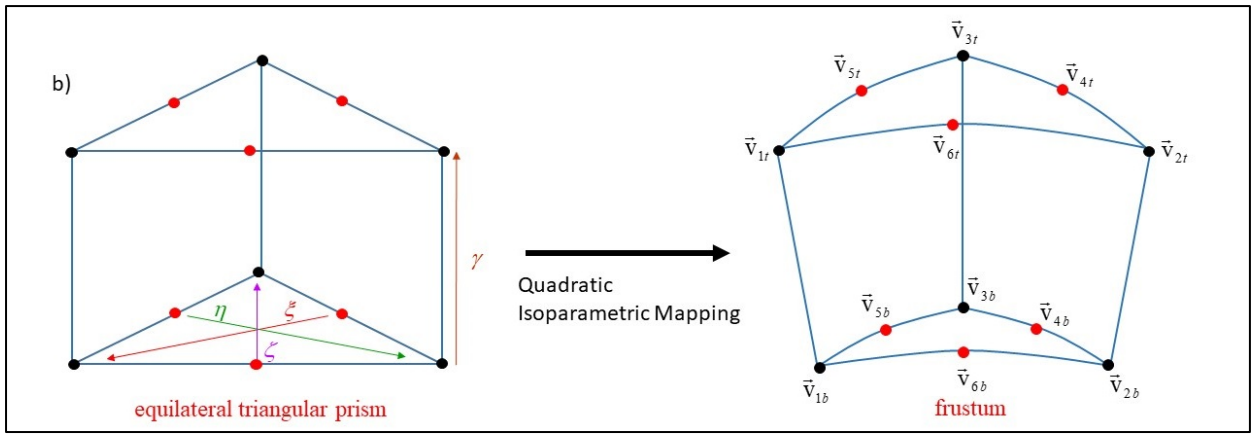

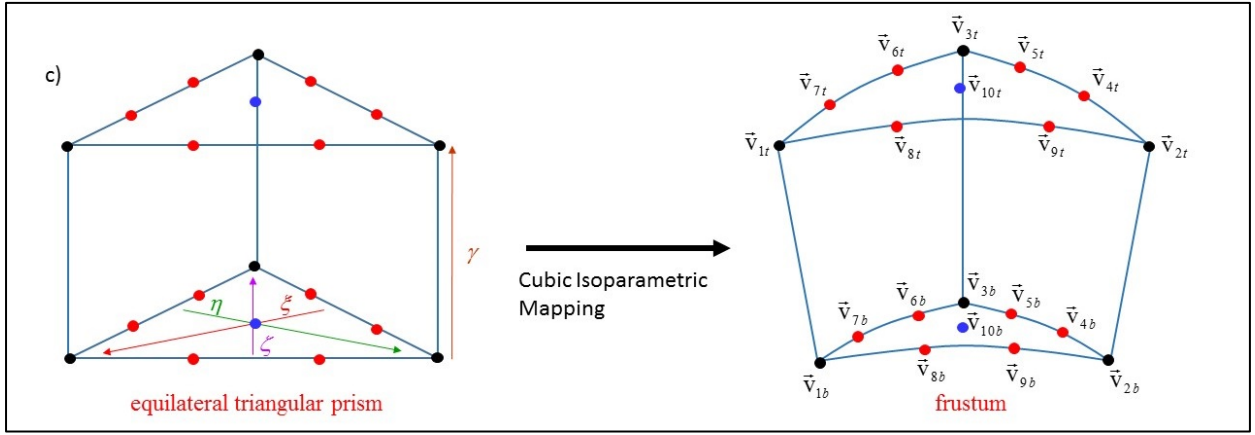

*Figs. 3a, 3b and 3c show isoparametric mappings at second, third and fourth order respectively from the reference equilateral triangular prism to the spherical frustum. Nodes at vertices are shown in black; nodes within the edges of triangles are shown in red; and nodes at the centroids of triangular faces are shown in blue.*

We will not document the details of various types of isoparametric mappings for two reasons. First, that has already been done in great detail by Zienkiewicz & Taylor (2000) and it is not the goal of this paper to repeat information from that textbook. Second, there is another paper (Florinski *et al*. 2019) which is indeed a compendium of underlying technologies for supporting high order calculations on geodesic meshes. It is, nevertheless, interesting to demonstrate to the reader how we get significant improvements in mapping the surface of the sphere as we go from second to third to fourth order accurate isoparametric mappings. To illustrate this, we consider a single, spherical, equilateral triangle that is centered at the polecap of a unit sphere. Each vertex of the triangle makes an angle of 5º relative to the centroid of the triangle. This triangular patch on the unit sphere was mapped with linear, quadratic and cubic polynomials, leading to second, third and fourth order accurate isoparametric mappings respectively. The deviation of the mapping from the unit radius of the sphere was plotted. Figs. 4a, 4b and 4c show the deviation from sphericity for linear, quadratic and cubic isoparametric mappings applied to a spherical triangle. The difference between the radius of the sphere and the radius as obtained from the isoparametric mapping is shown. Figs. 4a, 4b and 4c are shown on their own scales. Figs. 4d, 4e and 4f show the same results on the same scale. The latter three figures show that quadratic and cubic isoparametric mappings map the sphere (for this problem) up to one part in $10^6$ and one part in $10^7$ respectively.

Because of this test problem, we will prefer third and fourth order isoparametric mappings in the rest of this work.

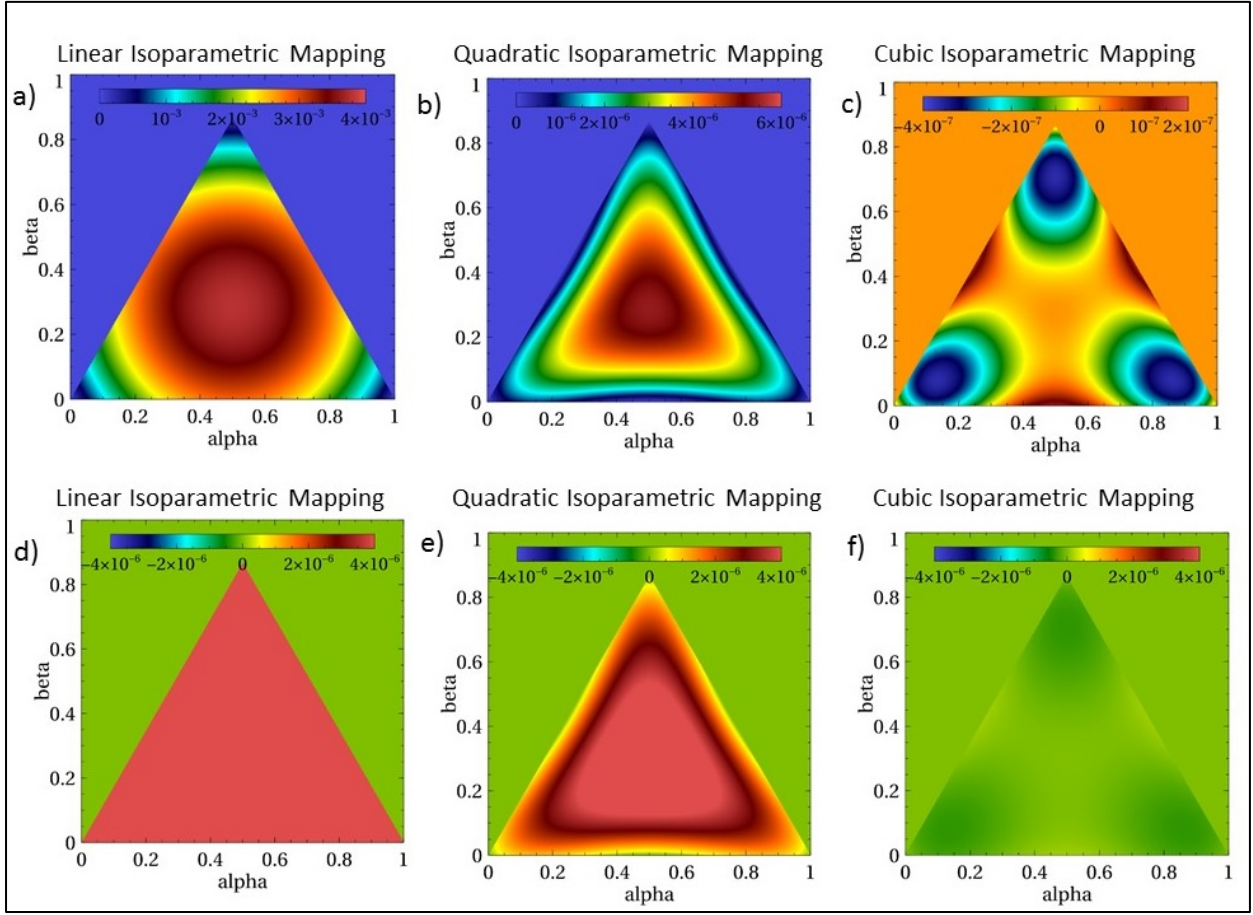

*Figs. 4a, 4b and 4c show the deviation from sphericity for linear, quadratic and cubic isoparametric mappings applied to a spherical triangle. The difference between the radius of the sphere and the radius as obtained from the isoparametric mapping is shown. Figs. 4a, 4b and 4c are shown on their own scales. Figs. 4d, 4e and 4f show the same results on the same scale. The latter three figures shown that quadratic and cubic isoparametric mappings map the sphere (for this problem) up to one part in $10^6$ and one part in $10^7$ respectively.*

Chapter 8 of Zienkiewicz & Taylor (2000) provides a great deal of detail about these isoparametric mappings. The process by which areal quadratures and volumetric cubatures can be obtained from these mappings is also described there. In Florinski *et al*. (2019) we also provide a one-stop-shop for many allied technologies for geodesic meshes and the reader who is interested in a single place where all this information is concatenated is welcome to visit that paper.

## II.c) Introductory Sketch of Divergence-Free MHD on Geodesic Meshes

The MHD equations can be formally written in flux form as

$$\frac{\partial \mathbf{U}}{\partial t} + \frac{\partial \mathbf{F}}{\partial x} + \frac{\partial \mathbf{G}}{\partial y} + \frac{\partial \mathbf{H}}{\partial z} = 0 \tag{2.1}$$

where $\mathbf{U}$ is the vector of conserved variables and $\mathbf{F}$, $\mathbf{G}$, $\mathbf{H}$ are the fluxes. Written explicitly, we have

$$\frac{\partial}{\partial t}\begin{pmatrix} \rho \\ \rho v_x \\ \rho v_y \\ \rho v_z \\ \varepsilon \\ B_x \\ B_y \\ B_z \end{pmatrix} + \frac{\partial}{\partial x}\begin{pmatrix} \rho v_x \\ \rho v_x^2 + P + \mathbf{B}^2/8\pi - B_x^2/4\pi \\ \rho v_x v_y - B_x B_y/4\pi \\ \rho v_x v_z - B_x B_z/4\pi \\ (\varepsilon + P + \mathbf{B}^2/8\pi)v_x - B_x(\mathbf{v}\cdot\mathbf{B})/4\pi \\ 0 \\ (v_x B_y - v_y B_x) \\ -(v_z B_x - v_x B_z) \end{pmatrix}$$

$$+ \frac{\partial}{\partial y}\begin{pmatrix} \rho v_y \\ \rho v_x v_y - B_x B_y/4\pi \\ \rho v_y^2 + P + \mathbf{B}^2/8\pi - B_y^2/4\pi \\ \rho v_y v_z - B_y B_z/4\pi \\ (\varepsilon + P + \mathbf{B}^2/8\pi)v_y - B_y(\mathbf{v}\cdot\mathbf{B})/4\pi \\ -(v_x B_y - v_y B_x) \\ 0 \\ (v_y B_z - v_z B_y) \end{pmatrix} + \frac{\partial}{\partial z}\begin{pmatrix} \rho v_z \\ \rho v_x v_z - B_x B_z/4\pi \\ \rho v_y v_z - B_y B_z/4\pi \\ \rho v_z^2 + P + \mathbf{B}^2/8\pi - B_z^2/4\pi \\ (\varepsilon + P + \mathbf{B}^2/8\pi)v_z - B_z(\mathbf{v}\cdot\mathbf{B})/4\pi \\ (v_z B_x - v_x B_z) \\ -(v_y B_z - v_z B_y) \\ 0 \end{pmatrix} = 0 \tag{2.2}$$

where $\rho$ is the fluid density, P is the fluid pressure, $v_x$, $v_y$, $v_z$ are the fluid velocities and $B_x$, $B_y$ and $B_z$ are the components of the magnetic field. The total energy is given by $\varepsilon = \rho v^2/2 + P/(\gamma-1) + \mathbf{B}^2/8\pi$. The first five components of eqn. (2.2) remind us to update the volume averaged fluid densities using area-averaged numerical fluxes. Since this form of conservative update is well-known, we do not focus further attention on it.

Owing to the electric field being defined by $\mathbf{E} = -\mathbf{v} \times \mathbf{B}$ (a factor involving the speed of light is omitted); the flux components have the following symmetries

$$E_x = H_7 = -G_8 \quad ; \quad E_y = F_8 = -H_6 \quad ; \quad E_z = G_6 = -F_7 \tag{2.3}$$

The symmetries in the flux terms are perfectly comprehensible when the last three rows of eqn. (2.2) are written in a format that makes Faraday's law explicit as follows:

$$\frac{\partial \mathbf{B}}{\partial t} + \nabla \times \mathbf{E} = 0 \tag{2.4}$$

Faraday's law ensures that if the magnetic field is initially divergence-free, $\nabla \cdot \mathbf{B} = 0$, it is so forever. Yee (1966) proposed a staggering of magnetic and electric fields, with the magnetic field components being area-averaged at the faces of the mesh and the electric field components being line-averaged at the edges of the same mesh. Such a Yee-type staggering gives the scheme a mimetic aspect, which ensures that the discrete magnetic field on the mesh will also remain divergence-free up to machine accuracy.

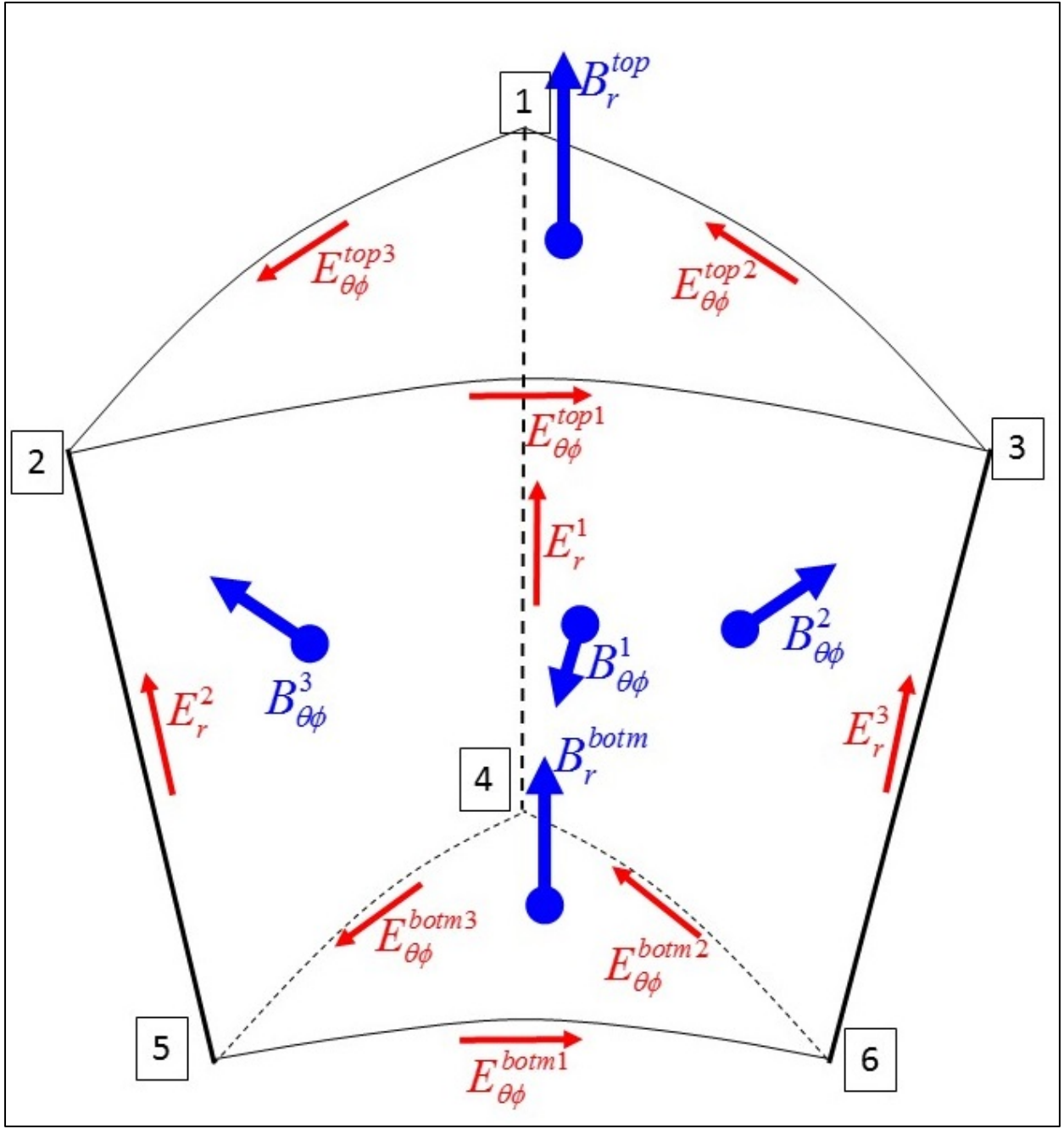

*Fig. 5 shows a single zone of an MHD mesh, showing the Yee-mesh type collocation of face-centered magnetic field components and the edge-centered electric field components on the frustum. The time evolution of the magnetic fields can be broadly conceptualized in the following three steps. First, the facial magnetic field components can be used to make a high order divergence-free reconstruction of the magnetic field in the interior of the zone. Second, once the variation of the zone-centered fluid variables and the reconstructed magnetic field variables are available within the volume of the frustum, they can be used to carry out the high order ADER predictor step. Third, the layout of the magnetic and electric fields shows that a Yee-type update of the facially-averaged magnetic field that is consistent with Faraday's law can even be achieved*

*on an isoparametrically mapped mesh. Multidimensional Riemann solvers are invoked at one or more locations on the edges of the mesh in order to obtain the high order edge-averaged electric fields, leading to a high order accurate update of the facial magnetic fields.*

Fig. 5 shows a single zone of an MHD mesh, showing the Yee-mesh type collocation of face-centered magnetic field components and the edge-centered electric field components on the frustum-shaped zone. The time evolution of the magnetic fields can be broadly conceptualized in the following three steps. First, the facial magnetic field components can be used to make a high order divergence-free reconstruction of the magnetic field in the interior of the zone. Second, once the variation of the zone-centered fluid variables and the reconstructed magnetic field variables are available within the volume of the frustum, they can be used to carry out the high order ADER predictor step. Third, the layout of the magnetic and electric fields in Fig. 5 shows that a Yee-type update of the facially-averaged magnetic field that is consistent with Faraday's law can even be achieved on an isoparametrically mapped mesh. One-dimensional Riemann solvers can be invoked at facial quadrature points in order to get numerical fluxes for conserved, fluid variables. Multidimensional (MuSIC) Riemann solvers are invoked at one or more locations on the edges of the mesh in order to obtain the high order edge-averaged electric fields, leading to a high order accurate update of the facial magnetic fields.

Now let us focus on the spherical triangular face $\Delta_{123}$ in Fig. 5 and denote its area as $A_{123}$. Denote the curved edge from vertex 1 to vertex 2 in Fig. 5 as $Edge12$. ( Line integrals along $Edge12$ should be taken from vertex 1 to vertex 2. A line integral along $Edge21$ would have a direction that is opposite to the line integral along $Edge12$. ) A similar naming convention applies to the other edges in Fig. 5. Let us now describe the time update from a time "$t$" to a time "$t + \Delta t$". Integrating Faraday's law over the two spatial dimensions of the spherical triangle, and also over time, gives

$$B_r^{top}(t+\Delta t) = B_r^{top}(t) - \frac{\Delta t}{A_{123}} \left\{ \int_t^{t+\Delta t} \left( \int_{Edge12} \vec{E}_{\theta\phi}^{top3} \cdot d\vec{\ell} \right) dt + \int_t^{t+\Delta t} \left( \int_{Edge23} \vec{E}_{\theta\phi}^{top1} \cdot d\vec{\ell} \right) dt + \int_t^{t+\Delta t} \left( \int_{Edge31} \vec{E}_{\theta\phi}^{top2} \cdot d\vec{\ell} \right) dt \right\}$$

(2.5)

A similar consideration applies to the annular face $\square_{5632}$ in Fig. 5, where we denote its area by $A_{5632}$. Integrating Faraday's law over the two spatial dimensions of the annular face, and also over time, gives

$$B^1_{\theta\phi}(t+\Delta t) = B^1_{\theta\phi}(t) - \frac{\Delta t}{A_{5632}} \left\{ \begin{array}{l} \int_{t}^{t+\Delta t}\left(\int_{Edge56} \vec{E}^{botm1}_{\theta\phi} \cdot d\vec{\ell}\right)dt + \int_{t}^{t+\Delta t}\left(\int_{Edge63} \vec{E}^{3}_{r} \cdot d\vec{\ell}\right)dt \\ + \int_{t}^{t+\Delta t}\left(\int_{Edge32} \vec{E}^{top1}_{\theta\phi} \cdot d\vec{\ell}\right)dt + \int_{t}^{t+\Delta t}\left(\int_{Edge25} \vec{E}^{2}_{r} \cdot d\vec{\ell}\right)dt \end{array} \right\} \quad (2.6)$$

The philosophy inherent in eqns. (2.5) and (2.6) can be applied to all the frustrum-like zones that make up the computational domain to get a globally divergence-free evolution of the magnetic field.

The above narrative, along with Fig. 5, serves to emphasize the important role played by the MuSIC Riemann solver. At the top and bottom edges of the frustum shown in Fig. 5, the edges will be curved, especially at higher order. Each such curved edge will be surrounded by four zones. Say that in each of those zones we have made a higher order spatial reconstruction of all the MHD variables and followed it up with an ADER predictor step within each zone. It is easy to see that the MuSIC Riemann solver can be called at any point along the curved edge. Specifically, it can be called at the three Simpson quadrature points along the edge. This would give us a fourth order accurate line integral for the edge-integrated electric fields. This enables us to obtain a fourth order accurate update for the area-averaged magnetic fields. At any of the straight edges of the frustum shown in Fig. 5, we will have six (or five at pentacorners) surrounding zones. The space-time information from those zones can again be used along with the MuSIC Riemann solver to get line integrals along the straight edges. We see now that the multidimensional Riemann solver plays a very important role in the divergence-free evolution of the magnetic field, especially when the curvature of the mesh boundaries has to be included in the calculation in order to obtain high order of accuracy.

**III) Very Brief Notes on WENO-AO Reconstruction**

The WENO-AO reconstruction (Balsara *et al*. 2019) is based on realizing that there is a favorable basis set, known as a Taylor series basis (Luo *et al*. 2008), in which it is very efficient to carry out weighted essentially non-oscillatory finite volume reconstruction. As with finite difference WENO-AO (Balsara, Garain & Shu 2016), the method is based on realizing that we can make a non-linear hybridization between a large, centered, very high accuracy stencil and a lower order central WENO scheme that is, nevertheless, very stable and capable of capturing physically meaningful extrema. The non-linear hybridization yields a class of adaptive order WENO schemes that work well on unstructured meshes. Since the geodesic meshing of the sphere produces triangulated meshes, the WENO-AO spatial reconstruction strategy is very well-matched to the task of producing a high order, oscillation-free reconstruction of the MHD variables. The scheme is efficient because it minimizes the number of evaluations of the smoothness indicator on the large, higher order, stencil. The smaller stencils consist of the set of CWENO-type stencils that are traditionally used for third order WENO calculations on unstructured meshes.

WENO-AO achieves a further modicum of efficiency because of a dexterous utilization and extension of the Parallel Axis Theorem from introductory mechanics. (The theorem, drawn from physics, states that the moment of inertia of a solid body about any other point is given by that same moment about the centroid plus the mass of the body times the square of the distance to the centroid.) By extending the Parallel Axis Theorem, we show that there is a significant simplification in the finite volume reconstruction. Instead of solving a constrained least squares problem, our method only requires the solution of a smaller least squares problem on each stencil. This also simplifies the matrix assembly and solution for each stencil. The evaluation of smoothness indicators is also simplified. In several tests, the WENO-AO reconstruction has shown itself to be almost twice as fast as a traditional WENO reconstruction. In Section IV of Balsara *et al*. (2019) we have shown that several efficiencies can be realized when a spherical, logarithmically-ratioed mesh is used. Since this is the typical/intended usage for spherical geodesic mesh codes, we find it very beneficial to use the WENO-AO algorithm in this work.

Section V of Balsara *et al*. (2019) provides a pointwise description of how a WENO-AO algorithm is to be implemented. Therefore, we do not present too much detail here. However, the basis functions are also very useful for the divergence-free reconstruction of magnetic fields, which is described in the next section. Therefore, we provide just enough detail about Taylor series basis, and expansion in that basis, in this section to make the next section accessible.

Let us consider a zone that is labeled "0" which has a zone-averaged flow variable $\bar{u}^0$. This zone has a characteristic length $l_0$. The reconstruction problem for this zone can be thought of as identifying a stencil of neighboring zones, where the zones are labelled by an index "$j$", and using the flow variables in that stencil to obtain all the higher moments of the fluid variable in zone "0". For the zone "0", we identify its center of mass and expand the reconstructed solution $u^0(x, y, z)$ in terms of the Taylor basis for that zone. We explicitly illustrate this process at third order by writing the three-dimensional reconstruction as

$$u^0(x,y,z) = \bar{u}^0 + u_x^0\left(\frac{x}{l_0}\right) + u_y^0\left(\frac{y}{l_0}\right) + u_z^0\left(\frac{z}{l_0}\right) + u_{xx}^0\left[\left(\frac{x}{l_0}\right)^2 - C_{xx}^0\right] + u_{yy}^0\left[\left(\frac{y}{l_0}\right)^2 - C_{yy}^0\right]$$
$$+ u_{zz}^0\left[\left(\frac{z}{l_0}\right)^2 - C_{zz}^0\right] + u_{xy}^0\left[\left(\frac{x}{l_0}\right)\left(\frac{y}{l_0}\right) - C_{xy}^0\right] + u_{yz}^0\left[\left(\frac{y}{l_0}\right)\left(\frac{z}{l_0}\right) - C_{yz}^0\right] + u_{xz}^0\left[\left(\frac{x}{l_0}\right)\left(\frac{z}{l_0}\right) - C_{xz}^0\right]$$
(3.1)

The terms $C_{xx}^0$, $C_{yy}^0$, $C_{zz}^0$, $C_{xy}^0$, $C_{yz}^0$ and $C_{xz}^0$ are just a generalization of the moments of inertia as defined in elementary classical mechanics. They are to be defined about the centroid (center of mass) of zone "0" and a few of them are explicitly catalogued below:-

$$C_{xx}^0 = \frac{1}{|V_0|}\iiint_{V_0}\left(\frac{x}{l_0}\right)^2 dx\, dy\, dz \quad ; \quad C_{yy}^0 = \frac{1}{|V_0|}\iiint_{V_0}\left(\frac{y}{l_0}\right)^2 dx\, dy\, dz \quad ;$$
$$C_{xy}^0 = \frac{1}{|V_0|}\iiint_{V_0}\left(\frac{x}{l_0}\right)\left(\frac{y}{l_0}\right) dx\, dy\, dz \quad ; \quad \text{with } |V_0| \equiv \iiint_{V_0} dx\, dy\, dz$$
(3.2)

Here $V_0$ is the volume of the zone "0". The moments, like the ones described above, are built once and stored for all the zones of the mesh. The previous definitions are easy to generalize and their generalization is given in Balsara *et al.* (2019). Note that in our usage, the moments do not depend on the fluid density within a zone but only respond to the geometry of the zone being considered. The above moments should be evaluated relative to the centroid (center of mass) of the zone in question. The centroid for a zone is the unique location within the zone for which we have:-

$$\frac{1}{|V_0|}\iiint_{V_0}\left(\frac{x}{l_0}\right)dx\,dy\,dz=0\quad;\quad\frac{1}{|V_0|}\iiint_{V_0}\left(\frac{y}{l_0}\right)dx\,dy\,dz=0\quad;\quad\frac{1}{|V_0|}\iiint_{V_0}\left(\frac{z}{l_0}\right)dx\,dy\,dz=0 \qquad (3.3)$$

The above equation serves to define the centroid of a zone.

The coefficients of eqn. (3.1) can be satisfied for a given stencil by making the following demand. Let "$j$" be one of the zones in the stencil that we are considering; and let that zone have a characteristic length $l_j$. We then demand that when eqn. (3.1) is volume-averaged over the volume of zone "$j$" we should retrieve the zone-averaged flow variable $\bar{u}^j$ for zone "$j$". This is an extension to any type of mesh of the concept of "reconstruction by primitive" first advocated in Colella & Woodward (1984) and Woodward & Colella (1984). Our extension of the Parallel Axis Theorem enables us to write that condition very simply as:-

$$u_x^0\left[\left(\frac{x_j}{l_0}\right)\right]+u_y^0\left[\left(\frac{y_j}{l_0}\right)\right]+u_z^0\left[\left(\frac{z_j}{l_0}\right)\right]+u_{xx}^0\left[\left(\frac{x_j}{l_0}\right)^2-C_{xx}^0+\left(\frac{l_j}{l_0}\right)^2 C_{xx}^j\right]+u_{yy}^0\left[\left(\frac{y_j}{l_0}\right)^2-C_{yy}^0+\left(\frac{l_j}{l_0}\right)^2 C_{yy}^j\right]$$

$$+u_{zz}^0\left[\left(\frac{z_j}{l_0}\right)^2-C_{zz}^0+\left(\frac{l_j}{l_0}\right)^2 C_{zz}^j\right]+u_{xy}^0\left[\left(\frac{x_j}{l_0}\right)\left(\frac{y_j}{l_0}\right)-C_{xy}^0+\left(\frac{l_j}{l_0}\right)^2 C_{xy}^j\right]$$

$$+u_{yz}^0\left[\left(\frac{y_j}{l_0}\right)\left(\frac{z_j}{l_0}\right)-C_{yz}^0+\left(\frac{l_j}{l_0}\right)^2 C_{yz}^j\right]+u_{xz}^0\left[\left(\frac{x_j}{l_0}\right)\left(\frac{z_j}{l_0}\right)-C_{xz}^0+\left(\frac{l_j}{l_0}\right)^2 C_{xz}^j\right]=\bar{u}^j-\bar{u}^0$$

(3.4)

In the above equation, $(x_j, y_j, z_j)$ gives the components of the displacement vector from the centroid of zone "0" to the centroid of zone "$j$". Because the moments of inertia for zones "0" and "$j$" have been pre-computed once and for all, all the terms in the square brackets of the above equation are easy to evaluate. The above equation, therefore, becomes one linear equation for the reconstruction coefficients in eqn. (2.1). If the stencil being considered has sufficiently many such zones "$j$", we get a system of linear equations. For a general stencil, we may have an overdetermined system with more equations than unknowns. The standard trick is to obtain the reconstruction coefficients in eqn. (2.1) via a process of least squares minimization. This gives us the higher order reconstructed polynomial for the stencil of neighboring zones that is being considered. Since the WENO-AO algorithm non-linearly combines the reconstructed solution from a sequence of suitably-chosen stencils, the reconstruction coefficients can be obtained for each of those stencils.

Sections II and III of Balsara *et al*. (2019) show how a smoothness indicator can be constructed for several stencils that can cover the zone "0" that is being considered. This can be used in a non-linearly hybridized fashion to obtain the WENO-AO reconstruction within that zone. The resulting reconstructed polynomial looks just like eqn. (3.1); however, the coefficients from the non-linear hybridization will be such as to produce a non-oscillatory reconstruction. Please see Section V of Balsara *et al*. (2019) for implementation-related details.

**IV) Higher Order Divergence-Free Reconstruction of the Magnetic Field for Isoparametrically Mapped Meshes**

The previous section has shown us how the zone-centered variables can always be reconstructed with high accuracy. In Balsara & Dumbser (2015a) a strategy was found for the divergence-free reconstruction of a constraint-preserving vector field that capitalizes on this high quality finite volume reconstruction. Balsara & Dumbser (2015a) did not document the extension of their reconstruction strategy to isoparametrically mapped meshes with curved boundaries; though the core ideas for doing so are implicit in that paper. Here we present a more economical version of that strategy which extends to isoparametrically mapped meshes with curved boundaries. The strategy is based on realizing that the full set of eight zone-centered conserved variables in eqn. (2.2) can be updated in a finite volume sense in the course of a timestep. Of course, the first five components of the vector of conserved variables are indeed the fluid variables; they are also the primal variables of the scheme and are updated as such. But the last three components of the vector of conserved variables are made up of the three zone-averaged magnetic fields. They are not the primal variables of the scheme and they can be used as auxiliary or helping variables. In other words, the primal variables for the magnetic field still remain the facially averaged, divergence-free components shown in Fig. 5. However, for a single timestep, the zone-centered magnetic field is indeed quite a good representation of the magnetic field as long as it is eventually regulated (overwritten) by the facially averaged primal magnetic fields. So the approach of Balsara & Dumbser (2015a) consists of first carrying out volume-based WENO-AO reconstruction on these zone-centered magnetic fields. This reconstruction at least has the virtue of being non-linearly hybridized; and we will soon put that beneficial attribute to good use.

Realize though that the actual divergence-free reconstruction of the magnetic field within each zone should be such that it matches the mean value of the facial magnetic field components and also as many moments of the magnetic field as possible within each face. Furthermore, since the magnetic field is divergence-free, the reconstructed magnetic field within each zone should also be divergence free. However, recall that the volumetrically reconstructed magnetic field from the previous paragraph does indeed have the virtue of being non-linearly hybridized. For the divergence-free reconstruction to also inherit that property, we require that the coefficients of the divergence-free reconstruction should be as close as possible to the coefficients of the volumetrically reconstructed magnetic field. This paragraph, therefore, provides a verbal sketch of the desirable attributes of a divergence-free reconstruction of the magnetic field. The picture can only be made precise if we illustrate it concretely at a given order. We do that in the subsequent paragraphs by focusing on the third order case in Sub-section IV.a. The description is sufficiently illustrative and can be extended to all orders. Sub-section IV.b provides a pointwise synopsis of the strategy that is suitable for implementation.

## IV.a) Instantiation of the Divergence-Free Reconstruction of the Magnetic Field at Third Order

Let us consider the divergence-free reconstruction of the magnetic field at third order. Within each zone we use the same local coordinate system that was used in Section III. We focus on zone "0". The coordinate system is centered at the centroid of each zone. The divergence-free $x$-component of the magnetic field should have all the moments associated with eqn. (3.1) in order to retain third order of accuracy. However, it will also have some additional moments in order to ensure that the magnetic field is divergence-free. The same is true for the other components.

Extending Balsara (2009) or Balsara *et al*. (2019) to use the same Taylor series basis as was used in eqn. (3.1), we can write the divergence-free $x$-component of the magnetic field as

$$B^x(x, y, z) = a_0 + a_x\left(\frac{x}{l_0}\right) + a_y\left(\frac{y}{l_0}\right) + a_z\left(\frac{z}{l_0}\right) + a_{xx}\left[\left(\frac{x}{l_0}\right)^2 - C_{xx}^0\right] + a_{yy}\left[\left(\frac{y}{l_0}\right)^2 - C_{yy}^0\right]$$

$$+ a_{zz}\left[\left(\frac{z}{l_0}\right)^2 - C_{zz}^0\right] + a_{xy}\left[\left(\frac{x}{l_0}\right)\left(\frac{y}{l_0}\right) - C_{xy}^0\right] + a_{yz}\left[\left(\frac{y}{l_0}\right)\left(\frac{z}{l_0}\right) - C_{yz}^0\right] + a_{xz}\left[\left(\frac{x}{l_0}\right)\left(\frac{z}{l_0}\right) - C_{xz}^0\right]$$

$$+ a_{xxx}\left[\left(\frac{x}{l_0}\right)^3 - C_{xxx}^0\right] + a_{xxy}\left[\left(\frac{x}{l_0}\right)^2\left(\frac{y}{l_0}\right) - C_{xxy}^0\right] + a_{xxz}\left[\left(\frac{x}{l_0}\right)^2\left(\frac{z}{l_0}\right) - C_{xxz}^0\right]$$

$$+ a_{xyy}\left[\left(\frac{x}{l_0}\right)\left(\frac{y}{l_0}\right)^2 - C_{xyy}^0\right] + a_{xzz}\left[\left(\frac{x}{l_0}\right)\left(\frac{z}{l_0}\right)^2 - C_{xzz}^0\right] + a_{xyz}\left[\left(\frac{x}{l_0}\right)\left(\frac{y}{l_0}\right)\left(\frac{z}{l_0}\right) - C_{xyz}^0\right]$$

(4.1)

Notice that the coefficients in eqn. (4.1) are expanded in exactly the same Taylor series basis functions as eqn. (3.1). Let the non-linearly hybridized WENO-AO reconstruction of the zone-centered *x*-component of the magnetic field be explicitly written as

$$\bar{\bar{B}}^x(x, y, z) = \bar{\bar{a}}_0 + \bar{\bar{a}}_x\left(\frac{x}{l_0}\right) + \bar{\bar{a}}_y\left(\frac{y}{l_0}\right) + \bar{\bar{a}}_z\left(\frac{z}{l_0}\right) + \bar{\bar{a}}_{xx}\left[\left(\frac{x}{l_0}\right)^2 - C_{xx}^0\right] + \bar{\bar{a}}_{yy}\left[\left(\frac{y}{l_0}\right)^2 - C_{yy}^0\right]$$

$$+ \bar{\bar{a}}_{zz}\left[\left(\frac{z}{l_0}\right)^2 - C_{zz}^0\right] + \bar{\bar{a}}_{xy}\left[\left(\frac{x}{l_0}\right)\left(\frac{y}{l_0}\right) - C_{xy}^0\right] + \bar{\bar{a}}_{yz}\left[\left(\frac{y}{l_0}\right)\left(\frac{z}{l_0}\right) - C_{yz}^0\right] + \bar{\bar{a}}_{xz}\left[\left(\frac{x}{l_0}\right)\left(\frac{z}{l_0}\right) - C_{xz}^0\right]$$

(4.2)

In order for the coefficients of the divergence-free *x*-component of the magnetic field from eqn. (4.1) to be as close as possible to the non-linearly limited coefficients from eqn. (4.2) we require that the following 16 equations should be minimized in a least squares sense

$$a_0 = \bar{\bar{a}}_0 \ ; \ a_x = \bar{\bar{a}}_x \ ; \ a_y = \bar{\bar{a}}_y \ ; \ a_z = \bar{\bar{a}}_z \ ; \ a_{xx} = \bar{\bar{a}}_{xx} \ ; \ a_{yy} = \bar{\bar{a}}_{yy} \ ; \ a_{zz} = \bar{\bar{a}}_{zz} \ ; \ a_{xy} = \bar{\bar{a}}_{xy} \ ;$$

$$a_{yz} = \bar{\bar{a}}_{yz} \ ; \ a_{xz} = \bar{\bar{a}}_{xz} \ ; \ a_{xxx} = 0 \ ; \ a_{xxy} = 0 \ ; \ a_{xxz} = 0 \ ; \ a_{xyy} = 0 \ ; \ a_{xzz} = 0 \ ; \ a_{xyz} = 0$$

(4.3)

The divergence-free *y*-component of the magnetic field can be written as

$$B^y(x,y,z) = b_0 + b_x\left(\frac{x}{l_0}\right) + b_y\left(\frac{y}{l_0}\right) + b_z\left(\frac{z}{l_0}\right) + b_{xx}\left[\left(\frac{x}{l_0}\right)^2 - C_{xx}^0\right] + b_{yy}\left[\left(\frac{y}{l_0}\right)^2 - C_{yy}^0\right]$$

$$+ b_{zz}\left[\left(\frac{z}{l_0}\right)^2 - C_{zz}^0\right] + b_{xy}\left[\left(\frac{x}{l_0}\right)\left(\frac{y}{l_0}\right) - C_{xy}^0\right] + b_{yz}\left[\left(\frac{y}{l_0}\right)\left(\frac{z}{l_0}\right) - C_{yz}^0\right] + b_{xz}\left[\left(\frac{x}{l_0}\right)\left(\frac{z}{l_0}\right) - C_{xz}^0\right]$$

$$+ b_{yyy}\left[\left(\frac{y}{l_0}\right)^3 - C_{yyy}^0\right] + b_{xyy}\left[\left(\frac{x}{l_0}\right)\left(\frac{y}{l_0}\right)^2 - C_{xyy}^0\right] + b_{yyz}\left[\left(\frac{y}{l_0}\right)^2\left(\frac{z}{l_0}\right) - C_{yyz}^0\right]$$

$$+ b_{xxy}\left[\left(\frac{x}{l_0}\right)^2\left(\frac{y}{l_0}\right) - C_{xxy}^0\right] + b_{yzz}\left[\left(\frac{y}{l_0}\right)\left(\frac{z}{l_0}\right)^2 - C_{yzz}^0\right] + b_{xyz}\left[\left(\frac{x}{l_0}\right)\left(\frac{y}{l_0}\right)\left(\frac{z}{l_0}\right) - C_{xyz}^0\right]$$

(4.4)

Let the non-linearly hybridized WENO-AO reconstruction of the zone-centered *y*-component of the magnetic field be explicitly written as

$$\bar{\bar{B}}^y(x,y,z) = \bar{\bar{b}}_0 + \bar{\bar{b}}_x\left(\frac{x}{l_0}\right) + \bar{\bar{b}}_y\left(\frac{y}{l_0}\right) + \bar{\bar{b}}_z\left(\frac{z}{l_0}\right) + \bar{\bar{b}}_{xx}\left[\left(\frac{x}{l_0}\right)^2 - C_{xx}^0\right] + \bar{\bar{b}}_{yy}\left[\left(\frac{y}{l_0}\right)^2 - C_{yy}^0\right]$$

$$+ \bar{\bar{b}}_{zz}\left[\left(\frac{z}{l_0}\right)^2 - C_{zz}^0\right] + \bar{\bar{b}}_{xy}\left[\left(\frac{x}{l_0}\right)\left(\frac{y}{l_0}\right) - C_{xy}^0\right] + \bar{\bar{b}}_{yz}\left[\left(\frac{y}{l_0}\right)\left(\frac{z}{l_0}\right) - C_{yz}^0\right] + \bar{\bar{b}}_{xz}\left[\left(\frac{x}{l_0}\right)\left(\frac{z}{l_0}\right) - C_{xz}^0\right]$$

(4.5)

In order for the coefficients of the divergence-free *y*-component of the magnetic field from eqn. (4.4) to be as close as possible to the non-linearly limited coefficients from eqn. (4.5) we require that the following 16 equations should be minimized in a least squares sense

$$b_0 = \bar{\bar{b}}_0 \;;\; b_x = \bar{\bar{b}}_x \;;\; b_y = \bar{\bar{b}}_y \;;\; b_z = \bar{\bar{b}}_z \;;\; b_{xx} = \bar{\bar{b}}_{xx} \;;\; b_{yy} = \bar{\bar{b}}_{yy} \;;\; b_{zz} = \bar{\bar{b}}_{zz} \;;\; b_{xy} = \bar{\bar{b}}_{xy} \;;$$
$$b_{yz} = \bar{\bar{b}}_{yz} \;;\; b_{xz} = \bar{\bar{b}}_{xz} \;;\; b_{yyy} = 0 \;;\; b_{xyy} = 0 \;;\; b_{yyz} = 0 \;;\; b_{xxy} = 0 \;;\; b_{yzz} = 0 \;;\; b_{xyz} = 0$$

(4.6)

The divergence-free *z*-component of the magnetic field can be written as

$$B^z(x,y,z) = c_0 + c_x\left(\frac{x}{l_0}\right) + c_y\left(\frac{y}{l_0}\right) + c_z\left(\frac{z}{l_0}\right) + c_{xx}\left[\left(\frac{x}{l_0}\right)^2 - C^0_{xx}\right] + c_{yy}\left[\left(\frac{y}{l_0}\right)^2 - C^0_{yy}\right]$$

$$+ c_{zz}\left[\left(\frac{z}{l_0}\right)^2 - C^0_{zz}\right] + c_{xy}\left[\left(\frac{x}{l_0}\right)\left(\frac{y}{l_0}\right) - C^0_{xy}\right] + c_{yz}\left[\left(\frac{y}{l_0}\right)\left(\frac{z}{l_0}\right) - C^0_{yz}\right] + c_{xz}\left[\left(\frac{x}{l_0}\right)\left(\frac{z}{l_0}\right) - C^0_{xz}\right]$$

$$+ c_{zzz}\left[\left(\frac{z}{l_0}\right)^3 - C^0_{zzz}\right] + c_{xzz}\left[\left(\frac{x}{l_0}\right)\left(\frac{z}{l_0}\right)^2 - C^0_{xzz}\right] + c_{yzz}\left[\left(\frac{y}{l_0}\right)\left(\frac{z}{l_0}\right)^2 - C^0_{yzz}\right]$$

$$+ c_{xxz}\left[\left(\frac{x}{l_0}\right)^2\left(\frac{z}{l_0}\right) - C^0_{xxz}\right] + c_{yyz}\left[\left(\frac{y}{l_0}\right)^2\left(\frac{z}{l_0}\right) - C^0_{yyz}\right] + c_{xyz}\left[\left(\frac{x}{l_0}\right)\left(\frac{y}{l_0}\right)\left(\frac{z}{l_0}\right) - C^0_{xyz}\right]$$

(4.7)

Let the non-linearly hybridized WENO-AO reconstruction of the zone-centered $z$-component of the magnetic field be explicitly written as

$$\bar{\bar{B}}^z(x,y,z) = \bar{\bar{c}}_0 + \bar{\bar{c}}_x\left(\frac{x}{l_0}\right) + \bar{\bar{c}}_y\left(\frac{y}{l_0}\right) + \bar{\bar{c}}_z\left(\frac{z}{l_0}\right) + \bar{\bar{c}}_{xx}\left[\left(\frac{x}{l_0}\right)^2 - C^0_{xx}\right] + \bar{\bar{c}}_{yy}\left[\left(\frac{y}{l_0}\right)^2 - C^0_{yy}\right]$$

$$+ \bar{\bar{c}}_{zz}\left[\left(\frac{z}{l_0}\right)^2 - C^0_{zz}\right] + \bar{\bar{c}}_{xy}\left[\left(\frac{x}{l_0}\right)\left(\frac{y}{l_0}\right) - C^0_{xy}\right] + \bar{\bar{c}}_{yz}\left[\left(\frac{y}{l_0}\right)\left(\frac{z}{l_0}\right) - C^0_{yz}\right] + \bar{\bar{c}}_{xz}\left[\left(\frac{x}{l_0}\right)\left(\frac{z}{l_0}\right) - C^0_{xz}\right]$$

(4.8)

In order for the coefficients of the divergence-free $z$-component of the magnetic field from eqn. (4.7) to be as close as possible to the non-linearly limited coefficients from eqn. (4.8) we require that the following 16 equations should be minimized in a least squares sense

$$c_0 = \bar{\bar{c}}_0 \ ; \ c_x = \bar{\bar{c}}_x \ ; \ c_y = \bar{\bar{c}}_y \ ; \ c_z = \bar{\bar{c}}_z \ ; \ c_{xx} = \bar{\bar{c}}_{xx} \ ; \ c_{yy} = \bar{\bar{c}}_{yy} \ ; \ c_{zz} = \bar{\bar{c}}_{zz} \ ; \ c_{xy} = \bar{\bar{c}}_{xy} \ ;$$
$$c_{yz} = \bar{\bar{c}}_{yz} \ ; \ c_{xz} = \bar{\bar{c}}_{xz} \ ; \ c_{zzz} = 0 \ ; \ c_{xzz} = 0 \ ; \ c_{yzz} = 0 \ ; \ c_{xxz} = 0 \ ; \ c_{yyz} = 0 \ ; \ c_{xyz} = 0$$

(4.9)

The number of coefficients that are represented in eqns. (4.1), (4.4) and (4.7) have been strategically determined to be the smallest number that is required. That is not the case for the formulation in Balsara & Dumbser (2015a), with the result that the present formulation is somewhat more efficient.

When the divergence-free condition given by
$\partial_x B^x(x,y,z) + \partial_y B^y(x,y,z) + \partial_z B^z(x,y,z) = 0$ is applied to eqns. (4.1), (4.4) and (4.7), the constraints are given by

$$a_x + b_y + c_z = 0 \; ; \; 2a_{xx} + b_{xy} + c_{xz} = 0 \; ; \; a_{xy} + 2b_{yy} + c_{yz} = 0 \; ; \; a_{xz} + b_{yz} + 2c_{zz} = 0 \; ;$$
$$3a_{xxx} + b_{xxy} + c_{xxz} = 0 \; ; \; a_{xyy} + 3b_{yyy} + c_{yyz} = 0 \; ; \; a_{xzz} + b_{yzz} + 3c_{zzz} = 0 \; ; \qquad (4.10)$$
$$a_{xyz} + 2b_{yyz} + 2c_{yzz} = 0 \; ; \; 2a_{xxz} + b_{xyz} + 2c_{xzz} = 0 \; ; \; 2a_{xxy} + 2b_{xyy} + c_{xyz} = 0$$

The first of the constraints is equivalent to the fact that the sum of the magnetic flux evaluated over all the faces of a zone equals zero. Therefore, it is not used. We require that the nine remaining constraints in eqn. (4.10) should be satisfied exactly.

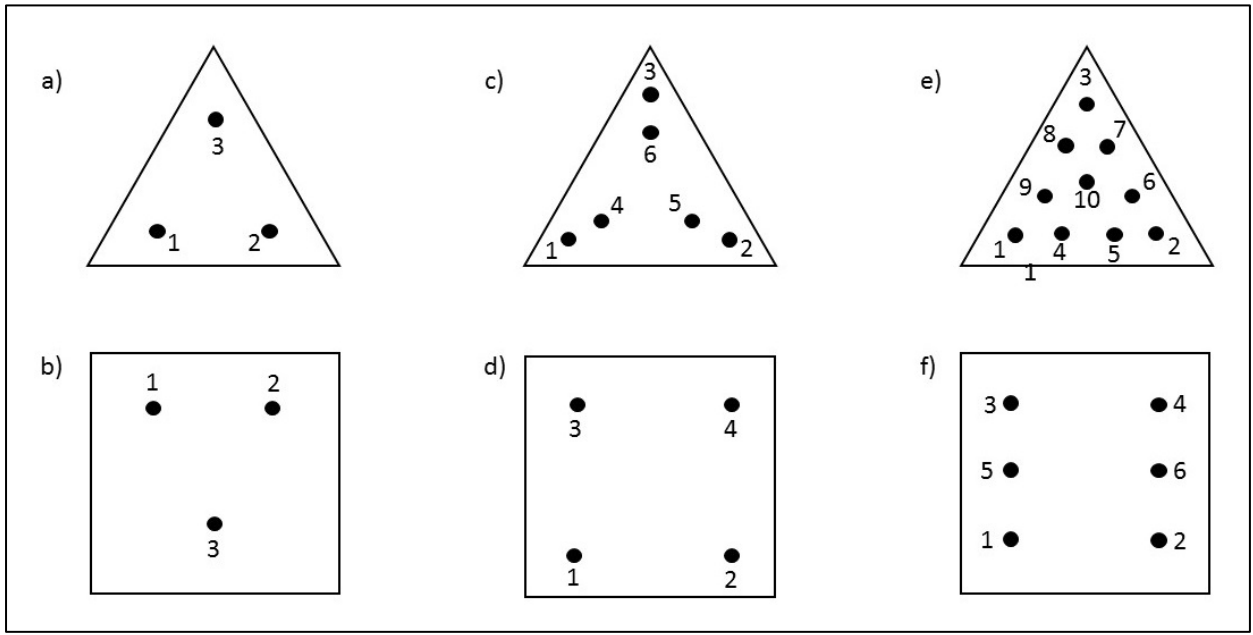

*Figs. 6a and 6b show the areal quadrature points on the reference equilateral triangle and a reference square at second order. Figs. 6c and 6d show the same at third order. Figs. 6e and 6f show the same at fourth order.*

For each face of the reference element, we can identify a set of areal quadrature points at that face. The faces of the reference element will be unit squares and equilateral triangles with unit sides. Figs. 6a and 6b show the areal quadrature points on the reference equilateral triangle and a reference square at second order. Figs. 6c and 6d show the same at third order. Figs. 6e and 6f show the same at fourth order. (If one wishes, a more efficient six-point quadrature rule is available from Dunavant 1985.) Appendix A gives the locations of the nodes within each of those faces as well as their weights. Isoparametric mapping can then be used to find the location of the same nodes on all the faces of the zones as well as their weights. The isoparametric mapping can also give us the direction of the unit outward pointing normal at each nodal point on each face of each zone.

Now consider one of the faces, labeled "$j$", of our mesh. A facially-averaged magnetic field component is collocated at that face because of the Yee-type mesh staggering. This facially-averaged normal component of the magnetic field, denoted by $\langle B_\perp^j \rangle$, is the primal variable in our mesh, so we should preserve its value all through the divergence-free reconstruction process. The face "$j$" will have two zones on either side of it. Eqns. (4.2), (4.5) and (4.8) give us the non-linearly hybridized, WENO-reconstruction-based magnetic fields in each of those two zones. Let us denote these two magnetic fields with subscripts "$L$" and "$R$". Therefore, at any point $(x, y, z)$ in any face, we can evaluate two magnetic fields $\bar{\bar{\mathbf{B}}}_L(x, y, z)$ and $\bar{\bar{\mathbf{B}}}_R(x, y, z)$ from the two zones that come together at that face. Since $\bar{\bar{\mathbf{B}}}_L(x, y, z)$ and $\bar{\bar{\mathbf{B}}}_R(x, y, z)$ have all the non-linearly hybridized modal information, they can potentially give us the non-linearly hybridized higher order modes in the face. In general, those two fields will not integrate to the facially-averaged normal component of the magnetic field, $\langle B_\perp^j \rangle$, at that face (which is indeed the primal variable of the scheme). Neither will those left-sided and right-sided values match at the boundary (though if the solution is very smooth, the two values from the two abutting sides will be pretty close). Our task in the next few paragraphs will be to make those two values consistent at each face and then use them for the divergence-free magnetic field reconstruction in the zone being considered.

At the face "$j$", Fig. 6 shows us that we will have $N_j$ facial nodes. The set of locations for those nodes is given by $\{(x_i^j, y_i^j, z_i^j) : i = 1, ..., N_j\}$; their corresponding areal quadrature weights are given by $\{w_i^j : i = 1, ..., N_j\}$ and the set of unit, normals at each of those nodes is given by $\{\hat{\mathbf{n}}_i^j : i = 1, ..., N_j\}$. Our method is designed to handle elements that may have curved faces. At each of the $N_j$ facial nodes for face "$j$", we define normal components to that face given by $\{B_{\perp;i}^j : i = 1,...,N_j\}$. These normal components are so designed that they pick out the smaller of the one-sided variations in the normal component at that facial nodal point. This is done by the following limiting procedure:

$$B_{\perp;i}^j = \langle B_{\perp}^j \rangle + MinMod\left(\hat{\mathbf{n}}_i^j \cdot \bar{\bar{\mathbf{B}}}_L(x_i, y_i, z_i) - \langle B_{\perp}^j \rangle, \hat{\mathbf{n}}_i^j \cdot \bar{\bar{\mathbf{B}}}_R(x_i, y_i, z_i) - \langle B_{\perp}^j \rangle\right) \quad \text{for} \ i = 1,...,N_j \quad (4.11)$$

Now realize that our set of normal components at the nodes, given by $\{B_{\perp;i}^j : i = 1,...,N_j\}$ from eqn. (4.11), will not integrate to $\langle B_{\perp}^j \rangle$. Since consistency with the primal variable should be an essential requirement in the scheme, we reset

$$B_{\perp;i}^j \to B_{\perp;i}^j + \left[\langle B_{\perp}^j \rangle - \left(\sum_{i=1}^{N_j} w_i^j B_{\perp;i}^j\right)\right] \quad (4.12)$$

After application of eqn. (4.12) we can be sure that the limited nodal values from eqn. (4.11) will also have the right area-weighted mean value:

$$\sum_{i=1}^{N_j} w_i^j B_{\perp;i}^j = \langle B_{\perp}^j \rangle \quad (4.13)$$

In the next two paragraphs, we will show how $\{B_{\perp;i}^j : i = 1,...,N_j\}$ can be used to carry out divergence-free reconstruction of magnetic fields within a zone that has some rather nice properties.

Our first requirement is that the divergence-free magnetic fields within a zone, given by eqns. (4.1), (4.4) and (4.7), should also integrate to the primal variable $\langle B_{\perp}^j \rangle$ at each face "$j$" that bounds the zone being considered. Thus for each of the five faces that bound the zone being considered we require

$$\sum_{i=1}^{N_j} \left\{w_i^j \ \hat{\mathbf{n}}_i^j \cdot \left[B^x(x_i^j, y_i^j, z_i^j)\hat{\mathbf{x}} + B^y(x_i^j, y_i^j, z_i^j)\hat{\mathbf{y}} + B^z(x_i^j, y_i^j, z_i^j)\hat{\mathbf{z}}\right]\right\} = \langle B_{\perp}^j \rangle \quad (4.14)$$

Explicitly using eqns. (4.1), (4.4) and (4.7) in eqn. (4.14) gives us a linear equation for the coefficients of eqns. (4.1), (4.4) and (4.7). Application of eqn. (4.14) at each of the five faces that bound the zone being considered gives us five constraints that we have to put on the divergence-free reconstruction. These five constraints are in addition to the nine constraints given in eqn. (4.9). We skip the first of the constraints given in eqn. (4.9) because it is linearly dependent with the five conditions shown in eqn. (4.14).

In addition to matching the primal variables, we would like to individually match the nodal values within each face "$j$", i.e. $\{B_{\perp;i}^{j} : i = 1,..., N_j\}$, as closely as possible. Note though that we cannot insist that they match exactly because some of the conditions can be linearly degenerate with eqn. (4.14). Consequently, for each face "$j$", we have the following $N_j$ linear equations

$$\hat{\mathbf{n}}_i^j \cdot \left[ B^x\left(x_i^j, y_i^j, z_i^j\right)\hat{\mathbf{x}} + B^y\left(x_i^j, y_i^j, z_i^j\right)\hat{\mathbf{y}} + B^z\left(x_i^j, y_i^j, z_i^j\right)\hat{\mathbf{z}} \right] = B_{\perp;i}^{j} \quad \text{for } i = 1,..., N_j \qquad (4.15)$$

Since we cannot put these equations in the set of constraints, we can at least demand that they be satisfied up to least squares minimization. Therefore, eqns. (4.15) applied at all the faces of the zone being considered combine with eqns. (4.3), (4.6) and (4.9) to give us a set of equations that need to undergo least squares minimization.

The previous two paragraphs show us that the problem of divergence-free reconstruction is a constrained, least squares minimization problem. Eqns. (4.9) and (4.14) provide the constraints which ensure that the reconstructed magnetic field within the zone of interest will be divergence-free and will match up with the primal magnetic field components in the faces. Eqns. (4.3), (4.6) and (4.9) will be minimized in a least squares sense to ensure that the divergence-free reconstruction that we arrive at will have coefficients that are as close to the non-linearly hybridized modes that we have obtained by our volumetric WENO-AO limiting procedure. Eqns. (4.15) at the facial nodes will also be minimized in a least squares sense to ensure that our divergence-free reconstruction will match as many higher order modes as we can have in the faces of the mesh.

This divergence-free reconstruction is very versatile and goes at least a little beyond the one suggested in Balsara & Dumbser (2015a) because it carefully accounts for isoparametrically-mapped zones with curved boundaries. To the best of our knowledge, this is the first time such a general divergence-free magnetic field reconstruction strategy has been presented for isoparametrically mapped meshes.

## IV.b) Stepwise Description of the Higher Order Divergence-Free Reconstruction of the Magnetic Field

The higher order divergence-free reconstruction follows the logic that was presented in Sub-section IV.a. Here we synopsize it in pointwise form to facilitate easy implementation. The hard step is the assembly of the Karush–Kuhn–Tucker (KKT) matrix for the solution of the constrained least squares problem; but we show that the cost of its assembly and inversion can be amortized over many zones. The steps go as follows:-

**1)** Using Fig. 6 and the facial nodes in the reference triangles and squares given in Appendix A, we use the isoparametric mapping to obtain facial nodes $\{(x_i^j, y_i^j, z_i^j) : i = 1,..., N_j\}$ in all the faces "$j$" of the physical zone that is being considered. We also obtain the corresponding facial weights $\{w_i^j : i = 1,..., N_j\}$ in all the faces "$j$" of the physical zone. We can also obtain the unit, normals $\{\hat{\mathbf{n}}_i^j : i = 1,..., N_j\}$ in all the faces "$j$" of the physical zone. This step only needs to be done once if the nodal locations, weights and normals are stored.

**2)** We apply eqns. (4.11) and (4.12) to obtain the normal components $\{B_{\perp;i}^j : i = 1,..., N_j\}$ in all the faces "$j$" of the physical zone.

**3)** The last nine linear constraints from eqn. (4.10) should be incorporated into the KKT matrix.

**4)** The five integral conditions from eqn. (4.14) are also incorporated into the KKT matrix. They also act as linear constraints.

**5)** Eqns. (4.3), (4.6) and (4.9) should be incorporated as part of the least squares linear system in the KKT matrix.

**6)** The nodal conditions from eqn. (4.15) at each node of each face "$j$" that bounds the zone in question should be incorporated as part of the least squares linear system in the KKT matrix. The assembly of the KKT matrix is now complete.

**7)** The KKT matrix can now be inverted. If one is using logarithmically ratioed meshes, which is the typical use case on a spherical mesh, this inversion of the KKT matrix only needs to be done once along each radial array of zones. See Section IV of Balsara *et al*. (2018) for the reasoning, which is based on the self-similarity of the zones.

**8)** As one proceeds through steps 3), 4), 5) and 6), it is also advisable to assemble the right hand side of our constrained least squares system.

**9)** Using the inverse of the KKT matrix and the right hand side from the previous step, we can obtain the coefficients for the divergence-free reconstruction of the magnetic field in eqns. (4.1), (4.4) and (4.7).

**10)** Overwrite the zone-centered magnetic field variables with the first terms of eqns. (4.1), (4.4) and (4.7). This ensures that the facial primal variables for the magnetic field eventually regulate the zone-centered magnetic field. In other words, the zone-centered magnetic field is made consistent with the primal, facially-averaged, normal components of the magnetic field.

This completes our description of the higher order divergence-free reconstruction of the magnetic field on isoparametrically mapped meshes.

## V) ADER Formulation on Mapped Elements

The ADER scheme we describe here is based on a continuous Galerkin representation in time (also known as ADER-CG). It is not suited for problems having stiff source terms, but most MHD applications are not required to handle stiff source terms. The upshot is that ADER-CG is suitable for our uses here; and it is the scheme that we describe. (The alternative would have been to use a discontinuous Galerkin representation in time resulting in a more expensive ADER-DG scheme which is suitable for use with stiff source terms.)

We split this discussion into four easy parts. In Sub-section V.a we show that the conservation law can be formulated in isoparametrically mapped coordinates. In Sub-section V.b we describe the construction of serendipity bases that are very useful for the construction of efficient ADER schemes on isoparametrically mapped elements. We also describe how the ADER scheme is made even more efficient when logarithmically ratioed meshes are used (see also Koldoba *et al*. 2002). Since this is the usual choice for spherical problems, it makes our ADER scheme even more efficient on such meshes. In Sub-section V.c we describe the formulation of the ADER method in isoparametrically mapped elements. In Sub-section V.d we provide a pointwise description of the ADER scheme, which should simplify its implementation.

## V.a) Formulating the Conservation Law in Isoparametrically Mapped Coordinates

Consider the conservation law

$$\frac{\partial \mathbf{U}}{\partial t} + \frac{\partial \mathbf{F}(\mathbf{U})}{\partial x} + \frac{\partial \mathbf{G}(\mathbf{U})}{\partial y} + \frac{\partial \mathbf{H}(\mathbf{U})}{\partial z} = \mathbf{S}(\mathbf{U}) \tag{5.1}$$

We want to set up an ADER method in a mapped coordinate system. Let $\mathbf{r} = x\hat{\mathbf{x}} + y\hat{\mathbf{y}} + z\hat{\mathbf{z}}$ be the physical coordinate vector. Let $(\xi, \eta, \zeta)$ be the coordinates in a reference element. In our case, the reference element is a triangular prism and the physical zone is a triangular frustum in spherical geometry. To map the curvature of the frustum, we assume that we have $N$ suitably defined nodes $\{\mathbf{r}_i; i = 1,...,N\}$ on the frustum. An isoparametric mapping from reference element to the frustum is defined by

$$\mathbf{r} = \mathbf{r}(\xi, \eta, \zeta) = \sum_{i=1}^{N} \mathbf{r}_i \psi_i(\xi, \eta, \zeta) \tag{5.2}$$

Within the reference element we have a set of $N$ suitably defined nodes $\{(\xi_i, \eta_i, \zeta_i), i = 1,...,N\}$ so that the Lagrange basis functions that define the isoparametric mapping in eqn. (5.2) satisfy

$$\psi_j(\xi_i, \eta_i, \zeta_i) = \delta_{ij} \tag{5.3}$$

Associated with the mapping in eqn. (5.2) we have the coordinate basis vectors

$$\mathbf{h}_\xi = \frac{\partial \mathbf{r}}{\partial \xi} \quad ; \quad \mathbf{h}_\eta = \frac{\partial \mathbf{r}}{\partial \eta} \quad ; \quad \mathbf{h}_\zeta = \frac{\partial \mathbf{r}}{\partial \zeta} \tag{5.4}$$

which do not have to be orthogonal. The above coordinate basis vectors will not be unit vectors in general. Our zones are constructed in such a way that their boundaries (faces) are surfaces of $\xi$, $\eta$ or $\zeta$, or some linear combination thereof.

We can define the three dimensional flux, which is a function of the conserved variables, as $\mathcal{F}(\mathbf{U}) = \mathbf{F}(\mathbf{U})\hat{\mathbf{x}} + \mathbf{G}(\mathbf{U})\hat{\mathbf{y}} + \mathbf{H}(\mathbf{U})\hat{\mathbf{z}}$. The volume integrals over the physical space transform into the following volume integrals in the reference space as follows

$$\int \mathbf{U}(x,y,z) dx\, dy\, dz = \int \mathbf{U}(\xi,\eta,\zeta) \left| (\mathbf{h}_\xi \times \mathbf{h}_\eta) \cdot \mathbf{h}_\zeta \right| d\xi\, d\eta\, d\zeta \tag{5.5}$$

In our case, the Jacobian $J \equiv \left| (\mathbf{h}_\xi \times \mathbf{h}_\eta) \cdot \mathbf{h}_\zeta \right|$ is independent of time; though it can vary with space. Taking $d\mathbf{A}_{\xi_0}$ to be an area element (vector) in the surface given by $\xi = \xi_0$ we can write

$$\int \mathcal{F}(x,y,z) \cdot d\mathbf{A}_{\xi_0} = \int (\mathbf{h}_\eta \times \mathbf{h}_\zeta) \cdot \mathcal{F}(\xi = \xi_0, \eta, \zeta) d\eta\, d\zeta \tag{5.6}$$

Similarly, taking $d\mathbf{A}_{\eta_0}$ to be an area element (vector) in the surface given by $\eta = \eta_0$ we can write

$$\int \mathcal{F}(x,y,z) \cdot d\mathbf{A}_{\eta_0} = \int (\mathbf{h}_\zeta \times \mathbf{h}_\xi) \cdot \mathcal{F}(\xi, \eta = \eta_0, \zeta) d\xi\, d\zeta \tag{5.7}$$

Likewise, taking $d\mathbf{A}_{\zeta_0}$ to be an area element (vector) in the surface given by $\zeta = \zeta_0$ we can write

$$\int \mathcal{F}(x,y,z) \cdot d\mathbf{A}_{\zeta_0} = \int (\mathbf{h}_\xi \times \mathbf{h}_\eta) \cdot \mathcal{F}(\xi, \eta, \zeta = \zeta_0) d\xi\, d\eta \tag{5.8}$$

The result is that the governing equation can be written as

$$\frac{\partial \mathbf{U}}{\partial t} + \frac{1}{J} \left\{ \frac{\partial \left[ (\mathbf{h}_\eta \times \mathbf{h}_\zeta) \cdot \mathcal{F}(\mathbf{U}) \right]}{\partial \xi} + \frac{\partial \left[ (\mathbf{h}_\zeta \times \mathbf{h}_\xi) \cdot \mathcal{F}(\mathbf{U}) \right]}{\partial \eta} + \frac{\partial \left[ (\mathbf{h}_\xi \times \mathbf{h}_\eta) \cdot \mathcal{F}(\mathbf{U}) \right]}{\partial \xi} \right\} = \mathbf{S}(\mathbf{U}) \tag{5.9}$$

Utilizing the time-independence of the Jacobian, the above equation can be written as

$$\frac{\partial \widehat{\mathbf{U}}}{\partial t} + \frac{\partial \widehat{\mathbf{F}}(\mathbf{U})}{\partial \xi} + \frac{\partial \widehat{\mathbf{G}}(\mathbf{U})}{\partial \eta} + \frac{\partial \widehat{\mathbf{H}}(\mathbf{U})}{\partial \zeta} = \widehat{\mathbf{S}} \tag{5.10}$$

with the definitions

$$\widehat{\mathbf{U}} \equiv J\, \mathbf{U} \;\; ; \;\; \widehat{\mathbf{F}}(\mathbf{U}) \equiv (\mathbf{h}_\eta \times \mathbf{h}_\zeta) \cdot \mathcal{F}(\mathbf{U}) \;\; ; \;\; \widehat{\mathbf{G}}(\mathbf{U}) \equiv (\mathbf{h}_\zeta \times \mathbf{h}_\xi) \cdot \mathcal{F}(\mathbf{U}) \;\; ;$$
$$\widehat{\mathbf{H}}(\mathbf{U}) \equiv (\mathbf{h}_\xi \times \mathbf{h}_\eta) \cdot \mathcal{F}(\mathbf{U}) \;\; ; \;\; \widehat{\mathbf{S}} = J\, \mathbf{S}(\mathbf{U}) \tag{5.11}$$

Notice the analogy between eqn. (5.10) and eqn. (5.1). Eqn. (5.10) suggests a simple ADER solution strategy. The idea would be to formulate the ADER solution strategy in the reference element using eqn. (5.10). Such an ADER scheme will naturally include the curvature terms in the isoparametric mapping by way of the flux and Jacobian definitions in eqn. (5.11).

Unlike the ADER formulation in Boscheri & Dumbser (2013, 2016, 2017) this ADER formulation achieves its simplification because we exploit the time-independence of the Jacobian.

**V.b) Serendipity Basis for Spherical Meshes and Efficient Processing on Logarithmically Ratioed Meshes**

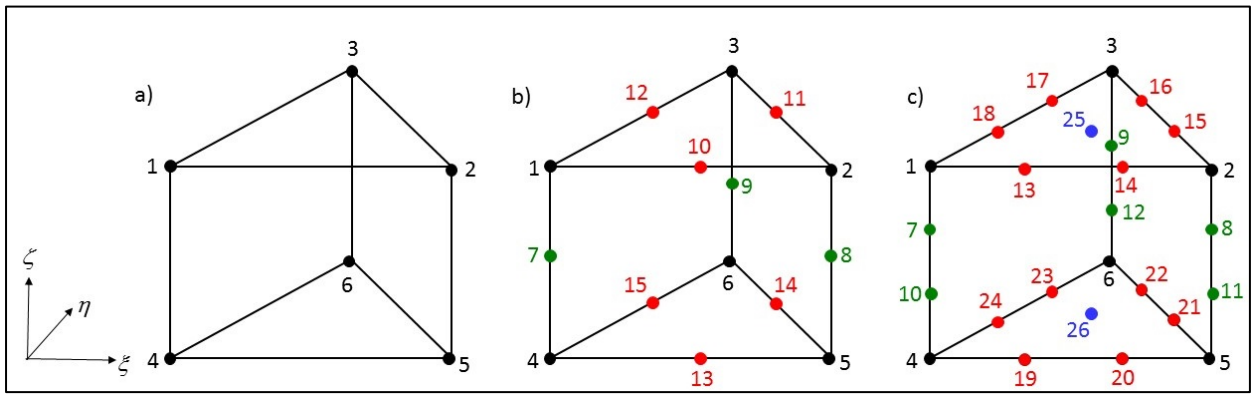

*Fig. 7 shows the nodes on the reference element consisting of a triangular prism at (a) second, (b) third and (c) fourth orders. The nodal points for the serendipity elements are shown, with the result that there are far fewer nodes than one would have in a tensor product element. Nodes at vertices are shown in black; nodes within the edges of triangles are shown in red; nodes within vertical edges are shown in green and nodes at the centroids of triangular faces are shown in blue. The edges are bisected/trisected in equidistant fashion in Figs. 7b and 7c.*

For simplicial elements, (triangles and tetrahedral) the element topology is such that one always obtains elements with the smallest number of nodal points. This fact was exploited to obtain very efficient nodal-based ADER schemes in Dumbser *et al*. (2008). However, this is not the case for spherical meshes. Using tensor product nodal points might work, but it would result in ADER schemes that utilize many more basis functions than the ones that are minimally needed for the order property. For this reason, we follow Zienkiewicz & Taylor (2000) and use the serendipity elements documented in their Fig. 8.23. A similar figure, with some very helpful colorization, is shown in Fig. 7 of this paper. Fig. 7 shows the nodes on the reference element consisting of a triangular prism at (a) second, (b) third and (c) fourth orders. The nodal points for the serendipity elements are shown, with the result that there are far fewer nodes than one would have in a tensor product element. Nodes at vertices are shown in black; nodes within the edges of triangles are shown in red; nodes within vertical edges are shown in green and nodes at the centroids of triangular faces are shown in blue. The edges are bisected/trisected in equidistant fashion in Figs. 7b and 7c. Notice that the second order reference element has 6 nodes instead of the maximum number of 8 which arises when tensor product bases are used. Similarly, the third order reference element has 15 nodes instead of a maximum of 27 nodes that would be obtained in a tensor product basis. Likewise, the fourth order reference element has 26 nodes instead of the maximum of 64 that would be needed in a tensor product basis. Using a smaller number of nodes results in a dramatic simplification and a substantial increase in the speed of the ADER algorithm. We would urge the usage of serendipity elements even in ADER schemes for structured, logically Cartesian meshes because that would result in substantial reduction in computational complexity; see Dumbser *et al*. (2013).

Once the nodes are specified on the reference triangular prism, and once a suitable polynomial expansion is chosen, it is possible to fully specify the Lagrange basis functions from eqn. (5.3). In Fig. 7, the nodes along each edge are always chosen to be equidistant and that is sufficient to specify the location of all the nodes on any reference triangular prism. Given the set of $N$ nodal points, $\{(\xi_i, \eta_i, \zeta_i), i = 1, ..., N\}$, a computer algebra system can easily be used to discover the coefficients of the Lagrange basis functions that have the property specified in eqn. (5.3). We provide the explicit form of the basis functions that should be used at various orders in the ensuing paragraphs.

At second order, the generic basis function has the form

$$\psi(\xi, \eta, \zeta) = a_{000} + a_{100}\xi + a_{010}\eta + a_{001}\zeta + a_{101}\xi\zeta + a_{011}\eta\zeta \tag{5.12}$$

Depending on the node chosen from Fig. 7a, eqn. (5.12) specifies one of the six second order Lagrange basis functions. If we make the basis function unity at the chosen node, while requiring it to be zero at the other five nodes, then the specification of the six coefficients in eqn. (5.12) is fully determined by the condition in eqn. (5.3). This process can be repeated for all the nodes in Fig. 7a.

At third order, the generic basis function has the form

$$\psi(\xi,\eta,\zeta) = a_{000} + a_{100}\xi + a_{010}\eta + a_{001}\zeta + a_{200}\xi^2 + a_{020}\eta^2 + a_{002}\zeta^2 + a_{110}\xi\eta + a_{101}\xi\zeta + a_{011}\eta\zeta$$
$$+ a_{111}\xi\eta\zeta + a_{201}\xi^2\zeta + a_{021}\eta^2\zeta + a_{102}\xi\zeta^2 + a_{012}\eta\zeta^2 \quad (5.13)$$

Depending on the node chosen from Fig. 7b, eqn. (5.13) specifies one of the fifteen third order Lagrange basis functions. If we make the basis function unity at the chosen node, while requiring it to be zero at the other fourteen nodes, then the specification of the fifteen coefficients in eqn. (5.13) is fully determined by the condition in eqn. (5.3). This process can be repeated for all the nodes in Fig. 7b.

At fourth order, the generic basis function has the form

$$\psi(\xi,\eta,\zeta) = a_{000} + a_{100}\xi + a_{010}\eta + a_{001}\zeta + a_{200}\xi^2 + a_{020}\eta^2 + a_{002}\zeta^2 + a_{110}\xi\eta + a_{101}\xi\zeta + a_{011}\eta\zeta$$
$$+ a_{300}\xi^3 + a_{030}\eta^3 + a_{003}\zeta^3 + a_{210}\xi^2\eta + a_{201}\xi^2\zeta + a_{120}\xi\eta^2 + a_{021}\eta^2\zeta + a_{102}\xi\zeta^2 + a_{012}\eta\zeta^2 + a_{111}\xi\eta\zeta$$
$$+ a_{103}\xi\zeta^3 + a_{013}\eta\zeta^3 + a_{301}\xi^3\zeta + a_{031}\eta^3\zeta + a_{211}\xi^2\eta\zeta + a_{121}\xi\eta^2\zeta$$

(5.14)

Depending on the node chosen from Fig. 7c, eqn. (5.14) specifies one of the twenty-six fourth order Lagrange basis functions. If we make the basis function unity at the chosen node, while requiring it to be zero at the other twenty-five nodes, then the specification of the twenty-six coefficients in eqn. (5.14) is fully determined by the condition in eqn. (5.3). This process can be repeated for all the nodes in Fig. 7c. This completes our description of the construction of serendipity bases for the ADER schemes that are described in this section.

The above equations have helped us define the *N* spatial basis functions; with *N*=6, 15 and 26 at second, third and fourth orders respectively. We also have to endow our basis functions with time-evolution. To that end, we choose *M* levels in time; with $M = 2, 3$ and 4 at second, third and fourth orders respectively. Since we are focusing on ADER-CG schemes that are optimized to treat MHD problems with non-stiff source terms, we let one of those time levels coincide with the starting time of the timestep. We take $\Delta t$ to be our timestep and let the scaled time $\tau = t/\Delta t$ have the range $\tau \in [0,1]$. At second order we therefore choose time points

$$\tau_1 = 0 \quad ; \quad \tau_2 = 1/2 \quad (5.15)$$

At third order, we choose time points

$$\tau_1 = 0 \quad ; \quad \tau_2 = \frac{1}{2} - \frac{1}{2\sqrt{3}} \quad ; \quad \tau_3 = \frac{1}{2} + \frac{1}{2\sqrt{3}} \tag{5.16}$$

At fourth order, we choose the time points

$$\tau_1 = 0 \quad ; \quad \tau_2 = \frac{1}{2} - \frac{1}{2}\sqrt{\frac{3}{5}} \quad ; \quad \tau_3 = \frac{1}{2} \quad ; \quad \tau_4 = \frac{1}{2} + \frac{1}{2}\sqrt{\frac{3}{5}} \tag{5.17}$$

This enables us to choose temporal Lagrange basis functions which satisfy the property

$$\phi_i(\tau_j) = \delta_{ij} \tag{5.18}$$

Notice that the time points in eqns. (5.15) to (5.17) have been chosen to coincide with Gaussian quadrature points in order to facilitate easy time-integration. (As an aside, it is worth mentioning that the ADER-DG scheme, which is suitable for the treatment of stiff source terms, can be defined in its most convenient form by using time points that are given in Balsara *et al*. 2018.)

Our spatial basis functions have been defined via eqn. (5.3) and our choice of spatial nodal points in Fig. 7. Our temporal basis functions have been defined via eqn. (5.18) and our choice of temporal nodal points in the above paragraph. We can, therefore, define $L \equiv N \times M$ space-time basis functions defined by

$$\theta_{i+(j-1)N}(\xi,\eta,\zeta,\tau) = \psi_i(\xi,\eta,\zeta)\phi_j(\tau) \quad \text{with} \quad i = 1,...,N \quad \text{and} \quad j = 1,...,M \tag{5.19}$$

These are the basis functions that we will use in the next section to formulate our ADER-CG scheme. (As an aside, it is also worth mentioning that an ADER-DG scheme would use the same spatial serendipity basis while using temporal basis from Balsara *et al*. 2018; that being the primary difference between ADER-DG and the ADER-CG documented here.)

Our eventual goal will be to have sets of zones on a geodesic mesh that are self-similar to a given zone. Let that given zone be indicated by a number "0" and let it have a scale length $l_0$. A zone labeled by number "$k$" that is self-similar to the original zone will have a scale length $l_k$. We wish to obtain an ADER scheme that re-uses the same update matrices from zone "0" by making a simple rescaling of the update matrices for the rest of the zones that are self-similar to it.

(An analogous plan was implemented for WENO-AO reconstruction in Balsara *et al.* 2018 resulting in dramatic speed-ups for the WENO algorithm on unstructured meshes and also geodesic meshes; please see Fig. 3 and Section IV of that paper.) To that end, the update eqn. (5.10) for the zone "*k*" can be written as

$$\frac{\partial \tilde{\mathbf{U}}}{\partial \tau} + \frac{\partial \tilde{\mathbf{F}}(\mathbf{U})}{\partial \xi} + \frac{\partial \tilde{\mathbf{G}}(\mathbf{U})}{\partial \eta} + \frac{\partial \tilde{\mathbf{H}}(\mathbf{U})}{\partial \zeta} = \tilde{\mathbf{S}} \qquad (5.20)$$

with the definitions that are slightly modified from eqn. (5.11) as

$$\tilde{\mathbf{U}} \equiv J\,\mathbf{U} \;\;;\;\; \tilde{\mathbf{F}}(\mathbf{U}) \equiv \frac{\Delta t\, l_0}{l_k}\left(\mathbf{h}_\eta \times \mathbf{h}_\zeta\right)\cdot \boldsymbol{\mathcal{F}}(\mathbf{U}) \;\;;\;\; \tilde{\mathbf{G}}(\mathbf{U}) \equiv \frac{\Delta t\, l_0}{l_k}\left(\mathbf{h}_\zeta \times \mathbf{h}_\xi\right)\cdot \boldsymbol{\mathcal{F}}(\mathbf{U}) \;\;;\;\;$$
$$\tilde{\mathbf{H}}(\mathbf{U}) \equiv \frac{\Delta t\, l_0}{l_k}\left(\mathbf{h}_\xi \times \mathbf{h}_\eta\right)\cdot \boldsymbol{\mathcal{F}}(\mathbf{U}) \;\;;\;\; \tilde{\mathbf{S}} = \Delta t\, J\, \mathbf{S}(\mathbf{U}) \qquad (5.21)$$

Eqns. (5.20) and (5.21) give us the final form for our governing equation on the reference element. The ADER matrices are unchanged from one zone to any other zone that is self-similar to it; with the only exception that the ADER fluxes are rescaled in the zone "*k*" by a factor of $(l_0/l_k)$ compared to the flux matrices in zone "0". In Balsara *et al.* (2018) we showed that this self-similarity can be achieved by using logarithmically scaled zones in the radial direction. We recommend that such scaled zones be used because it results in dramatic efficiencies in storage and processing not just for the WENO reconstruction but also for the ADER update.

**V.c) ADER-CG in Serendipity Bases on Isoparametrically Mapped Meshes**

The ADER scheme we describe here is based on a continuous Galerkin representation in time (also known as ADER-CG) which is not suited for problems having stiff source terms. However, most MHD applications are not required to handle stiff source terms. The scheme operates on the nodal bases described in eqns. (5.3), (5.18) and (5.19). The governing equations that are treated on the reference element are given in eqn. (5.20). The inclusion of the cross products of the coordinate vectors in eqn. (5.21), and their subsequent differentiation in eqn. (5.20), ensures that curvature-inducing non-linearities in the isoparametric mapping are incorporated in the update. This is equivalent to saying that the role of Christoffel symbols is implicitly accommodated.

Eqn. (5.19) gives us a total of $L \equiv N \times M$ space-time basis functions in which we can represent the solution vector as

$$\tilde{\mathbf{U}}(\xi,\eta,\zeta,\tau) = \sum_{l=1}^{L} \hat{\mathbf{U}}_l \, \theta_l(\xi,\eta,\zeta,\tau) \tag{5.22}$$

The degrees of freedom are given by $\{\hat{\mathbf{U}}_l : l = 1,...,L\}$. Note that the first "$N$" of these degrees of freedom are specified at $\tau_1 = 0$ and are, therefore, non-evolutionary. Equations similar to the one above can be formulated for the $\xi$-flux $\tilde{\mathbf{F}}(\xi,\eta,\zeta,\tau)$ which depends on $\{\hat{\mathbf{F}}_l : l = 1,...,L\}$, the $\eta$-flux $\tilde{\mathbf{G}}(\xi,\eta,\zeta,\tau)$ which depends on $\{\hat{\mathbf{G}}_l : l = 1,...,L\}$, the $\zeta$-flux $\tilde{\mathbf{H}}(\xi,\eta,\zeta,\tau)$ which depends on $\{\hat{\mathbf{H}}_l : l = 1,...,L\}$ and source term $\tilde{\mathbf{S}}(\xi,\eta,\zeta,\tau)$ which depends on $\{\hat{\mathbf{S}}_l : l = 1,...,L\}$. Because the basis functions are nodal, and because the Jacobians and coordinate vectors do not depend on time, the nodal values of $\hat{\mathbf{U}}_l$ can be used to obtain the nodal values of $\hat{\mathbf{F}}_l$, $\hat{\mathbf{G}}_l$, $\hat{\mathbf{H}}_l$ and $\hat{\mathbf{S}}_l$. To illustrate the situation for $\hat{\mathbf{F}}_l$ we can write

$$\hat{\mathbf{F}}_l = \frac{\Delta t \, l_0}{l_k} \left( \mathbf{h}_\eta \times \mathbf{h}_\zeta \right)_l \cdot \mathcal{F}\left( \frac{1}{J_l} \hat{\mathbf{U}}_l \right) \tag{5.23}$$

Here $J_l$ is the Jacobian, shown immediately after eqn. (5.5), and evaluated at the nodal point "$l$" and $\left( \mathbf{h}_\eta \times \mathbf{h}_\zeta \right)_l$ is the cross product of the coordinate basis vectors shown in eqn. (5.4) evaluated at the same nodal point "$l$". With the above changes in problem specification, the formulation of the ADER-CG scheme runs entirely parallel to Section 3.2 of Dumbser *et al.* (2008) or Section 3.1 of Balsara *et al.* (2009). We only specify it as briefly as possible below for the sake of completeness, and because it enables us to provide a pointwise description of the scheme and its implementation in the next Sub-section.

Applying the Galerkin projection to eqn. (5.20) gives

$$\left\langle \theta_j, \frac{\partial \theta_l}{\partial \tau} \right\rangle \hat{\mathbf{U}}_l + \left\langle \theta_j, \frac{\partial \theta_l}{\partial \xi} \right\rangle \hat{\mathbf{F}}_l + \left\langle \theta_j, \frac{\partial \theta_l}{\partial \eta} \right\rangle \hat{\mathbf{G}}_l + \left\langle \theta_j, \frac{\partial \theta_l}{\partial \zeta} \right\rangle \hat{\mathbf{H}}_l = \left\langle \theta_j, \theta_l \right\rangle \hat{\mathbf{S}}_l \tag{5.24}$$

The angled brackets in the above equation indicate space-time integration over the reference element. The Einstein summation convention is operative. Denoting $\hat{\mathbf{U}} = \{\hat{U}_1, \hat{U}_2, ..., \hat{U}_L\}^T$ and so on, we can write the above equation in matrix notation as

$$\mathbf{K}_\tau \hat{\mathbf{U}} + \mathbf{K}_\xi \hat{\mathbf{F}} + \mathbf{K}_\eta \hat{\mathbf{G}} + \mathbf{K}_\zeta \hat{\mathbf{H}} = \mathbf{M} \hat{\mathbf{S}} \qquad (5.25)$$

where, consistent with the usual terminology for Galerkin schemes, $\mathbf{M}$ is the mass matrix, $\mathbf{K}_\tau$ is the time-stiffness matrix and $\mathbf{K}_\xi, \mathbf{K}_\eta, \mathbf{K}_\zeta$ are the flux-stiffness matrices. The $(j,l)^{th}$ elements of the above-mentioned matrices can be made explicit and they are easily evaluated using a computer algebra system. They are written explicitly as

$$\mathbf{K}_{\tau;j,l} = \left\langle \theta_j, \frac{\partial \theta_l}{\partial \tau} \right\rangle \,;\, \mathbf{K}_{\xi;j,l} = \left\langle \theta_j, \frac{\partial \theta_l}{\partial \xi} \right\rangle \,;\, \mathbf{K}_{\eta;j,l} = \left\langle \theta_j, \frac{\partial \theta_l}{\partial \eta} \right\rangle \,;\, \mathbf{K}_{\zeta;j,l} = \left\langle \theta_j, \frac{\partial \theta_l}{\partial \zeta} \right\rangle \,;\, \mathbf{M}_{j,l} = \left\langle \theta_j, \theta_l \right\rangle$$

(5.26)

Having specified the basic formulation of ADER-CG scheme in the above two equations, we now turn our attention to its efficient processing in the next paragraph.

Notice that only the last $L-N$ components of $\hat{\mathbf{U}}$ change as the ADER-CG iteration proceeds; the first "$N$" component of $\hat{\mathbf{U}}$ are evaluated only at the beginning of the iteration and are subsequently left unchanged. The same is true for $\hat{\mathbf{F}}$, $\hat{\mathbf{G}}$, $\hat{\mathbf{H}}$ and $\hat{\mathbf{S}}$. We thus write $\hat{\mathbf{U}} = \{\hat{\mathbf{U}}^0, \hat{\mathbf{U}}^1\}^T$ where $\hat{\mathbf{U}}^0$ has the first "$N$" non-evolutionary degrees of freedom and $\hat{\mathbf{U}}^1$ has the subsequent $L-N$ evolutionary degrees of freedom. A similar split can be made for $\hat{\mathbf{F}}$, $\hat{\mathbf{G}}$, $\hat{\mathbf{H}}$ and $\hat{\mathbf{S}}$. A similar split can be made for the mass matrix and the stiffness matrices so that all the matrices in eqn. (5.26) can be written as

$$\mathbf{M} = \begin{bmatrix} \mathbf{M}^{00} & \mathbf{M}^{01} \\ \mathbf{M}^{10} & \mathbf{M}^{11} \end{bmatrix} \,,\, \mathbf{K}_\alpha = \begin{bmatrix} \mathbf{K}_\alpha^{00} & \mathbf{K}_\alpha^{01} \\ \mathbf{K}_\alpha^{10} & \mathbf{K}_\alpha^{11} \end{bmatrix} \qquad (5.27)$$

where $\alpha$ can be $\xi$, $\eta$, $\zeta$ or $\tau$ in the above equation. Only the last $L-N$ components of eqn. (5.25) are useful and yield the equation

$$\hat{\mathbf{U}}^1 = \left(\hat{\mathbf{K}}_\tau^0 \hat{\mathbf{U}}^0 + \hat{\mathbf{M}}^0 \hat{\mathbf{S}}^0 - \hat{\mathbf{K}}_\xi^0 \hat{\mathbf{F}}^0 - \hat{\mathbf{K}}_\eta^0 \hat{\mathbf{G}}^0 - \hat{\mathbf{K}}_\zeta^0 \hat{\mathbf{H}}^0\right) + \hat{\mathbf{M}} \hat{\mathbf{S}}^1 - \hat{\mathbf{K}}_\xi \hat{\mathbf{F}}^1 - \hat{\mathbf{K}}_\eta \hat{\mathbf{G}}^1 - \hat{\mathbf{K}}_\zeta \hat{\mathbf{H}}^1 \tag{5.28}$$

The matrices in the above equation are given by

$$\hat{\mathbf{K}}_\xi = \left(\mathbf{K}_\tau^{11}\right)^{-1} \mathbf{K}_\xi^{11} \;;\; \hat{\mathbf{K}}_\eta = \left(\mathbf{K}_\tau^{11}\right)^{-1} \mathbf{K}_\eta^{11} \;;\; \hat{\mathbf{K}}_\zeta = \left(\mathbf{K}_\tau^{11}\right)^{-1} \mathbf{K}_\zeta^{11} \;;\; \hat{\mathbf{M}} = \left(\mathbf{K}_\tau^{11}\right)^{-1} \mathbf{M}^{11} \;;$$

$$\hat{\mathbf{K}}_\xi^0 = \left(\mathbf{K}_\tau^{11}\right)^{-1} \mathbf{K}_\xi^{10} \;;\; \hat{\mathbf{K}}_\eta^0 = \left(\mathbf{K}_\tau^{11}\right)^{-1} \mathbf{K}_\eta^{10} \;;\; \hat{\mathbf{K}}_\zeta^0 = \left(\mathbf{K}_\tau^{11}\right)^{-1} \mathbf{K}_\zeta^{10} \;;\; \hat{\mathbf{M}}^0 = \left(\mathbf{K}_\tau^{11}\right)^{-1} \mathbf{M}^{10} \;; \tag{5.29}$$

$$\hat{\mathbf{K}}_\tau^0 = -\left(\mathbf{K}_\tau^{11}\right)^{-1} \mathbf{K}_\tau^{10} \;;$$

Eqn. (5.28) shows explicitly that the terms inside the round brackets only need to be evaluated once, leading to the efficiency of the ADER-CG scheme. The remaining terms on the right hand side of eqn. (5.28) change with each successive ADER-CG iteration; however, their evaluation can be lagged in the iteration sequence, which each prior iterate supplying improved approximations for the fluxes and source terms that contribute to the current iterate.

Further computational efficiencies are found by evaluating the matrices in eqn. (5.29) explicitly by using a computer algebra system. In particular, realize that eqns. (5.15) to (5.17) specify discrete time levels. Those time levels are reflected in the update eqn. (5.28) so that the update at a given time level depends on gradients of the fluxes evaluated at all the time levels. However, the expressions for the gradients of the fluxes at any time level are completely similar in form to the analogous expressions at all other time levels. This formal similarity should also be exploited for efficient processing.

This completes our description of the ADER-CG method for isoparametrically mapped meshes. We have also identified the techniques and tricks for efficient processing of such an ADER scheme when it is formulated in nodal basis on the reference element.

**V.d) Stepwise Description of the ADER Algorithm; Suitable for Implementation**

We now describe the step-by-step implementation of ADER-CG on an isoparametrically mapped mesh. We break the description into three pieces, dealing with pre-computation; initialization and iteration.

**<u>Pre-Computation Steps</u>**

**1)** Using a computer algebra system, evaluate the space-time basis functions using eqns. (5.3), (5.18) and (5.19). Use Fig. 7 to identify the nodal points for the serendipity elements. Appendix B provides some more helpful detail.

**2)** Using a computer algebra system, evaluate eqn. (5.26). Then make the split in eqn. (5.27) and use it in eqns. (5.29) and (5.28). Appendices B and C provide some more helpful detail.

**3)** Eqn. (5.28) gives the ADER-CG explicitly and most computer algebra systems can be made to write out the corresponding code. However, realize that the discrete time levels from eqn. (5.15) to (5.17) will be reflected in the update eqn. (5.28) so that the update at a given time level depends on gradients of the fluxes evaluated at all the time levels. The formal similarity in the expressions for the gradients of the fluxes can be used to obtain additional efficiencies.

**4)** First start with the zone "0" that may be similar to the other zones. For that zone, identify the isoparametric mapping in eqn. (5.2). At each serendipity node "$l$" in the reference element, evaluate the coordinate basis vectors from eqn. (5.4). Use those coordinate basis vectors to evaluate the Jacobian $J_l = \left|\left(\mathbf{h}_\xi \times \mathbf{h}_\eta\right)\cdot \mathbf{h}_\zeta\right|_l$ and the cross products $\left(\mathbf{h}_\eta \times \mathbf{h}_\zeta\right)_l$, $\left(\mathbf{h}_\zeta \times \mathbf{h}_\xi\right)_l$ and $\left(\mathbf{h}_\xi \times \mathbf{h}_\eta\right)_l$. This enables us to incorporate the curvature of the element in the mapping. These terms are costly to evaluate and can be evaluated once in zone "0" and reused in all zones that are self-similar to zone "0". As long as the scaling factor $\left(l_0/l_k\right)$ is used in the definition of the fluxes, this approach will work. However, we will revisit this issue in more detail in a subsequent step.

**Initialization Steps**

**5)** At each nodal point, use the spatially reconstructed solution from any WENO or DG step to evaluate $\hat{\mathbf{U}}_l$ for $l=1,...,N$ at $\tau_1 = 0$. Note from eqn. (5.21) that $\hat{\mathbf{U}}_l$ includes the Jacobian $J_l$. However, that Jacobian should be divided out when physical fluxes are evaluated, as in the right hand side of eqn. (5.23). Therefore, obtain $\hat{\mathbf{F}}_l$, $\hat{\mathbf{G}}_l$, $\hat{\mathbf{H}}_l$ and $\hat{\mathbf{S}}_l$ for $l=1,...,N$. By observing eqns. (5.21) and (5.23), it is good to be cognizant that these fluxes and source terms include geometrical effects. From the first "$N$" nodal values of the state, fluxes and source terms, we can populate all "$L$" nodal values of the same variables; because this is an initialization step.

**6)** Now evaluate the round bracket in eqn. (5.28) at each nodal point "*l*" for $l = 1, ..., N$. This completes the initialization steps.

**Iteration Steps**

**7)** Now for the nodes $l = N+1, ..., L$ compute the remaining terms on the right hand side of eqn. (5.28). Use them to obtain the vector $\hat{\mathbf{U}}^1$. This is the new iterate.

**8)** For the nodes $l = N+1, ..., L$, use the new iterate $\hat{\mathbf{U}}^1$ to evaluate improved values for $\hat{\mathbf{F}}^1$, $\hat{\mathbf{G}}^1$, $\hat{\mathbf{H}}^1$ and $\hat{\mathbf{S}}^1$. As discussed before, we should be mindful of the role of the geometrical terms, as shown in eqn. (5.23) and also (5.21). We should also be mindful of the scaling factor $(l_0/l_k)$ and incorporate it in our evaluation of the fluxes if we are working in a zone "*k*" that is self-similar to zone "0".

**9)** This iteration can be repeated till $\hat{\mathbf{U}}^1$ has come close to its value from the previous iteration. In practice, at second order, two iterations are sufficient. At third order, three iterations are sufficient; and at fourth order, four iterations are sufficient. Recall that the iterate needs to converge only up to the point where it achieves the desired order of accuracy; it is not required to converge up to machine precision. There is some theory associated with Picard iteration that supports the choices of number of iterations described in this step (Dumbser *et al*. 2008).

**Finalization Step**

**10)** Notice that the space integration and the time integration on the right hand sides of the update equations, eqn. (2.5) and (2.6), can be made to switch places. (A similar consideration can be made for the fluxes in the update for the conserved variables.) As a result, the only desirable output from the ADER scheme is a set of time-averaged variables. This is easily done because the time levels in eqns. (5.15), (5.16) and (5.17) correspond to Gaussian quadrature points in time. We can even obtain area and line averages from the ADER scheme that would result in a quadrature-free evaluation of the numerical fluxes at high order. Ideas on quadrature-free evaluation of the numerical fluxes were first discussed by Atkins & Shu (1998) and Dumbser *et al*. (2007); and they are also documented in a pedagogic fashion in Section V.3.ii of Balsara (2017).

This completes our pointwise description of the ADER-CG scheme at all orders on mapped isoparametric meshes. Appendix B provides an explicit catalogue of the ADER-CG update, i.e. eqn. (5.28), at second order on the reference equilateral prism. Appendix C does the same at third order. Eqns. (C.19) to (C.21) indeed show how the ADER iteration can be structured very compactly for easy implementation. We also catalogue all the nodal bases at third order in Appendix C. This should help people to check whether their ADER-CG formulation is correct. Appendix D catalogues the nodal locations and other relevant information at fourth order.

**VI) Pointwise Description of the Full Divergence-Free MHD Scheme on Geodesic Meshes**

Below we provide a pointwise description of the time-update for higher order, divergence-free MHD on geodesic meshes.

**1)** We assume that all geometrical aspects of the mesh have been built. See Florinski *et al*. (2018). This includes a specification via isoparametric mapping of quadrature points and weights at the faces of each zone and cubature points and weights within the volume of each zone. We also assume that all the moments of the zones, as illustrated in eqn. (3.2), have been built. This gives us the Taylor series basis within each zone. This basis will be used for reconstruction.

**2)** From the previous timestep, we have evolved all eight components of the MHD solution vector in zone-averaged fashion. We use this information, along with the steps in Section V of Balsara *et al*. (2018), to obtain high order, finite volume WENO-AO reconstruction.

**3)** From the previous timestep, we have also evolved facially-averaged components of the magnetic field. Using the steps described at the end of Section IV of this paper, we obtain a high order, divergence-free reconstruction of the magnetic field within each zone. The magnetic modes within the zone, including their mean values, are overwritten by the values that are obtained from the divergence-free reconstruction of the magnetic field. In this fashion, the facially-averaged magnetic fields are the primal variables of the method. The zone-averaged magnetic field, because it is overwritten in every timestep, is just a helping variable.

**4)** Using the ADER-CG steps described at the end of Section V of this paper, we obtain a high order, space-time reconstruction of the solution.

**5)** Using this space-time reconstruction, and the information about quadratures and cubatures, we call one-dimensional Riemann solvers in the faces of the mesh. This information can be used to obtain facially-averaged numerical fluxes that are needed for the finite volume update of all eight components of the MHD solution vector in zone-averaged fashion. We also call multidimensional MuSIC Riemann solvers at the edges of the mesh. This information can be used to obtain edge-averaged (and multidimensionally upwinded) line integrals for the electric field, as shown in eqns. (2.5) and (2.6). The facially-averaged components of the magnetic field can then be updated consistent with Faraday's law and the Yee-type staggering of variables. This completes the description of the time-update for higher order, divergence-free MHD on geodesic meshes.

## VII) Accuracy Tests

We present a couple of truly multidimensional test problems which show that the code preserves the order property. In other words, it meets its designed order of accuracy. The flows described in this section are smooth and do not have shock discontinuities, with the result that they are good test problems for demonstrating the higher order of accuracy of our numerical method. Please also recall that for our purposes the angular resolution of a geodesic mesh will refer to the mean central angle subtended by an edge.

Some of the test problems in this section have supersonic outflow at the outer boundary. Such boundary conditions were handled in the usual way that they are handled in astrophysical codes – i.e. with continuative outflow. The Riemann solvers, coupled with the upwinding in the reconstruction, are adequate for allowing the flow to exit the computational domain without any complications. (For accretion problems, an inflow boundary condition that is imposed at the outer boundary, along with an outflow boundary condition at the inner boundary where the flow becomes supersonic, was also found to work well.) This is also the typical usage for astrophysics and space physics codes. So we did not implement any special boundary conditions.

### VII.a) Solar Wind on Spherical Meshes

This test problem has been adopted from Balsara *et al*. (2018). For the description of the problem, readers are referred to the original reference. For the present simulation, we set up this problem on a domain of radial extent $[2.0, 3.5]$. Stopping time for this problem is 0.25.

Tables Ia, Ib and Ic show the accuracy results for second, third and fourth order ADER-WENO schemes, respectively. For the second order ADER-WENO scheme, we use second order accurate isoparametric mapping for mapping the triangular prisms to spherical frustrums. Similarly, for the third and fourth order ADER-WENO schemes, we use third and fourth order accurate isoparametric mapping respectively. On the coarsest level, we use a spherical mesh having angular resolution of $4.33^o$ and 16 logarithmically ratioed radial zones. On the finest level, the angular resolution is $0.54^o$ and 128 logarithmically ratioed radial zones are used.

We see from Tables Ia, Ib and Ic that the methods reach their design accuracies. We have used a flattener algorithm from Balsara (2012b) and on the coarsest meshes the flattener detects flow features as discontinuities. This explains why the coarsest meshes in Table I have large errors. However, the point of the demonstration is to show that on meshes that are even slightly well-resolved, the effect of the flattener is non-existent and the order of accuracy is retrieved. The flattener algorithm is entirely unnecessary for this test, however, it can be very valuable for more stringent test problems.

**Table Ia shows the accuracy analysis for the second-order ADER-WENO scheme for the solar wind on spherical meshes. A CFL of 0.25 was used. The errors and accuracy in the density ($\rho$) and x-momentum ($m_x$) are shown.**

| Angular Size (in degrees) | $\rho$ $L_1$ error | $\rho$ $L_1$ accuracy | $\rho$ $L_{inf}$ error | $\rho$ $L_{inf}$ accuracy |
|---|---|---|---|---|
| 4.33 | 4.7439E-04 | | 5.6854E-04 | |
| 2.16 | 5.3099E-06 | 6.48 | 1.5091E-05 | 5.24 |
| 1.08 | 1.3824E-06 | 1.94 | 5.6741E-06 | 1.41 |
| 0.54 | 3.5140E-07 | 1.98 | 1.9502E-06 | 1.54 |
| Angular Size (in degrees) | $m_x$ $L_1$ error | $m_x$ $L_1$ accuracy | $m_x$ $L_{inf}$ error | $m_x$ $L_{inf}$ accuracy |

| Angular Size (in degrees) | | | | |
|---|---|---|---|---|
| 4.33 | 8.5761E-04 | | 1.3565E-03 | |
| 2.16 | 1.3337E-05 | 6.01 | 3.9998E-05 | 5.08 |
| 1.08 | 3.3891E-06 | 1.98 | 1.5068E-05 | 1.41 |
| 0.54 | 8.5388E-07 | 1.99 | 5.1868E-06 | 1.54 |

**Table Ib shows the accuracy analysis for the third-order ADER-WENO scheme for the solar wind on spherical meshes. A CFL of 0.25 was used. The errors and accuracy in the density ( $\rho$ ) and x-momentum ($m_x$) are shown.**

| Angular Size (in degrees) | $\rho$ $L_1$ error | $\rho$ $L_1$ accuracy | $\rho$ $L_{inf}$ error | $\rho$ $L_{inf}$ accuracy |
|---|---|---|---|---|
| 4.33 | 4.6591E-04 | | 5.5838E-04 | |
| 2.16 | 8.9824E-07 | 9.02 | 1.7546E-06 | 8.31 |
| 1.08 | 1.1428E-07 | 2.97 | 3.4173E-07 | 2.36 |
| 0.54 | 1.4396E-08 | 2.99 | 5.7124E-08 | 2.58 |
| Angular Size (in degrees) | $m_x$ $L_1$ error | $m_x$ $L_1$ accuracy | $m_x$ $L_{inf}$ error | $m_x$ $L_{inf}$ accuracy |
| 4.33 | 8.1760E-04 | | 1.3250E-03 | |
| 2.16 | 1.4330E-06 | 9.16 | 3.3316E-06 | 8.64 |
| 1.08 | 1.8222E-07 | 2.98 | 6.8504E-07 | 2.28 |
| 0.54 | 2.3037E-08 | 2.98 | 1.1562E-07 | 2.57 |

Table Ic shows the accuracy analysis for the fourth-order ADER-WENO scheme for the solar wind on spherical meshes. A CFL of 0.25 was used. The errors and accuracy in the density ($\rho$) and x-momentum ($m_x$) are shown.

| Angular Size (in degrees) | $\rho$ L$_1$ error | $\rho$ L$_1$ accuracy | $\rho$ L$_{inf}$ error | $\rho$ L$_{inf}$ accuracy |
|---|---|---|---|---|
| 4.33 | 4.6984E-04 | | 5.6026E-04 | |
| 2.16 | 2.9347E-08 | 13.97 | 2.2295E-07 | 11.30 |
| 1.08 | 2.0861E-09 | 3.81 | 2.5449E-08 | 3.13 |
| 0.54 | 1.3826E-10 | 3.92 | 2.8885E-09 | 3.14 |
| Angular Size (in degrees) | $m_x$ L$_1$ error | $m_x$ L$_1$ accuracy | $m_x$ L$_{inf}$ error | $m_x$ L$_{inf}$ accuracy |
| 4.33 | 8.2934E-04 | | 1.3331E-03 | |
| 2.16 | 5.1020E-08 | 13.99 | 5.3216E-07 | 11.29 |
| 1.08 | 3.2654E-09 | 3.97 | 6.0601E-08 | 3.13 |
| 0.54 | 2.1082E-10 | 3.95 | 6.8918E-09 | 3.14 |

**VII.b) Manufactured Solution on Spherical Meshes**

This test problem has been adopted from Ivan *et al.* (2015). For the description of the problem, readers are referred to the original reference.

Tables IIa, IIb and IIc show the accuracy results for second, third and fourth order ADER-WENO schemes, respectively. For the second order ADER-WENO scheme, we use second order accurate isoparametric mapping for mapping the triangular prisms to spherical frustrums. Similarly, for the third and fourth order ADER-WENO schemes, we use third and fourth order

accurate isoparametric mapping respectively. On the coarsest level, we use a spherical mesh having angular resolution of $4.33^o$ and 16 logarithmically ratioed radial zones. On the finest level, the angular resolution is $0.54^o$ and 128 logarithmically ratioed radial zones are used.

We see from Tables IIa, IIb and IIc that the methods reach their design accuracies. For this problem we did not use the flattener. As a result, all meshes display their optimal order of accuracy from the coarsest mesh to the finest mesh. We see, therefore, that the WENO-AO schemes with the divergence-free reconstruction and the ADER-CG predictor step actually do reach their design accuracies very quickly on relatively coarse meshes.

**Table IIa shows the accuracy analysis for the second-order ADER-WENO scheme for the manufactured solution on spherical meshes. A CFL of 0.25 was used. The errors and accuracy in the density ($\rho$) and x-magnetic field ($B_x$) are shown.**

| Angular Size (in degrees) | $\rho$ $L_1$ error | $\rho$ $L_1$ accuracy | $\rho$ $L_{inf}$ error | $\rho$ $L_{inf}$ accuracy |
|---|---|---|---|---|
| 4.33 | 3.5397E-05 | | 7.8658E-05 | |
| 2.16 | 8.7766E-06 | 2.01 | 2.0197E-05 | 1.96 |
| 1.08 | 2.1946E-06 | 2.00 | 5.1163E-06 | 1.98 |
| 0.54 | 5.4884E-07 | 2.00 | 1.4912E-06 | 1.78 |
| Angular Size (in degrees) | $B_x$ $L_1$ error | $B_x$ $L_1$ accuracy | $B_x$ $L_{inf}$ error | $B_x$ $L_{inf}$ accuracy |
| 4.33 | 2.2628E-04 | | 9.0744E-04 | |
| 2.16 | 5.6825E-05 | 1.99 | 2.5493E-04 | 1.83 |
| 1.08 | 1.4271E-05 | 1.99 | 7.0561E-05 | 1.85 |
| 0.54 | 3.5792E-06 | 2.00 | 2.0015E-05 | 1.82 |

Table IIb shows the accuracy analysis for the third-order ADER-WENO scheme for the manufactured solution on spherical meshes. A CFL of 0.25 was used. The errors and accuracy in the density ($\rho$) and x-magnetic field ($B_x$) are shown.

| Angular Size (in degrees) | $\rho$ $L_1$ error | $\rho$ $L_1$ accuracy | $\rho$ $L_{inf}$ error | $\rho$ $L_{inf}$ accuracy |
|---|---|---|---|---|
| 4.33 | 1.2648E-05 | | 2.3020E-05 | |
| 2.16 | 1.5999E-06 | 2.98 | 2.9878E-06 | 2.95 |
| 1.08 | 2.0097E-07 | 2.99 | 4.8646E-07 | 2.62 |
| 0.54 | 2.5186E-08 | 3.00 | 8.0617E-08 | 2.59 |
| Angular Size (in degrees) | $B_x$ $L_1$ error | $B_x$ $L_1$ accuracy | $B_x$ $L_{inf}$ error | $B_x$ $L_{inf}$ accuracy |
| 4.33 | 2.3509E-05 | | 1.1016E-04 | |
| 2.16 | 3.0428E-06 | 2.95 | 2.2537E-05 | 2.29 |
| 1.08 | 3.9110E-07 | 2.96 | 4.5833E-06 | 2.30 |
| 0.54 | 4.9865E-08 | 2.97 | 9.0948E-07 | 2.33 |

Table IIc shows the accuracy analysis for the fourth-order ADER-WENO scheme for the manufactured solution on spherical meshes. A CFL of 0.25 was used. The errors and accuracy in the density ($\rho$) and x-magnetic field ($B_x$) are shown.

| Angular Size (in degrees) | $\rho$ $L_1$ error | $\rho$ $L_1$ accuracy | $\rho$ $L_{inf}$ error | $\rho$ $L_{inf}$ accuracy |
|---|---|---|---|---|
| 4.33 | 7.8731E-07 | | 1.9460E-06 | |

| Angular Size (in degrees) | $B_x$ $L_1$ error | $B_x$ $L_1$ accuracy | $B_x$ $L_{inf}$ error | $B_x$ $L_{inf}$ accuracy |
|---|---|---|---|---|
| 2.16 | 3.9774E-08 | 4.31 | 1.2650E-07 | 3.94 |
| 1.08 | 2.3222E-09 | 4.10 | 1.4481E-08 | 3.13 |
| 0.54 | 1.4537E-10 | 4.00 | 1.7813E-09 | 3.02 |
| 4.33 | 2.0282E-06 | | 1.8006E-05 | |
| 2.16 | 1.1520E-07 | 4.14 | 2.0084E-06 | 3.16 |
| 1.08 | 7.1716E-09 | 4.01 | 2.0044E-07 | 3.32 |
| 0.54 | 4.6395E-10 | 3.95 | 1.7091E-08 | 3.55 |

## VIII) Test Problems

All the 3D simulations shown here were run with a CFL of 0.25 where the diameter of the in-sphere of a zone was used to restrict the timestep. The maximum permissible CFL for this type of simulation would have been 0.33. As a result, the simulations were run without making a significant compromise with respect to an analogous CFL on a Cartesian mesh.

For all our present test problems, the mesh does not need to extend to r=0, i.e. to the origin. It is possible that for some applications the mesh does need to extend to the origin. The zones that make contact with the origin degenerate from frustums to tetrahedra. ADER schemes have been formulated for tetrahedra (Dumbser *et al*. 2008) and the MHD reconstruction described here has also been formulated for tetrahedra (Balsara & Dumbser 2015a). As a result, there are well-developed algorithmic capabilities that do exist for applications that extend all the way to the origin.

### VIII.a) Rotor Test Problem

This test problem has been taken from Balsara & Spicer (1999) and Balsara (2004). However, the problem set up has been slightly changed so that this simulation can be performed on a spherical sector. Therefore, instead of initializing a rapidly spinning cylinder, we initialized a rapidly spinning sphere (the rotor) of radius 0.8 at the center of our computational domain. The rotation axis of the sphere is chosen to be along a line joining the origin of the coordinate system and the center of the spherical sector. The rotor has a density of 10 whereas the ambient has a density of 1. A uniform magnetic field with a magnitude of 2.5 is initialized along the perpendicular direction to the rotation axis. The ratio of specific heat is set to 5/3. The rotor has a constant angular velocity $\omega = 1$. Following the above references, we applied a taper of six radial zone-size on the density and angular velocity.

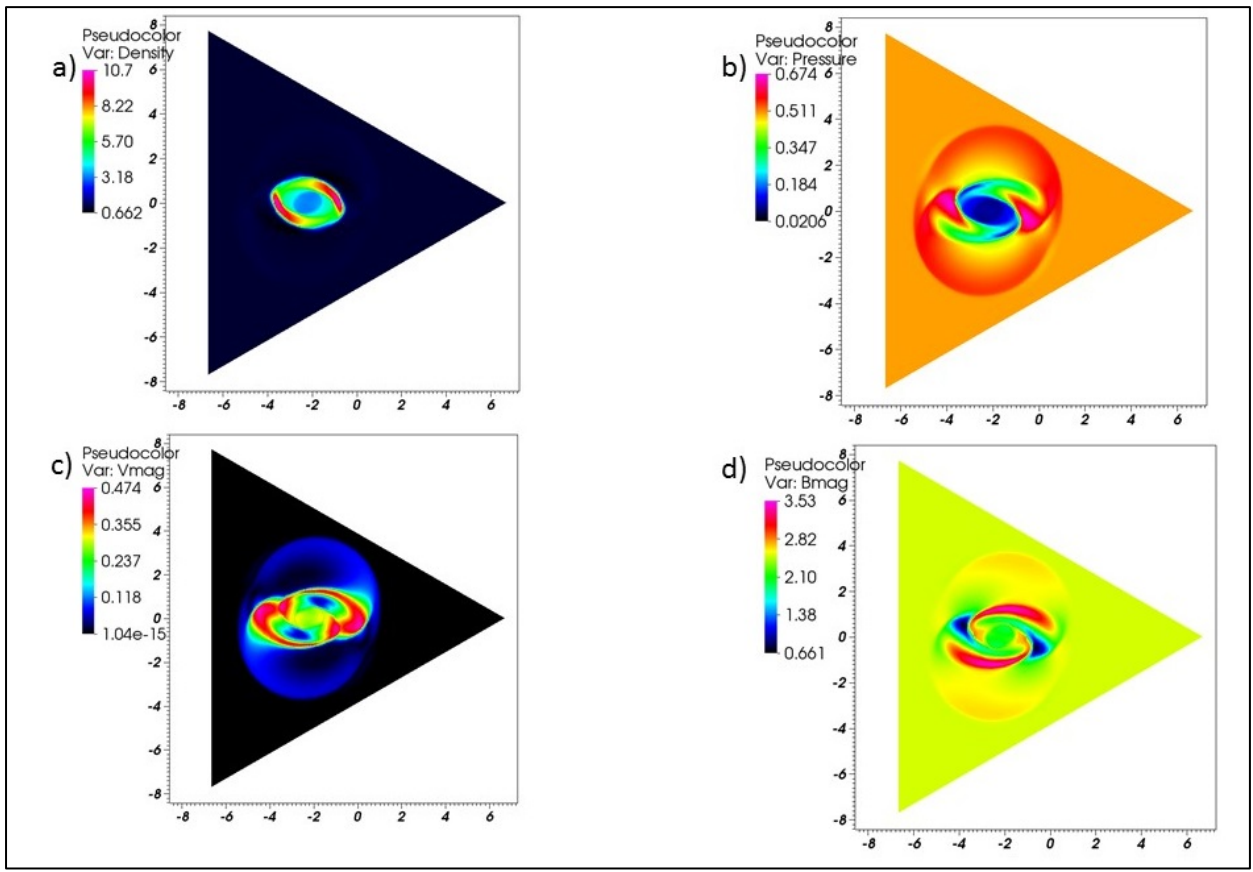

*Fig. 8 shows the results of the rotor test problem at a final time of 2.5. Fig 8a shows the final density, 8b shows the final pressure, 8c shows the total velocity and 8d shows the magnitude of the magnetic field.*

The simulation has been performed on a spherical sector with a radial extent of $[7,16]$. Thus, the center of the rotor is placed at the radius of 11.5. We used a spherical geodesic mesh with angular resolution of $0.27^o$ and 256 logarithmically ratioed zones in the radial direction. A fourth order accurate ADER-WENO scheme is chosen for the results shown in Figure 8. The simulation has been run to a final time of 2.5. Fig 8a shows the final density, 8b shows the final pressure, 8c shows the total velocity and 8d shows the magnitude of the magnetic field. We see that all the flow features are well resolved and are similar to the two-dimensional problem, as expected.

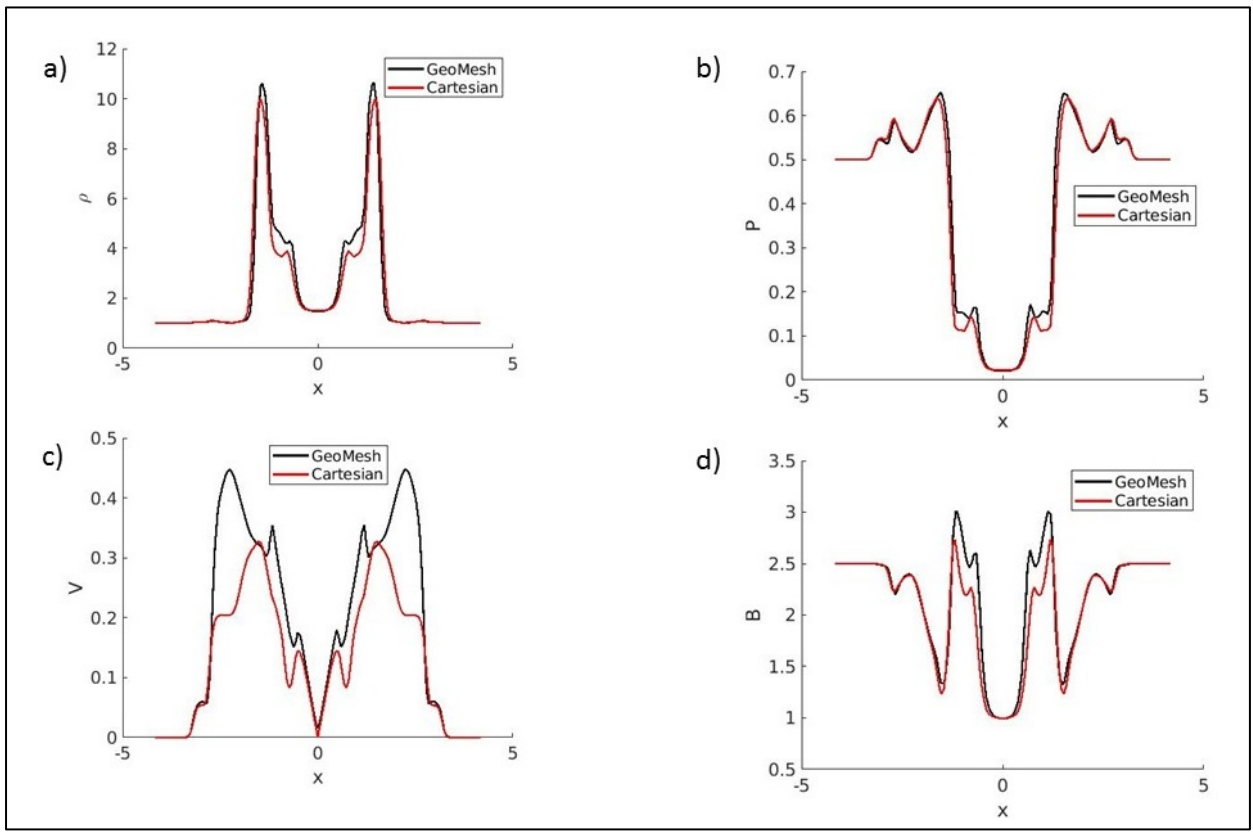

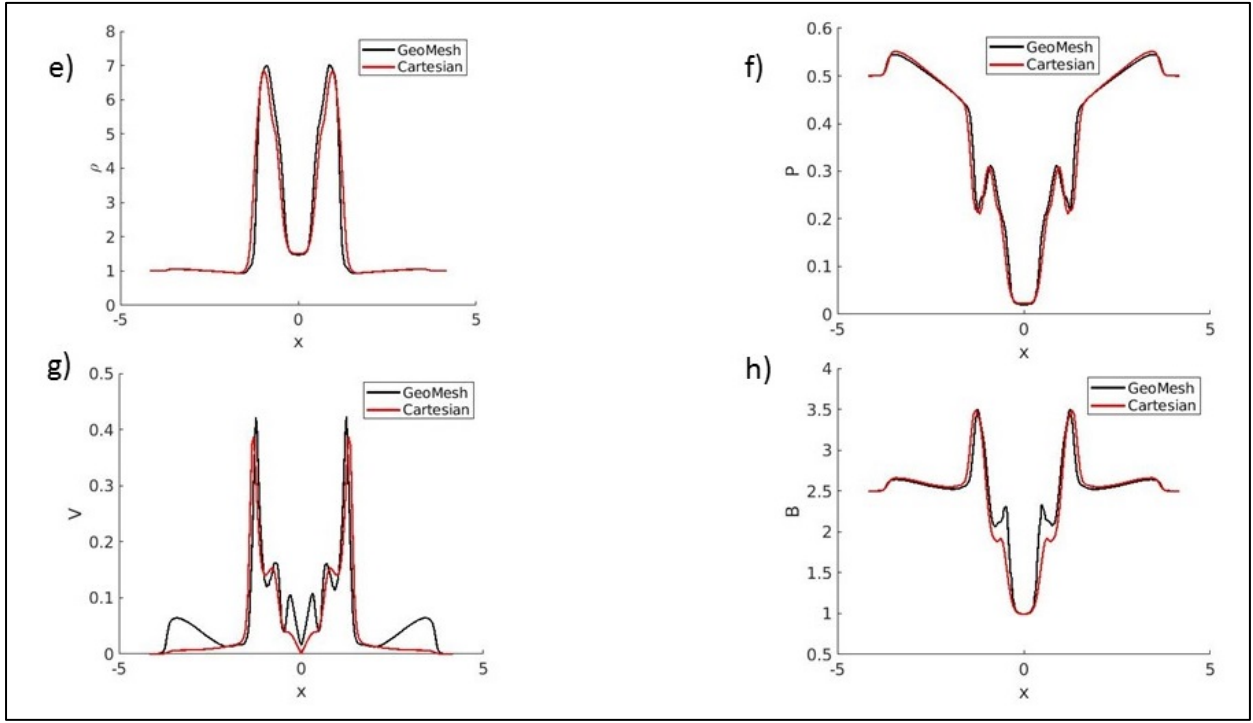

*For the flow variables shown in Fig. 8, we made cuts in the directions along the magnetic field and transverse to the magnetic field and overplotted the results from the geodesic and Cartesian meshes. Figs. 9a to 9d plot the density, pressure, total velocity and total magnetic field along the direction of the original magnetic field. Figs. 9e to 9h plot the density, pressure, total velocity and total magnetic field perpendicular to the direction of the original magnetic field. The solid black and red lines in Fig. 9 show the results from the geodesic mesh and the Cartesian mesh respectively.*

It is interesting to show that the same physical simulation on a geodesic mesh produces results that are closely comparable to the results that were produced on a 3D Cartesian mesh. For that reason, we ran the same rotor problem on a $149^3$ zone Cartesian mesh. The two meshes were chosen so that their effective resolution was similar. (As seen from Fig. 8, the rotor expands out to form an inscribed circular region on the triangular mesh; therefore, there are some unutilized zones in the geodesic mesh calculation.) For the flow variables shown in Fig. 8, we made cuts in the directions along the magnetic field and transverse to the magnetic field and overplotted the results from the geodesic and Cartesian meshes. Figs. 9a to 9d plot the density, pressure, total velocity and total magnetic field along the direction of the original magnetic field. Figs. 9e to 9h plot the density, pressure, total velocity and total magnetic field perpendicular to the direction of the original magnetic field. The solid black and red lines in Fig. 9 show the results from the geodesic mesh and the Cartesian mesh respectively. We see that the solid and dashed curves track one another, showing that despite the use of curved zones, the results from the geodesic mesh are closely concordant with the results from the Cartesian mesh. The density, pressure and magnetic fields track one another especially closely. The velocity does show some inevitable differences because a uniform Cartesian mesh will have some natural advantages in propagating shocks along mesh lines. However, please note that any mesh that is based on ratioed spherical or cylindrical coordinates will also show some differences in shock propagation compared to a uniform Cartesian mesh.

**VIII.b) Blast Test Problem**

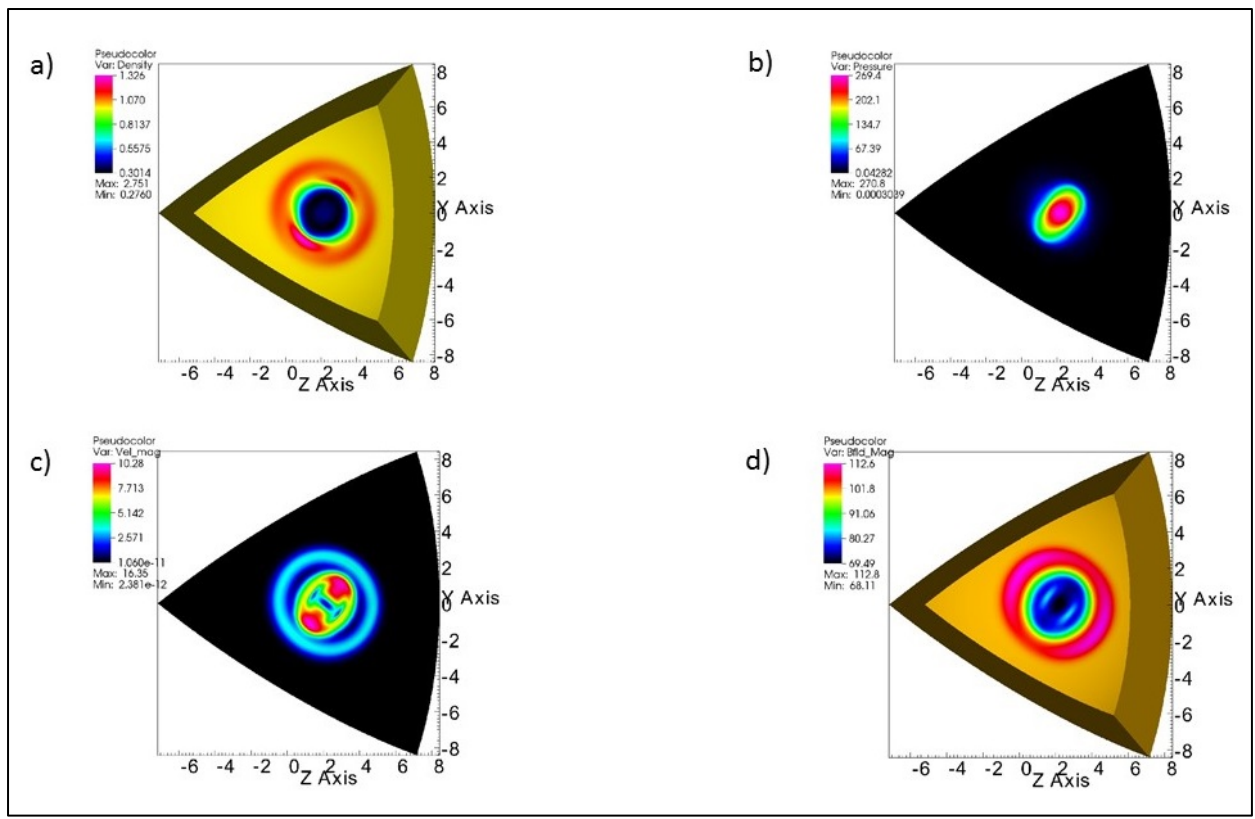

*Fig. 10 shows the results of the blast test problem at a final time of 0.06. Fig 10a shows the final density, 10b shows the final pressure, 10c shows the total velocity and 10d shows the magnitude of the magnetic field. The relevant flow variables are shown for a spherical surface having a radius of 11.26. The rest of the frustum is also shown.*

In this test problem, we simulated a three dimensional MHD blast problem on a spherical sector with a radial extent of $[7,16]$. A spherical explosion zone of unit radius with a high pressure of 1000 is initialized around the radius of 11.5 at the center of our computational domain. The rest of the computational domain has a pressure of 0.1. The initial density has a uniform value of 1 all over the domain. The initial velocity is set to zero. The magnitude of the initial, uniform magnetic field is set to 100. Each of the three components of the magnetic field have a magnitude of $100/\sqrt{3}$. The ratio of specific heat is 1.4 for this problem set up.

The above set up is run on a single sector of the spherical geodesic mesh with an angular resolution of $0.54^o$ and 180 logarithmically ratioed radial zones. We stopped the simulation at a time of 0.06. The relevant flow variables are shown for a spherical surface having a radius of 11.26. For the results shown in Figure 10, we used a third order accurate ADER-WENO scheme. Fig 10a shows the final density, 10b shows the final pressure, 10c shows the total velocity and 10d shows the magnitude of the magnetic field. We see that all flow features in this stringent blast problem are well-resolved.

### VIII.c) Spherical MHD Shock Tubes

Let us begin our discussion of spherical Riemann problems by providing a clarification. We state at the onset that a Riemann problem is a self-similar solution of a hyperbolic PDE. In one-dimension, and on a Cartesian mesh, it arises when there is a discontinuity in the initial conditions at a single position. Usually, that position is taken to be the center of the computational domain, but the discontinuous solution can be initialized anywhere in the one-dimensional domain. Researchers sometimes build Riemann problems in spherical geometry, where the initial conditions assume one set of values within some radius and another set of values outside that radius. A simple example of such a situation is given by the well-known Sod shock problem, transcribed to spherical geometry. For this problem, we take a spherical mesh with radial extent $[2,3]$. We then initialize the problem as

$$(\rho_L, P_L, \mathbf{v}_L) = (1,1,0) \quad \text{and} \quad (\rho_R, P_R, \mathbf{v}_R) = (0.125, 0.1, 0)$$

Here $(\rho_L, P_L, \mathbf{v}_L)$ pertains to the variables with $r \leq 2.5$ and $(\rho_R, P_R, \mathbf{v}_R)$ pertains to variable with $r > 2.5$. Fig. 11a, 11b and 11c shows the density, pressure and x-velocity from a one-dimensional Sod shock problem in Cartesian geometry while Fig. 11d, 11e and 11f shows the density, pressure and radial velocity in spherical geometry. In Figs. 11d, 11e and 11f the black dots show the actual data points while the overlaid thin solid curve shows the analytical result for the spherical Riemann problem; it is satisfying to note that the numerical and analytical results track each other very well. Both results are shown at a time of 0.2 and both results used a mesh with 300 zones. We see that the analogous fluid variables track one another quite well. It is for this reason that we refer to the variables in Figs. 11d, 11e and 11f as a spherical Riemann problem. We make this statement even though we realize that Figs. 11d, 11e and 11f do not truly evolve in a self-similar fashion; i.e. they are not truly Riemann problems in the sense of having a self-similar evolution. Therefore, it is important to realize that spherical Riemann problems will show some of the signatures of an actual one-dimensional Riemann problem in Cartesian geometry, even though a spherical Riemann problem is not strictly-speaking a Riemann problem. Viewed physically, the spherical geometry is like a diverging nozzle (Blandford & Rees 1974) and outwardly-propagating flow features move at different speeds from inwardly-propagating flow features.

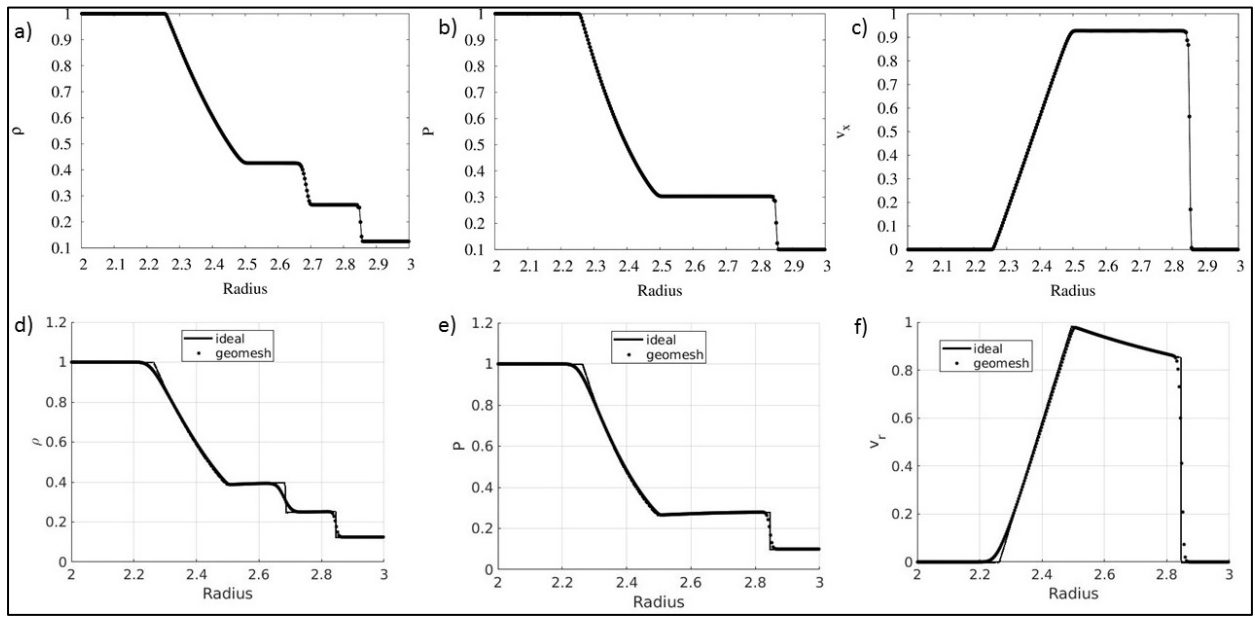

*Fig. 11a, 11b and 11c shows the density, pressure and x-velocity from a one-dimensional Sod shock problem in Cartesian geometry while Fig. 11d, 11e and 11f shows the density, pressure and radial velocity in spherical geometry. Both results are shown at a time of 0.2. The solid line in Figs. 11d, 11e and 11f shows the reference solution.*

That distinction becomes especially relevant when dealing with spherical Riemann problems for MHD. It becomes relevant because in spherical geometry we can only initialize a radial magnetic field whose magnitude falls off as the reciprocal of the square of the radius. This has three consequences. First, any magnetic field with radial variation also has a radial variation in magnetic pressure and is, therefore, not steady state. Therefore, we should accept that even without any jump in the other variables, there is no steady state. Second, this has consequences for the imposition of boundary conditions at the inner and outer boundaries. If constant fluid variables have no option to evolve because of a radial magnetic field, then we have no hope of asserting static boundary conditions at the inner and outer radial boundaries of the mesh. The best we can do is to pick a large enough computational domain in the radial direction and show that the interior solution for a spherical Riemann problem mimics the corresponding solution for a Riemann problem in Cartesian geometry. Third, the strong radial variation in the radially-oriented magnetic field changes the timestep as well as the stopping time in the problem.

There is a further complication in setting up the variation in the transverse magnetic field variables. Such variables would have a toroidal geometry and having a strong toroidal magnetic field at the polecaps of a spherical mesh would produce numerical instability. For that reason, we choose to vary the toroidal magnetic field so that it achieves its full value only on the equator while smoothly going to zero at the poles. Consequently, the spherical MHD Riemann problems will only be analogous to the one-dimensional Riemann problems in Cartesian geometry when the variables are plotted out at the equator. Therefore, all the solutions that we show here are plotted in the equatorial plane. Let "$r_m$" be the radial location where we wish to have a variation in the value of the toroidal magnetic field (in the vicinity of the equatorial plane). This can be set up using a magnetic vector potential of the form

$$A_r(r,\theta) = \frac{1}{2} B_{\phi L} r \cos\theta \left[ 1 - \tanh\left(\frac{r-r_m}{\delta}\right) \right] + \frac{1}{2} B_{\phi R} r \cos\theta \left[ 1 + \tanh\left(\frac{r-r_m}{\delta}\right) \right]$$

Such a magnetic vector potential produces a toroidal magnetic field of the form

$$B_\phi(r,\theta) = \frac{1}{2} B_{\phi L} \sin\theta \left[ 1 - \tanh\left(\frac{r-r_m}{\delta}\right) \right] + \frac{1}{2} B_{\phi R} \sin\theta \left[ 1 + \tanh\left(\frac{r-r_m}{\delta}\right) \right]$$

Here "$\delta$" is the small distance over which the field varies from "$B_{\phi L}$" to "$B_{\phi R}$" and "$r_m$" is the radial location where this variation takes place. Usually, we set "$\delta$" to be a value that is half a zone size or so. The advantage of introducing the taper in the above two equations is that we can then integrate the vector potential over the edges of the mesh in order to obtain the facial components of the magnetic field for all the zones. No such taper is needed, or used, for the fluid variables.

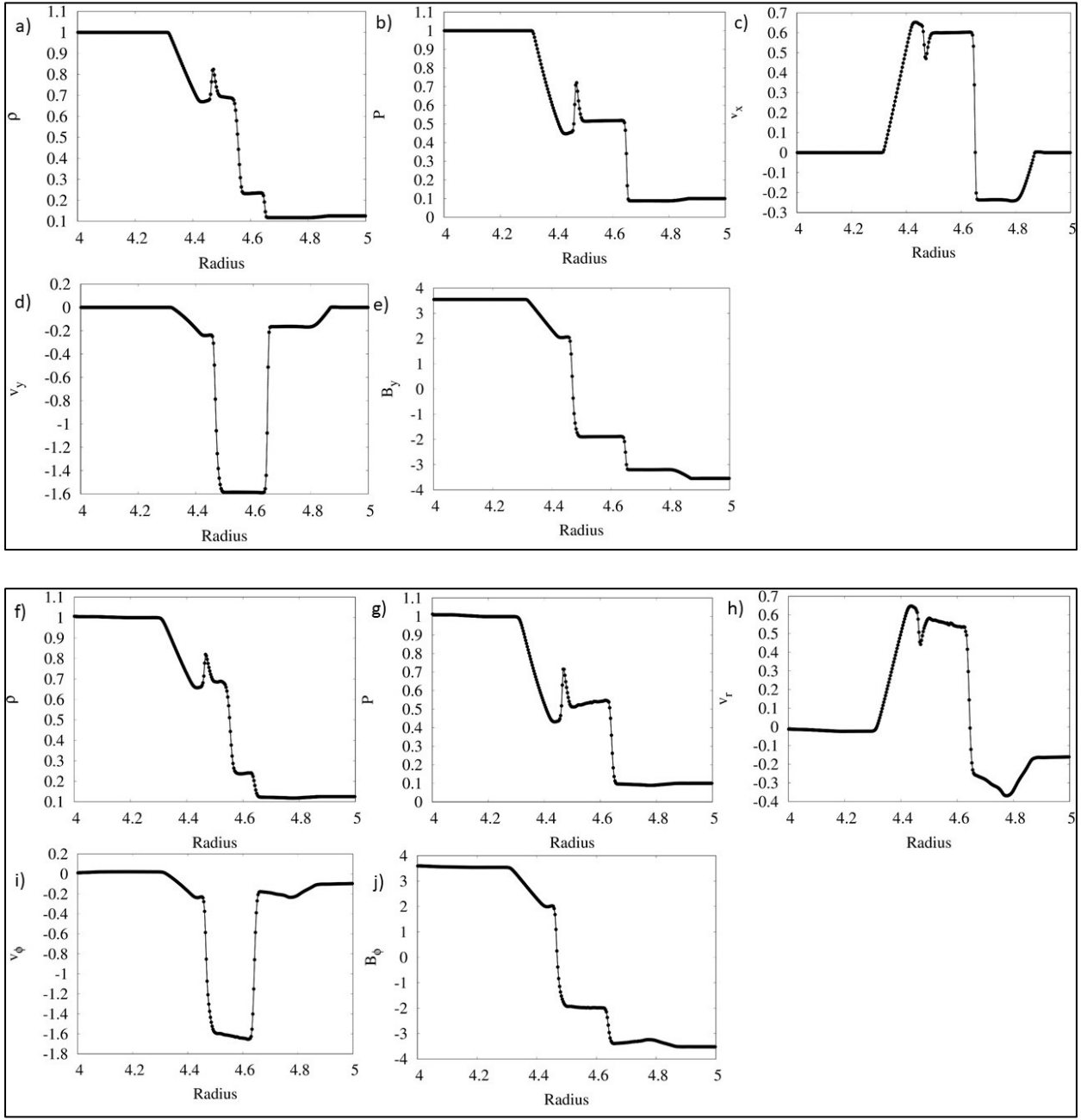

*In Figs. 12a, 12b, 12c, 12d, 12e we plot out the density, pressure, x-velocity, y-velocity and y-magnetic field for the Cartesian version of the Brio-Wu test problem. In Figs. 12f, 12g, 12h, 12i, 12j we plot out the density, pressure, radial velocity, toroidal velocity and toroidal magnetic field in the equatorial plane of a spherical mesh.*

Our first spherical MHD Riemann problem is adapted from Brio & Wu (1988). The problem has a radial extent of $[4,6]$. We have

$$(\rho_L, P_L, v_{rL}, v_{\theta L}, v_{\phi L}, B_{\theta L}, B_{\phi L}) = (1, 1, 0, 0, 0, 0, \sqrt{4\pi}) \quad \text{and}$$
$$(\rho_R, P_R, v_{rR}, v_{\theta R}, v_{\phi R}, B_{\theta R}, B_{\phi LR}) = (0.125, 0.1, 0, 0, 0, 0, -\sqrt{4\pi})$$

For the radial magnetic field we set

$$B_r(r,\theta) = 0.75\sqrt{4\pi}\sin\theta\left(\frac{4.5}{r}\right)^2$$

We use a ratio of specific heats given by 2.0. Notice that at the equator, i.e. at $\theta = \pi/2$, and at a radial location of 4.5 we have the same variation in the MHD variables as in the conventional Brio-Wu test problem in Cartesian geometry.

We plot out the result of our spherical Brio-Wu Riemann problem at a time of 0.1 and restrict our plot to the radial extent given by $[4,5]$, which corresponds to the inner 300 zones. The Cartesian version of this problem was run to a final time of 0.1 and also had 300 zones. To facilitate comparison, we plot out the density, pressure, x-velocity, y-velocity and y-magnetic field for the Cartesian version of the Brio-Wu test problem in Figs. 12a, 12b, 12c, 12d, 12e. In Figs. 12f, 12g, 12h, 12i, 12j we plot out the density, pressure, radial velocity, toroidal velocity and toroidal magnetic field in the equatorial plane of a spherical mesh. We see that the densities are closely analogous and even show the presence of a compound wave. The contact discontinuity is not as sharp in the spherical case because it does not evolve self-similarly and, therefore, cannot establish a crisp profile. The pressure in the outward-propagating shock for the spherical Riemann problem shows a radial variation, with a steepening of the pressure as a function of radius, this is expected because the pressure is propagating into a region with progressively lower magnetic pressure. We also see the formation of a rotational discontinuity in the magnetic field, consistent with the presence of a compound wave. We see, therefore, that many of the features in the Cartesian Riemann problem are replicated in the spherical Riemann problem and the points of deviation are also explained by the presence of a spherical geometry.

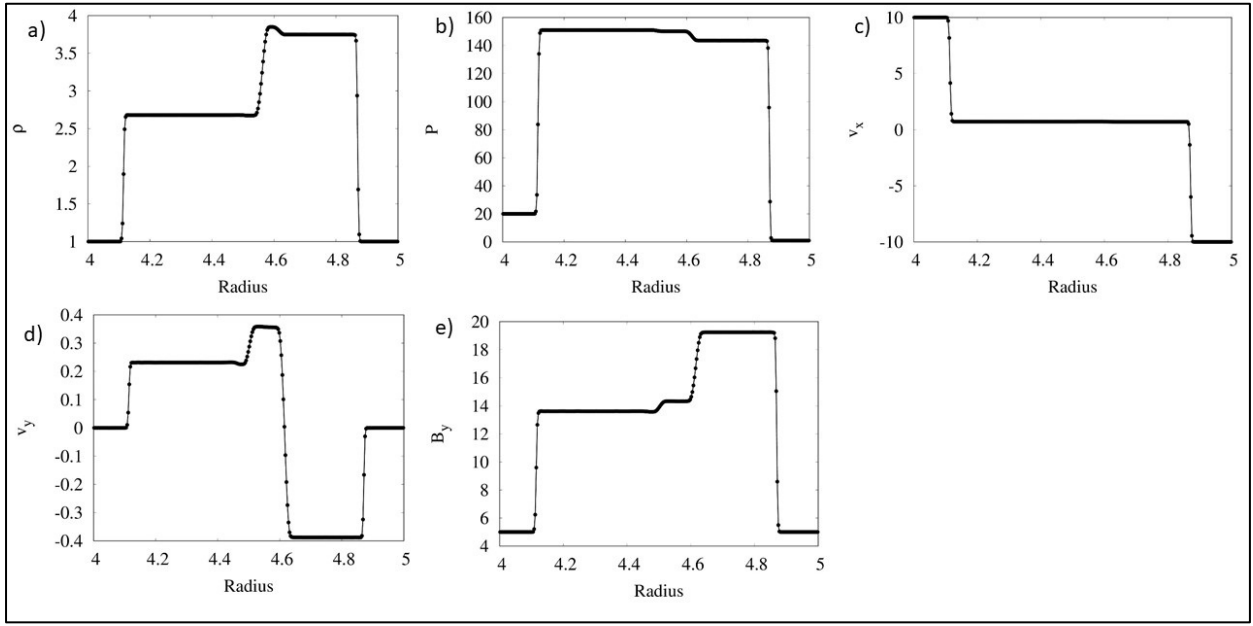

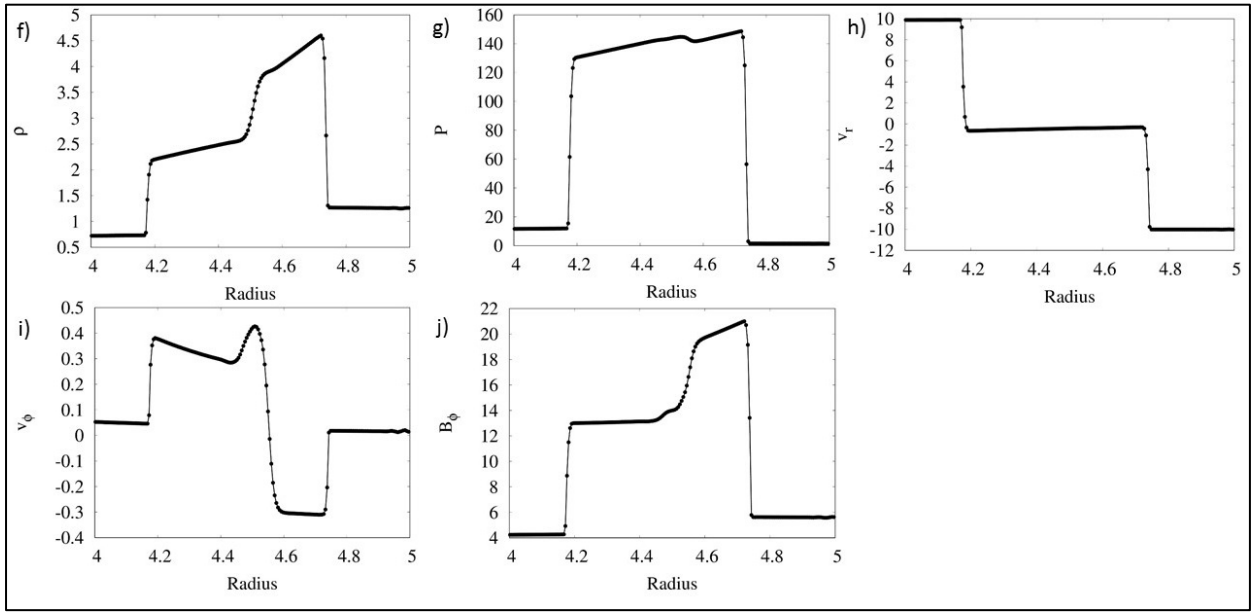

*In Figs. 13a, 13b, 13c, 13d, 13e we plot out the density, pressure, x-velocity, y-velocity and y-magnetic field for the Cartesian version of one of the Ryu and Jones test problems. In Figs. 13f, 13g, 13h, 13i, 13j we plot out the density, pressure, radial velocity, toroidal velocity and toroidal magnetic field in the equatorial plane of a spherical mesh.*

Our second spherical MHD Riemann problem is adapted from Ryu & Jones (1995). This problem has a radial extent of [3,6] . We have

$$\left(\rho_L, P_L, v_{rL}, v_{\theta L}, v_{\phi L}, B_{\theta L}, B_{\phi L}\right) = (1, 20, 10, 0, 0, 0, 5) \quad \text{and}$$
$$\left(\rho_R, P_R, v_{rR}, v_{\theta R}, v_{\phi R}, B_{\theta R}, B_{\phi LR}\right) = (1, 1, -10, 0, 0, 0, 5)$$

For the radial magnetic field we set

$$B_r(r, \theta) = 5 \sin \theta \left(\frac{4.5}{r}\right)^2$$

We use a ratio of specific heats given by 5/3. As in the previous problem, notice that at the equator, i.e. at $\theta = \pi/2$, and at a radial location of 4.5 we have the same variation in the MHD variables as in the conventional test problem presented in the above reference in Cartesian geometry.

We plot out the result of this Riemann problem at a time of 0.06 and restrict our plot to the radial extent given by [4,5] which corresponds to the central 300 zones. The Cartesian version of this problem was run to a final time of 0.08. To facilitate comparison, we plot out the density, pressure, x-velocity, y-velocity and y-magnetic field for the Cartesian version of this Riemann problem in Figs. 13a, 13b, 13c, 13d, 13e. In Figs. 13f, 13g, 13h, 13i, 13j we plot out the density, pressure, radial velocity, toroidal velocity and toroidal magnetic field in the equatorial plane of a spherical mesh. We see that the inward- and outward-propagating shocks have travelled at different speeds. The density variables also show interesting differences.

## IX) Sustained PetaScale Performance

In today's research environment, it is very beneficial to demonstrate that an astrophysical algorithm/code can also support sustained PetaScale Performance. To that end, we present a weak scalability study of the geodesic mesh version of the *RIEMANN* code. In such a study one keeps the number of zones per processor the same but one increases the problem size while proportionally increasing the number of processing cores on a modern supercomputer. The scalability study was carried out on the Blue Waters supercomputer at NCSA by sequentially doubling the angular resolution on the surface of the sphere and then doubling the radial resolution.

With every doubled angular resolution on the surface of the sphere, we have a four-fold increase in the number of triangles; see Fig. 1. As a result, every time the angular resolution was doubled, we timed the same problem with a four-fold increase in the number of cores. Every time the radial resolution was doubled, we timed the same problem with a two-fold increase in the number of cores.

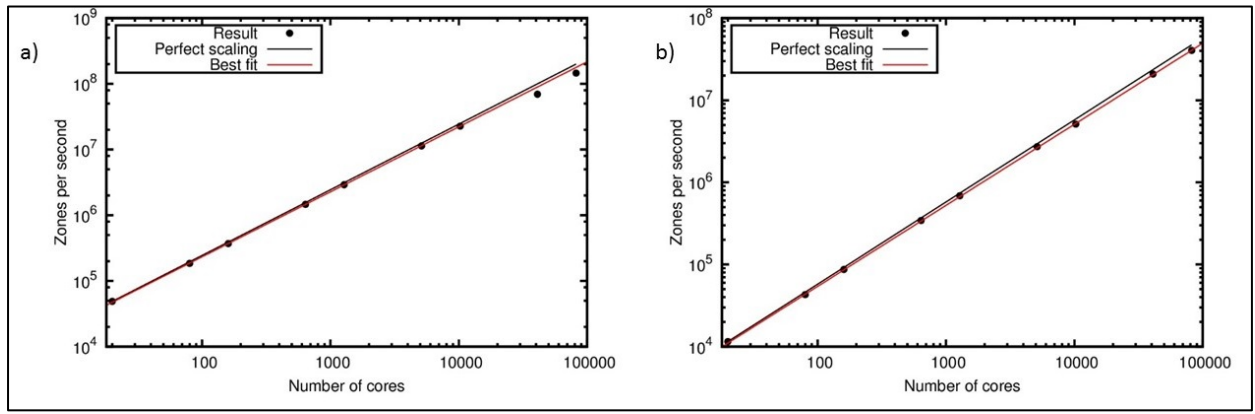

*Fig. 14a shows the scalability study for a second order ADER-WENO simulation with linear, isoparametric, mapping to the geometry. Fig. 14b shows the scalability study for a third order ADER-WENO simulation with quadratic, isoparametric, mapping to the geometry.*

Fig. 14a shows the scalability study for a second order ADER-WENO simulation with linear, isoparametric, mapping to the geometry. Fig. 14b shows the scalability study for a third order ADER-WENO simulation with quadratic, isoparametric, mapping to the geometry. We have found that increasing the geometric complexity of the mapping has almost no effect on the speed of the ADER algorithm which means that one can always have an optimal mapping to the curved spherical surface without loss of speed. We see that the lower order and higher order algorithms both have superlative scalability. This is attributed to the fact that the ADER algorithm provides a single stage update which requires only one synchronization across processors per timestep. The larger stencil in the higher order scheme does not degrade the scalability to any noticeable extent. The third order algorithm is about 3 times slower than the second order algorithm; but this in keeping with analogous findings in Balsara *et al*. (2009).

It is also worth documenting that the scalability of the geodesic mesh code is virtually comparable to the scalability of Cartesian mesh-based astrophysical codes. The interested reader can compare the scalability in Fig. 14 to the results from Garain, Balsara & Reid (2015) which show the corresponding scalability of a Cartesian mesh-based astrophysical code. We see that despite the use of zones that are logically equivalent to triangular prisms, the two codes have comparable scalability. The reason is that we use well-designed message packing and unpacking strategies to ensure that the geodesic mesh-based code exchanges data as efficiently as a Cartesian mesh-based code on a parallel supercomputer. In other words, though the mesh looks like an unstructured, triangulated mesh, the messaging is as efficient as the messaging in a structured mesh code.

## X) Conclusions

In this paper we have presented many novel algorithmic elements that contribute to the design of higher order divergence-free MHD schemes for isoparametrically mapped meshes. The geodesically mapped meshes on spheres can be regarded as one of the very specific use cases of these novel algorithms. Several application areas in space-physics, astrophysics and other areas of science and engineering have need for such algorithms. By developing these algorithms for a 3D geodesic meshing of the sphere, we demonstrate that these algorithms all work together to produce highly accurate results. They are also shown to be robust performers when strong discontinuities are present in the MHD flow. The above-mentioned algorithms have all been implemented in the geodesic mesh version of the *RIEMANN* code.

The fluid variables are reconstructed using a WENO-AO algorithm in Taylor series basis (Balsara *et al*. 2018). We use a Yee-type collocation of facially-averaged magnetic fields along with edge-integrated electric fields in order to achieve a high order accurate numerical treatment of Faraday's law. The facially-averaged normal components of the magnetic field at each face of a frustum-shaped zone, therefore, constitute the primal magnetic field variables in our scheme. The Cauchy problem for any PDE requires that we should have a complete representation of the spatial variation of the solution in order to extend that solution in the time direction. For this reason, we extend the divergence-free magnetic reconstruction strategies from Balsara (2001, 2004, 2009) and Balsara & Dumbser (2015a) so that they can be adapted to isoparametrically mapped meshes. A stepwise description of the divergence-free reconstruction algorithm for magnetic fields is given in Sub-section IV.b in order to facilitate easy implementation.

Once we have the spatial variation of all the MHD flow variables at all locations on the mesh, we wish to make a temporal extension of the same. This is very useful because, if done properly, such a predictor step will enable us to design a one-step update strategy. The advantage of such strategies is that they can be parallelized on a parallel supercomputer with only one messaging step per timestep. The predictor step that we develop is based on a modification of the ADER algorithm, where the algorithm is formulated so that it can function seamlessly on isoparametrically mapped meshes. A further innovation consists of formulating the ADER algorithm using serendipity elements, thereby reducing the computational complexity of the algorithm. A stepwise description of the isoparametrically mapped ADER algorithm in serendipity basis is given in Sub-section V.d in order to facilitate easy implementation.

Once the predictor step has provided us with the space and time evolution of the solution "in the small" within each zone, we are ready for the corrector step. The MHD corrector step, which consists of applying one-dimensional Riemann solvers at facial quadrature points and multidimensional Riemann solvers at the edges of the mesh, is reduced to a single step operation. The entire scheme is sketched in Sub-section II.c and a stepwise implementation strategy is presented in Section VI.

Several tests are presented in Section VII to show that the method achieves its design accuracy. Stringent test problems are also presented in Section VIII to show that the method can simultaneously handle strong shocks while retaining high order of accuracy in regions of smooth flow.

**Acknowledgements**

DSB acknowledges support via NSF grants NSF-ACI-1533850, NSF-DMS-1622457 , NSF_ACI-1713765 and NSF-DMS-1821242. Support from a grant by Notre Dame International is also acknowledged. VAF acknowledges support via NSF grant NSF-DMS-1361197 and NASA grant NNX17AB85G. KFG acknowledges support from NSF grant NSF-DMS-1361197 and Simons foundation grant 245237. Several simulations were performed on a cluster at UND that is run by the Center for Research Computing.  Computer support on NSF's XSEDE and Blue Waters computing resources is also acknowledged.

# Supplement: Efficient, Divergence-Free, High Order MHD on 3D Spherical Meshes with Optimal Geodesic Meshing

## By

**Dinshaw S. Balsara[1,2], Vladimir Florinski[3], Sudip Garain[2], Sethupathy Subramanian[2], Katharine F. Gurski[4]**

(dbalsara@nd.edu, sgarain@nd.edu, ssubrama@nd.edu) [1]ACMS Department and [2]Physics Department, University of Notre Dame

(vaf0001@uah.edu) [3]Space Physics, University of Alabama, Huntsville,

(kgurski@howard.edu) [4]Department of Mathematics, Howard University

**Appendix A**

The triangles and squares in Fig. 6 are assumed to be centered at the origin of the coordinate system. We specify the three facial nodes and the corresponding weights for the triangle in Fig. 6a by

$$(x_1, y_1) = \left(\frac{1}{4}, \frac{1}{4\sqrt{3}}\right) \quad ; w_1 = 1/3$$
$$(x_2, y_2) = \left(\frac{3}{4}, \frac{1}{4\sqrt{3}}\right) \quad ; w_2 = 1/3 \quad \text{(A.1)}$$
$$(x_3, y_3) = \left(\frac{1}{2}, \frac{1}{\sqrt{3}}\right) \quad ; w_3 = 1/3$$

The three facial nodes and the corresponding weights for the square in Fig. 6b are given by

$$(x_1, y_1) = \left(\frac{1-1/\sqrt{3}}{2}, \frac{1+1/\sqrt{3}}{2}\right) ; \quad w_1 = \frac{1}{4}$$
$$(x_2, y_2) = \left(\frac{1+1/\sqrt{3}}{2}, \frac{1+1/\sqrt{3}}{2}\right) ; \quad w_2 = \frac{1}{4} \quad \text{(A.2)}$$
$$(x_3, y_3) = \left(\frac{1}{2}, \frac{1-1/\sqrt{3}}{2}\right) ; \quad w_3 = \frac{1}{2}$$

The six facial nodes and the corresponding weights for the triangle in Fig. 6c are given by

$$(x_1, y_1) = \left(\frac{1}{8}, \frac{1}{8\sqrt{3}}\right) ; \quad w_1 = \frac{1}{8}$$

$$(x_2, y_2) = \left(\frac{7}{8}, \frac{1}{8\sqrt{3}}\right) ; \quad w_2 = \frac{1}{8}$$

$$(x_3, y_3) = \left(\frac{1}{2}, \frac{5}{4\sqrt{3}}\right) ; \quad w_3 = \frac{1}{8}$$

$$(x_4, y_4) = \left(\frac{3}{8}, \frac{\sqrt{3}}{8}\right) ; \quad w_4 = \frac{5}{24} \quad (A.3)$$

$$(x_5, y_5) = \left(\frac{5}{8}, \frac{\sqrt{3}}{8}\right) ; \quad w_5 = \frac{5}{24}$$

$$(x_6, y_6) = \left(\frac{1}{2}, \frac{\sqrt{3}}{4}\right) ; \quad w_6 = \frac{5}{24}$$

The four facial nodes and the corresponding weights for the square in Fig. 6d are given by

$$(x_1, y_1) = \left(\frac{1-1/\sqrt{3}}{2}, \frac{1-1/\sqrt{3}}{2}\right) ; \quad w_1 = \frac{1}{4}$$

$$(x_2, y_2) = \left(\frac{1+1/\sqrt{3}}{2}, \frac{1-1/\sqrt{3}}{2}\right) ; \quad w_2 = \frac{1}{4}$$

$$(x_3, y_3) = \left(\frac{1-1/\sqrt{3}}{2}, \frac{1+1/\sqrt{3}}{2}\right) ; \quad w_3 = \frac{1}{4} \quad (A.4)$$

$$(x_4, y_4) = \left(\frac{1+1/\sqrt{3}}{2}, \frac{1+1/\sqrt{3}}{2}\right) ; \quad w_4 = \frac{1}{4}$$

The ten facial nodes and the corresponding weights for the triangle in Fig. 6e are given by

$$(x_1, y_1) = \left(\frac{1}{8}, \frac{\sqrt{3}}{24}\right) ; \quad w_1 = \frac{32}{405}$$

$$(x_2, y_2) = \left(\frac{7}{8}, \frac{\sqrt{3}}{24}\right) ; \quad w_2 = \frac{32}{405}$$

$$(x_3, y_3) = \left(\frac{1}{2}, \frac{5\sqrt{3}}{12}\right) ; \quad w_3 = \frac{32}{405}$$

$$(x_4, y_4) = \left(\frac{3}{8}, \frac{\sqrt{3}}{24}\right) ; \quad w_4 = \frac{14}{135}$$

$$(x_5, y_5) = \left(\frac{5}{8}, \frac{\sqrt{3}}{24}\right) ; \quad w_5 = \frac{14}{135}$$

$$(x_6, y_6) = \left(\frac{3}{4}, \frac{\sqrt{3}}{6}\right) ; \quad w_6 = \frac{14}{135}$$

$$(x_7, y_7) = \left(\frac{5}{8}, \frac{7\sqrt{3}}{24}\right) ; \quad w_7 = \frac{14}{135}$$

$$(x_8, y_8) = \left(\frac{3}{8}, \frac{7\sqrt{3}}{24}\right) ; \quad w_8 = \frac{14}{135}$$

$$(x_9, y_9) = \left(\frac{1}{4}, \frac{\sqrt{3}}{6}\right) ; \quad w_9 = \frac{14}{135}$$

$$(x_{10}, y_{10}) = \left(\frac{1}{2}, \frac{\sqrt{3}}{6}\right) ; \quad w_{10} = \frac{19}{135}$$

(A.5)

The six facial nodes and the corresponding weights for the square in Fig. 6f are given by

$$(x_1, y_1) = \left(\frac{1-1/\sqrt{3}}{2}, \frac{1-\sqrt{3}/\sqrt{5}}{2}\right) \; ; \quad w_1 = \frac{5}{36}$$

$$(x_2, y_2) = \left(\frac{1+1/\sqrt{3}}{2}, \frac{1-\sqrt{3}/\sqrt{5}}{2}\right) \; ; \quad w_2 = \frac{5}{36}$$

$$(x_3, y_3) = \left(\frac{1-1/\sqrt{3}}{2}, \frac{1+\sqrt{3}/\sqrt{5}}{2}\right) \; ; \quad w_3 = \frac{5}{36} \tag{A.6}$$

$$(x_4, y_4) = \left(\frac{1+1/\sqrt{3}}{2}, \frac{1+\sqrt{3}/\sqrt{5}}{2}\right) \; ; \quad w_4 = \frac{5}{36}$$

$$(x_5, y_5) = \left(\frac{1-1/\sqrt{3}}{2}, \frac{1}{2}\right) \; ; \quad w_5 = \frac{2}{9}$$

$$(x_6, y_6) = \left(\frac{1+1/\sqrt{3}}{2}, \frac{1}{2}\right) \; ; \quad w_6 = \frac{2}{9}$$

**Appendix B**

This Appendix provides an explicit catalogue of the ADER-CG update, i.e. eqn. (5.28), at second order on the reference equilateral prism. All edges of the prism have unit length. In this Appendix, we denote the spatial nodes with numbers, as shown in Fig. 7a. The time levels are subscripted with letters just to make it easy to distinguish between different time levels. The time levels in eqn. (5.15) are, therefore, labeled as

$$\tau_a = 0 \; ; \quad \tau_b = 1/2 \tag{B.1}$$

The locations of the spatial nodes on the 3D reference triangular prism in Fig. 7a (with all edges having unit length, and centroid at the origin) are given by

$$(\xi_1, \eta_1, \zeta_1) = \left(0, 0, \frac{1}{2}\right) \tag{B.2}$$

$$(\xi_2, \eta_2, \zeta_2) = \left(1, 0, \frac{1}{2}\right) \tag{B.3}$$

$$(\xi_3, \eta_3, \zeta_3) = \left(\frac{1}{2}, \frac{\sqrt{3}}{2}, \frac{1}{2}\right) \tag{B.4}$$

$$(\xi_4, \eta_4, \zeta_4) = \left(0, 0, -\frac{1}{2}\right) \quad (B.5)$$

$$(\xi_5, \eta_5, \zeta_5) = \left(1, 0, -\frac{1}{2}\right) \quad (B.6)$$

$$(\xi_6, \eta_6, \zeta_6) = \left(\frac{1}{2}, \frac{\sqrt{3}}{2}, -\frac{1}{2}\right) \quad (B.7)$$

The solution vector at all space and time locations in the reference element can then be written explicitly using eqn. (5.22) as

$$\begin{aligned}
\tilde{\mathbf{U}}(\xi,\eta,\zeta,\tau) &= \hat{\mathbf{U}}_{a;1}\psi_1(\xi,\eta,\zeta)\phi_a(\tau) + \hat{\mathbf{U}}_{a;2}\psi_2(\xi,\eta,\zeta)\phi_a(\tau) + \hat{\mathbf{U}}_{a;3}\psi_3(\xi,\eta,\zeta)\phi_a(\tau) \\
&+ \hat{\mathbf{U}}_{a;4}\psi_4(\xi,\eta,\zeta)\phi_a(\tau) + \hat{\mathbf{U}}_{a;5}\psi_5(\xi,\eta,\zeta)\phi_a(\tau) + \hat{\mathbf{U}}_{a;6}\psi_6(\xi,\eta,\zeta)\phi_a(\tau) \\
&+ \hat{\mathbf{U}}_{b;1}\psi_1(\xi,\eta,\zeta)\phi_b(\tau) + \hat{\mathbf{U}}_{b;2}\psi_2(\xi,\eta,\zeta)\phi_b(\tau) + \hat{\mathbf{U}}_{b;3}\psi_3(\xi,\eta,\zeta)\phi_b(\tau) \\
&+ \hat{\mathbf{U}}_{b;4}\psi_4(\xi,\eta,\zeta)\phi_b(\tau) + \hat{\mathbf{U}}_{b;5}\psi_5(\xi,\eta,\zeta)\phi_b(\tau) + \hat{\mathbf{U}}_{b;6}\psi_6(\xi,\eta,\zeta)\phi_b(\tau)
\end{aligned} \quad (B.8)$$

Analogous expressions can be written for $\tilde{\mathbf{F}}(\xi,\eta,\zeta,\tau)$, $\tilde{\mathbf{G}}(\xi,\eta,\zeta,\tau)$, $\tilde{\mathbf{H}}(\xi,\eta,\zeta,\tau)$ and $\tilde{\mathbf{S}}(\xi,\eta,\zeta,\tau)$ from eqn. (5.21). The six update equations for the solution at time level "$b$" are given by first making explicit the matrix that gives us the contribution of the source terms from any one level to the state at the dynamically active level "$b$". The matrix is given by

$$\mathbf{R} = \begin{pmatrix} 0 & 0 \\ -1/6 & 2/3 \end{pmatrix} \quad (B.9)$$

Once the six nodes are specified, any computer algebra system can be used to obtain the nodal basis in the reference element. The six nodal basis functions can be explicitly written as,

$$\psi_1(\xi,\eta,\zeta) = \left(\frac{1}{6}\right) + \left(\frac{-1}{2}\right)\xi + \left(\frac{-1}{2\sqrt{3}}\right)\eta + \left(\frac{1}{3}\right)\zeta + (-1)\xi\zeta + \left(\frac{-1}{\sqrt{3}}\right)\eta\zeta \quad (B.10)$$

$$\psi_2(\xi,\eta,\zeta) = \left(\frac{1}{6}\right) + \left(\frac{1}{2}\right)\xi + \left(\frac{-1}{2\sqrt{3}}\right)\eta + \left(\frac{1}{3}\right)\zeta + \xi\zeta + \left(\frac{-1}{\sqrt{3}}\right)\eta\zeta \quad (B.11)$$

$$\psi_3(\xi,\eta,\zeta) = \left(\frac{1}{6}\right) + \left(\frac{1}{\sqrt{3}}\right)\eta + \left(\frac{1}{3}\right)\zeta + \left(\frac{2}{\sqrt{3}}\right)\eta\zeta \quad (B.12)$$

$$\psi_4(\xi,\eta,\zeta) = \left(\frac{1}{6}\right) + \left(\frac{-1}{2}\right)\xi + \left(\frac{-1}{2\sqrt{3}}\right)\eta + \left(\frac{-1}{3}\right)\zeta + \xi\zeta + \left(\frac{1}{\sqrt{3}}\right)\eta\zeta \tag{B.13}$$

$$\psi_5(\xi,\eta,\zeta) = \left(\frac{1}{6}\right) + \left(\frac{1}{2}\right)\xi + \left(\frac{-1}{2\sqrt{3}}\right)\eta + \left(\frac{-1}{3}\right)\zeta + (-1)\xi\zeta + \left(\frac{1}{\sqrt{3}}\right)\eta\zeta \tag{B.14}$$

$$\psi_6(\xi,\eta,\zeta) = \left(\frac{1}{6}\right) + \left(\frac{1}{\sqrt{3}}\right)\eta + \left(\frac{-1}{3}\right)\zeta + \left(\frac{-2}{\sqrt{3}}\right)\eta\zeta \tag{B.15}$$

The six update equations at time level $t_b$ are explicitly written as,

$$\begin{aligned}
\hat{\mathbf{U}}_{b;1} = \hat{\mathbf{U}}_{a;1} &+ (-1/6)\hat{\mathbf{S}}_{a;1} + (2/3)\hat{\mathbf{S}}_{b;1} \\
&- (-1/6)(-\hat{\mathbf{F}}_{a;1} + \hat{\mathbf{F}}_{a;2} - \hat{\mathbf{G}}_{a;1}/\sqrt{3} - \hat{\mathbf{G}}_{a;2}/\sqrt{3} + 2\hat{\mathbf{G}}_{a;3}/\sqrt{3} + \hat{\mathbf{H}}_{a;1} - \hat{\mathbf{H}}_{a;4}) \\
&- (\ 2/3)(-\hat{\mathbf{F}}_{b;1} + \hat{\mathbf{F}}_{b;2} - \hat{\mathbf{G}}_{b;1}/\sqrt{3} - \hat{\mathbf{G}}_{b;2}/\sqrt{3} + 2\hat{\mathbf{G}}_{b;3}/\sqrt{3} + \hat{\mathbf{H}}_{b;1} - \hat{\mathbf{H}}_{b;4})
\end{aligned} \tag{B.16}$$

$$\begin{aligned}
\hat{\mathbf{U}}_{b;2} = \hat{\mathbf{U}}_{a;2} &+ (-1/6)\hat{\mathbf{S}}_{a;2} + (2/3)\hat{\mathbf{S}}_{b;2} \\
&- (-1/6)(-\hat{\mathbf{F}}_{a;1} + \hat{\mathbf{F}}_{a;2} - \hat{\mathbf{G}}_{a;1}/\sqrt{3} - \hat{\mathbf{G}}_{a;2}/\sqrt{3} + 2\hat{\mathbf{G}}_{a;3}/\sqrt{3} + \hat{\mathbf{H}}_{a;2} - \hat{\mathbf{H}}_{a;5}) \\
&- (\ 2/3)(-\hat{\mathbf{F}}_{b;1} + \hat{\mathbf{F}}_{b;2} - \hat{\mathbf{G}}_{b;1}/\sqrt{3} - \hat{\mathbf{G}}_{b;2}/\sqrt{3} + 2\hat{\mathbf{G}}_{b;3}/\sqrt{3} + \hat{\mathbf{H}}_{b;2} - \hat{\mathbf{H}}_{b;5})
\end{aligned} \tag{B.17}$$

$$\begin{aligned}
\hat{\mathbf{U}}_{b;3} = \hat{\mathbf{U}}_{a;3} &+ (-1/6)\hat{\mathbf{S}}_{a;3} + (2/3)\hat{\mathbf{S}}_{b;3} \\
&- (-1/6)(-\hat{\mathbf{F}}_{a;1} + \hat{\mathbf{F}}_{a;2} - \hat{\mathbf{G}}_{a;1}/\sqrt{3} - \hat{\mathbf{G}}_{a;2}/\sqrt{3} + 2\hat{\mathbf{G}}_{a;3}/\sqrt{3} + \hat{\mathbf{H}}_{a;3} - \hat{\mathbf{H}}_{a;6}) \\
&- (\ 2/3)(-\hat{\mathbf{F}}_{b;1} + \hat{\mathbf{F}}_{b;2} - \hat{\mathbf{G}}_{b;1}/\sqrt{3} - \hat{\mathbf{G}}_{b;2}/\sqrt{3} + 2\hat{\mathbf{G}}_{b;3}/\sqrt{3} + \hat{\mathbf{H}}_{b;3} - \hat{\mathbf{H}}_{b;6})
\end{aligned} \tag{B.18}$$

$$\begin{aligned}
\hat{\mathbf{U}}_{b;4} = \hat{\mathbf{U}}_{a;4} &+ (-1/6)\hat{\mathbf{S}}_{a;4} + (2/3)\hat{\mathbf{S}}_{b;4} \\
&- (-1/6)(-\hat{\mathbf{F}}_{a;4} + \hat{\mathbf{F}}_{a;5} - \hat{\mathbf{G}}_{a;4}/\sqrt{3} - \hat{\mathbf{G}}_{a;5}/\sqrt{3} + 2\hat{\mathbf{G}}_{a;6}/\sqrt{3} + \hat{\mathbf{H}}_{a;1} - \hat{\mathbf{H}}_{a;4}) \\
&- (\ 2/3)(-\hat{\mathbf{F}}_{b;4} + \hat{\mathbf{F}}_{b;5} - \hat{\mathbf{G}}_{b;4}/\sqrt{3} - \hat{\mathbf{G}}_{b;5}/\sqrt{3} + 2\hat{\mathbf{G}}_{b;6}/\sqrt{3} + \hat{\mathbf{H}}_{b;1} - \hat{\mathbf{H}}_{b;4})
\end{aligned} \tag{B.19}$$

$$\begin{aligned}
\hat{\mathbf{U}}_{b;5} = \hat{\mathbf{U}}_{a;5} &+ (-1/6)\hat{\mathbf{S}}_{a;5} + (2/3)\hat{\mathbf{S}}_{b;5} \\
&- (-1/6)(-\hat{\mathbf{F}}_{a;4} + \hat{\mathbf{F}}_{a;5} - \hat{\mathbf{G}}_{a;4}/\sqrt{3} - \hat{\mathbf{G}}_{a;5}/\sqrt{3} + 2\hat{\mathbf{G}}_{a;6}/\sqrt{3} + \hat{\mathbf{H}}_{a;2} - \hat{\mathbf{H}}_{a;5}) \\
&- (\ 2/3)(-\hat{\mathbf{F}}_{b;4} + \hat{\mathbf{F}}_{b;5} - \hat{\mathbf{G}}_{b;4}/\sqrt{3} - \hat{\mathbf{G}}_{b;5}/\sqrt{3} + 2\hat{\mathbf{G}}_{b;6}/\sqrt{3} + \hat{\mathbf{H}}_{b;2} - \hat{\mathbf{H}}_{b;5})
\end{aligned} \tag{B.20}$$

$$\hat{\mathbf{U}}_{b;6} = \hat{\mathbf{U}}_{a;6} + (-1/6)\hat{\mathbf{S}}_{a;6} + (2/3)\hat{\mathbf{S}}_{b;6}$$
$$-(-1/6)(-\hat{\mathbf{F}}_{a;4} + \hat{\mathbf{F}}_{a;5} - \hat{\mathbf{G}}_{a;4}/\sqrt{3} - \hat{\mathbf{G}}_{a;5}/\sqrt{3} + 2\hat{\mathbf{G}}_{a;6}/\sqrt{3} + \hat{\mathbf{H}}_{a;3} - \hat{\mathbf{H}}_{a;6})$$
$$-(2/3)(-\hat{\mathbf{F}}_{b;4} + \hat{\mathbf{F}}_{b;5} - \hat{\mathbf{G}}_{b;4}/\sqrt{3} - \hat{\mathbf{G}}_{b;5}/\sqrt{3} + 2\hat{\mathbf{G}}_{b;6}/\sqrt{3} + \hat{\mathbf{H}}_{b;3} - \hat{\mathbf{H}}_{b;6})$$
(B.21)

**Appendix C**

This Appendix provides an explicit catalogue of the ADER-CG update, i.e. eqn. (5.28), at third order on the reference equilateral prism. All edges of the prism have unit length. In this Appendix, we denote the spatial nodes with numbers, as shown in Fig. 7b. The time levels are subscripted with letters just to make it easy to distinguish between different time levels. The time levels in eqn. (5.16) are, therefore, labeled as

$$\tau_a = 0 \quad ; \quad \tau_b = \frac{1}{2} - \frac{1}{2\sqrt{3}} \quad ; \quad \tau_c = \frac{1}{2} + \frac{1}{2\sqrt{3}} \tag{C.1}$$

The locations of the spatial nodes on the 3D reference triangular prism in Fig. 7b (with all edges having unit length, and centroid at the origin) are given by

$$(\xi_{01}, \eta_{01}, \zeta_{01}) = \left(-\frac{1}{2}, -\frac{1}{2\sqrt{3}}, \frac{1}{2}\right) \tag{C.2}$$

$$(\xi_{02}, \eta_{02}, \zeta_{02}) = \left(\frac{1}{2}, -\frac{1}{2\sqrt{3}}, \frac{1}{2}\right) \tag{C.3}$$

$$(\xi_{03}, \eta_{03}, \zeta_{03}) = \left(0, \frac{1}{\sqrt{3}}, \frac{1}{2}\right) \tag{C.4}$$

$$(\xi_{04}, \eta_{04}, \zeta_{04}) = \left(-\frac{1}{2}, -\frac{1}{2\sqrt{3}}, -\frac{1}{2}\right) \tag{C.5}$$

$$(\xi_{05}, \eta_{05}, \zeta_{05}) = \left(\frac{1}{2}, -\frac{1}{2\sqrt{3}}, -\frac{1}{2}\right) \tag{C.6}$$

$$(\xi_{06}, \eta_{06}, \zeta_{06}) = \left(0, \frac{1}{\sqrt{3}}, -\frac{1}{2}\right) \tag{C.7}$$

$$(\xi_{07}, \eta_{07}, \zeta_{07}) = \left(-\frac{1}{2}, -\frac{1}{2\sqrt{3}}, 0\right) \tag{C.8}$$

$$(\xi_{08}, \eta_{08}, \zeta_{08}) = \left(\frac{1}{2}, -\frac{1}{2\sqrt{3}}, 0\right) \tag{C.9}$$

$$(\xi_{09}, \eta_{09}, \zeta_{09}) = \left(0, \frac{1}{\sqrt{3}}, 0\right) \tag{C.10}$$

$$(\xi_{10}, \eta_{10}, \zeta_{10}) = \left(0, -\frac{1}{2\sqrt{3}}, \frac{1}{2}\right) \tag{C.11}$$

$$(\xi_{11}, \eta_{11}, \zeta_{11}) = \left(\frac{1}{4}, \frac{1}{4\sqrt{3}}, \frac{1}{2}\right) \tag{C.12}$$

$$(\xi_{12}, \eta_{12}, \zeta_{12}) = \left(-\frac{1}{4}, \frac{1}{4\sqrt{3}}, \frac{1}{2}\right) \tag{C.13}$$

$$(\xi_{13}, \eta_{13}, \zeta_{13}) = \left(0, -\frac{1}{2\sqrt{3}}, -\frac{1}{2}\right) \tag{C.14}$$

$$(\xi_{14}, \eta_{14}, \zeta_{14}) = \left(\frac{1}{4}, \frac{1}{4\sqrt{3}}, -\frac{1}{2}\right) \tag{C.15}$$

$$(\xi_{15}, \eta_{15}, \zeta_{15}) = \left(-\frac{1}{4}, \frac{1}{4\sqrt{3}}, -\frac{1}{2}\right) \tag{C.16}$$

The solution vector at all space and time locations in the reference element can then be written explicitly using eqn. (5.22) as

$$\begin{aligned}
\tilde{\mathbf{U}}(\xi,\eta,\zeta,\tau) = &\hat{\mathbf{U}}_{a;01}\psi_{01}(\xi,\eta,\zeta)\phi_a(\tau) + \hat{\mathbf{U}}_{a;02}\psi_{02}(\xi,\eta,\zeta)\phi_a(\tau) + \hat{\mathbf{U}}_{a;03}\psi_{03}(\xi,\eta,\zeta)\phi_a(\tau) \\
&+ \hat{\mathbf{U}}_{a;04}\psi_{04}(\xi,\eta,\zeta)\phi_a(\tau) + \hat{\mathbf{U}}_{a;05}\psi_{05}(\xi,\eta,\zeta)\phi_a(\tau) + \hat{\mathbf{U}}_{a;06}\psi_{06}(\xi,\eta,\zeta)\phi_a(\tau) \\
&+ \hat{\mathbf{U}}_{a;07}\psi_{07}(\xi,\eta,\zeta)\phi_a(\tau) + \hat{\mathbf{U}}_{a;08}\psi_{08}(\xi,\eta,\zeta)\phi_a(\tau) + \hat{\mathbf{U}}_{a;09}\psi_{09}(\xi,\eta,\zeta)\phi_a(\tau) \\
&+ \hat{\mathbf{U}}_{a;10}\psi_{10}(\xi,\eta,\zeta)\phi_a(\tau) + \hat{\mathbf{U}}_{a;11}\psi_{11}(\xi,\eta,\zeta)\phi_a(\tau) + \hat{\mathbf{U}}_{a;12}\psi_{12}(\xi,\eta,\zeta)\phi_a(\tau) \\
&+ \hat{\mathbf{U}}_{a;13}\psi_{13}(\xi,\eta,\zeta)\phi_a(\tau) + \hat{\mathbf{U}}_{a;14}\psi_{14}(\xi,\eta,\zeta)\phi_a(\tau) + \hat{\mathbf{U}}_{a;15}\psi_{15}(\xi,\eta,\zeta)\phi_a(\tau) \\
&+ \hat{\mathbf{U}}_{b;01}\psi_{01}(\xi,\eta,\zeta)\phi_b(\tau) + \hat{\mathbf{U}}_{b;02}\psi_{02}(\xi,\eta,\zeta)\phi_b(\tau) + \hat{\mathbf{U}}_{b;03}\psi_{03}(\xi,\eta,\zeta)\phi_b(\tau) \\
&+ \hat{\mathbf{U}}_{b;04}\psi_{04}(\xi,\eta,\zeta)\phi_b(\tau) + \hat{\mathbf{U}}_{b;05}\psi_{05}(\xi,\eta,\zeta)\phi_b(\tau) + \hat{\mathbf{U}}_{b;06}\psi_{06}(\xi,\eta,\zeta)\phi_b(\tau) \\
&+ \hat{\mathbf{U}}_{b;07}\psi_{07}(\xi,\eta,\zeta)\phi_b(\tau) + \hat{\mathbf{U}}_{b;08}\psi_{08}(\xi,\eta,\zeta)\phi_b(\tau) + \hat{\mathbf{U}}_{b;09}\psi_{09}(\xi,\eta,\zeta)\phi_b(\tau) \\
&+ \hat{\mathbf{U}}_{b;10}\psi_{10}(\xi,\eta,\zeta)\phi_b(\tau) + \hat{\mathbf{U}}_{b;11}\psi_{11}(\xi,\eta,\zeta)\phi_b(\tau) + \hat{\mathbf{U}}_{b;12}\psi_{12}(\xi,\eta,\zeta)\phi_b(\tau) \\
&+ \hat{\mathbf{U}}_{b;13}\psi_{13}(\xi,\eta,\zeta)\phi_b(\tau) + \hat{\mathbf{U}}_{b;14}\psi_{14}(\xi,\eta,\zeta)\phi_b(\tau) + \hat{\mathbf{U}}_{b;15}\psi_{15}(\xi,\eta,\zeta)\phi_b(\tau) \\
&+ \hat{\mathbf{U}}_{c;01}\psi_{01}(\xi,\eta,\zeta)\phi_c(\tau) + \hat{\mathbf{U}}_{c;02}\psi_{02}(\xi,\eta,\zeta)\phi_c(\tau) + \hat{\mathbf{U}}_{c;03}\psi_{03}(\xi,\eta,\zeta)\phi_c(\tau) \\
&+ \hat{\mathbf{U}}_{c;04}\psi_{04}(\xi,\eta,\zeta)\phi_c(\tau) + \hat{\mathbf{U}}_{c;05}\psi_{05}(\xi,\eta,\zeta)\phi_c(\tau) + \hat{\mathbf{U}}_{c;06}\psi_{06}(\xi,\eta,\zeta)\phi_c(\tau) \\
&+ \hat{\mathbf{U}}_{c;07}\psi_{07}(\xi,\eta,\zeta)\phi_c(\tau) + \hat{\mathbf{U}}_{c;08}\psi_{08}(\xi,\eta,\zeta)\phi_c(\tau) + \hat{\mathbf{U}}_{c;09}\psi_{09}(\xi,\eta,\zeta)\phi_c(\tau) \\
&+ \hat{\mathbf{U}}_{c;10}\psi_{10}(\xi,\eta,\zeta)\phi_c(\tau) + \hat{\mathbf{U}}_{c;11}\psi_{11}(\xi,\eta,\zeta)\phi_c(\tau) + \hat{\mathbf{U}}_{c;12}\psi_{12}(\xi,\eta,\zeta)\phi_c(\tau) \quad \text{(C.17)} \\
&+ \hat{\mathbf{U}}_{c;13}\psi_{13}(\xi,\eta,\zeta)\phi_c(\tau) + \hat{\mathbf{U}}_{c;14}\psi_{14}(\xi,\eta,\zeta)\phi_c(\tau) + \hat{\mathbf{U}}_{c;15}\psi_{15}(\xi,\eta,\zeta)\phi_c(\tau)
\end{aligned}$$

Analogous expressions can be written for $\tilde{\mathbf{F}}(\xi,\eta,\zeta,\tau)$, $\tilde{\mathbf{G}}(\xi,\eta,\zeta,\tau)$, $\tilde{\mathbf{H}}(\xi,\eta,\zeta,\tau)$ and $\tilde{\mathbf{S}}(\xi,\eta,\zeta,\tau)$ from eqn. (5.21). The fifteen update equations for the solution at time level "$b$", and the fifteen update equations for the solution at time level "$c$", are given by first specifying a matrix. That matrix gives us the contribution of the source terms from any one level to the state at any of the two dynamically active levels, i.e. "$b$" and "$c$". The matrix is given by

$$\mathbf{R} = \begin{pmatrix} 0 & 0 & 0 \\ \dfrac{-6+\sqrt{3}}{30} & \dfrac{18+5\sqrt{3}}{60} & \dfrac{24-17\sqrt{3}}{60} \\ \dfrac{-6-\sqrt{3}}{30} & \dfrac{24+17\sqrt{3}}{60} & \dfrac{18-5\sqrt{3}}{60} \end{pmatrix} \quad \text{(C.18)}$$

Once the fifteen nodes are specified, any computer algebra system can be used to obtain the nodal basis in the reference element. The fifteen nodal basis functions can be explicitly written as,

$$\psi_{01}(\xi,\eta,\zeta) = \left(\frac{-2}{9}\right) + \left(\frac{1}{3}\right)\xi + \left(\frac{1}{3\sqrt{3}}\right)\eta + \left(\frac{-1}{9}\right)\zeta + \xi^2 + \left(\frac{1}{3}\right)\eta^2 + \left(\frac{2}{3}\right)\zeta^2$$
$$+ \left(\frac{2}{\sqrt{3}}\right)\xi\eta + \left(\frac{-1}{3\sqrt{3}}\right)\eta\zeta + \left(\frac{-1}{3}\right)\xi\zeta + \left(\frac{4}{\sqrt{3}}\right)\xi\eta\zeta \quad \text{(C.19)}$$
$$+ (-2)\xi\zeta^2 + \left(\frac{-2}{\sqrt{3}}\right)\eta\zeta^2 + (2)\xi^2\zeta + \left(\frac{2}{3}\right)\eta^2\zeta$$

$$\psi_{02}(\xi,\eta,\zeta) = \left(\frac{-2}{9}\right) + \left(\frac{-1}{3}\right)\xi + \left(\frac{1}{3\sqrt{3}}\right)\eta + \left(\frac{-1}{9}\right)\zeta + \xi^2 + \left(\frac{1}{3}\right)\eta^2 + \left(\frac{2}{3}\right)\zeta^2$$
$$+ \left(\frac{-2}{\sqrt{3}}\right)\xi\eta + \left(\frac{-1}{3\sqrt{3}}\right)\eta\zeta + \left(\frac{1}{3}\right)\xi\zeta + \left(\frac{-4}{\sqrt{3}}\right)\xi\eta\zeta \quad \text{(C.20)}$$
$$+ (2)\xi\zeta^2 + \left(\frac{-2}{\sqrt{3}}\right)\eta\zeta^2 + (2)\xi^2\zeta + \left(\frac{2}{3}\right)\eta^2\zeta$$

$$\psi_{03}(\xi,\eta,\zeta) = \left(\frac{-2}{9}\right) + \left(\frac{-2}{3\sqrt{3}}\right)\eta + \left(\frac{-1}{9}\right)\zeta + \left(\frac{4}{3}\right)\eta^2 + \left(\frac{2}{3}\right)\zeta^2$$
$$+ \left(\frac{2}{3\sqrt{3}}\right)\eta\zeta + \left(\frac{4}{\sqrt{3}}\right)\eta\zeta^2 + \left(\frac{8}{3}\right)\eta^2\zeta \quad \text{(C.21)}$$

$$\psi_{04}(\xi,\eta,\zeta) = \left(\frac{-2}{9}\right) + \left(\frac{1}{3}\right)\xi + \left(\frac{1}{3\sqrt{3}}\right)\eta + \left(\frac{1}{9}\right)\zeta + (1)\xi^2 + \left(\frac{1}{3}\right)\eta^2 + \left(\frac{2}{3}\right)\zeta^2$$
$$+ \left(\frac{2}{\sqrt{3}}\right)\xi\eta + \left(\frac{1}{3\sqrt{3}}\right)\eta\zeta + \left(\frac{1}{3}\right)\xi\zeta + \left(\frac{-4}{\sqrt{3}}\right)\xi\eta\zeta \quad \text{(C.22)}$$
$$+ (-2)\xi\zeta^2 + \left(\frac{-2}{\sqrt{3}}\right)\eta\zeta^2 + (-2)\xi^2\zeta + \left(\frac{-2}{3}\right)\eta^2\zeta$$

$$\psi_{05}(\xi,\eta,\zeta) = \left(\frac{-2}{9}\right) + \left(\frac{-1}{3}\right)\xi + \left(\frac{1}{3\sqrt{3}}\right)\eta + \left(\frac{1}{9}\right)\zeta + (1)\xi^2 + \left(\frac{1}{3}\right)\eta^2 + \left(\frac{2}{3}\right)\zeta^2$$
$$+ \left(\frac{-2}{\sqrt{3}}\right)\xi\eta + \left(\frac{1}{3\sqrt{3}}\right)\eta\zeta + \left(\frac{-1}{3}\right)\xi\zeta + \left(\frac{4}{\sqrt{3}}\right)\xi\eta\zeta \quad \text{(C.23)}$$
$$+ (2)\xi\zeta^2 + \left(\frac{-2}{\sqrt{3}}\right)\eta\zeta^2 + (-2)\xi^2\zeta + \left(\frac{-2}{3}\right)\eta^2\zeta$$

$$\psi_{06}(\xi,\eta,\zeta) = \left(\frac{-2}{9}\right) + \left(\frac{-2}{3\sqrt{3}}\right)\eta + \left(\frac{1}{9}\right)\zeta + \left(\frac{4}{3}\right)\eta^2 + \left(\frac{2}{3}\right)\zeta^2$$
$$+ \left(\frac{-2}{3\sqrt{3}}\right)\eta\zeta + \left(\frac{4}{\sqrt{3}}\right)\eta\zeta^2 + \left(\frac{-8}{3}\right)\eta^2\zeta \tag{C.24}$$

$$\psi_{07}(\xi,\eta,\zeta) = \left(\frac{1}{3}\right) + (-1)\xi + \left(\frac{-1}{\sqrt{3}}\right)\eta + \left(\frac{-4}{3}\right)\zeta^2 + (4)\xi\zeta^2 + \left(\frac{4}{\sqrt{3}}\right)\eta\zeta^2 \tag{C.25}$$

$$\psi_{08}(\xi,\eta,\zeta) = \left(\frac{1}{3}\right) + (1)\xi + \left(\frac{-1}{\sqrt{3}}\right)\eta + \left(\frac{-4}{3}\right)\zeta^2 + (-4)\xi\zeta^2 + \left(\frac{4}{\sqrt{3}}\right)\eta\zeta^2 \tag{C.26}$$

$$\psi_{09}(\xi,\eta,\zeta) = \left(\frac{1}{3}\right) + \left(\frac{2}{\sqrt{3}}\right)\eta + \left(\frac{-4}{3}\right)\zeta^2 + \left(\frac{-8}{\sqrt{3}}\right)\eta\zeta^2 \tag{C.27}$$

$$\psi_{10}(\xi,\eta,\zeta) = \left(\frac{2}{9}\right) + \left(\frac{-4}{3\sqrt{3}}\right)\eta + \left(\frac{4}{9}\right)\zeta + (-2)\xi^2 + \left(\frac{2}{3}\right)\eta^2$$
$$+ \left(\frac{-8}{3\sqrt{3}}\right)\eta\zeta + (-4)\xi^2\zeta + \left(\frac{4}{3}\right)\eta^2\zeta \tag{C.28}$$

$$\psi_{11}(\xi,\eta,\zeta) = \left(\frac{2}{9}\right) + \left(\frac{2}{3}\right)\xi + \left(\frac{2}{3\sqrt{3}}\right)\eta + \left(\frac{4}{9}\right)\zeta + \left(\frac{-4}{3}\right)\eta^2$$
$$+ \left(\frac{4}{\sqrt{3}}\right)\xi\eta + \left(\frac{4}{3\sqrt{3}}\right)\eta\zeta + \left(\frac{4}{3}\right)\xi\zeta + \left(\frac{8}{\sqrt{3}}\right)\xi\eta\zeta + \left(\frac{-8}{3}\right)\eta^2\zeta \tag{C.29}$$

$$\psi_{12}(\xi,\eta,\zeta) = \left(\frac{2}{9}\right) + \left(\frac{-2}{3}\right)\xi + \left(\frac{2}{3\sqrt{3}}\right)\eta + \left(\frac{4}{9}\right)\zeta + \left(\frac{-4}{3}\right)\eta^2$$
$$+ \left(\frac{-4}{\sqrt{3}}\right)\xi\eta + \left(\frac{4}{3\sqrt{3}}\right)\eta\zeta + \left(\frac{-4}{3}\right)\xi\zeta + \left(\frac{-8}{\sqrt{3}}\right)\xi\eta\zeta + \left(\frac{-8}{3}\right)\eta^2\zeta \tag{C.30}$$

$$\psi_{13}(\xi,\eta,\zeta) = \left(\frac{2}{9}\right) + \left(\frac{-4}{3\sqrt{3}}\right)\eta + \left(\frac{-4}{9}\right)\zeta + (-2)\xi^2 + \left(\frac{2}{3}\right)\eta^2$$
$$+ \left(\frac{8}{3\sqrt{3}}\right)\eta\zeta + (4)\xi^2\zeta + \left(\frac{-4}{3}\right)\eta^2\zeta \tag{C.31}$$

$$\psi_{14}(\xi,\eta,\zeta) = \left(\frac{2}{9}\right) + \left(\frac{2}{3}\right)\xi + \left(\frac{2}{3\sqrt{3}}\right)\eta + \left(\frac{-4}{9}\right)\zeta + \left(\frac{-4}{3}\right)\eta^2$$
$$+ \left(\frac{4}{\sqrt{3}}\right)\xi\eta + \left(\frac{-4}{3\sqrt{3}}\right)\eta\zeta + \left(\frac{-4}{3}\right)\xi\zeta + \left(\frac{-8}{\sqrt{3}}\right)\xi\eta\zeta + \left(\frac{8}{3}\right)\eta^2\zeta \quad \text{(C.32)}$$

$$\psi_{15}(\xi,\eta,\zeta) = \left(\frac{2}{9}\right) + \left(\frac{-2}{3}\right)\xi + \left(\frac{2}{3\sqrt{3}}\right)\eta + \left(\frac{-4}{9}\right)\zeta + \left(\frac{-4}{3}\right)\eta^2$$
$$+ \left(\frac{-4}{\sqrt{3}}\right)\xi\eta + \left(\frac{-4}{3\sqrt{3}}\right)\eta\zeta + \left(\frac{4}{3}\right)\xi\zeta + \left(\frac{8}{\sqrt{3}}\right)\xi\eta\zeta + \left(\frac{8}{3}\right)\eta^2\zeta \quad \text{(C.33)}$$

The construction of the update equations can be automated as follows. At time level "$a$" and at nodal point "$i$" we can write the gradient of the flux terms very compactly as

$$(\Delta \mathbf{F})_{a;i} \equiv \sum_{j=1}^{15} \left[ \hat{\mathbf{F}}_{a;j} \frac{\partial \psi_j(\xi_i,\eta_i,\zeta_i)}{\partial \xi} + \hat{\mathbf{G}}_{a;j} \frac{\partial \psi_j(\xi_i,\eta_i,\zeta_i)}{\partial \eta} + \hat{\mathbf{H}}_{a;j} \frac{\partial \psi_j(\xi_i,\eta_i,\zeta_i)}{\partial \zeta} \right] \quad \text{(C.34)}$$

Exactly analogous equations to the one above can be written with $a \to b$ and $b \to c$. This gives us the gradient of the flux terms at other time levels. The evolutionary equation at time level "$b$" at any nodal point "$i$" can be written as

$$\mathbf{U}_{b;i} = \left(\mathbf{U}_{a;i} + R_{ba}\mathbf{S}_{a;i} - R_{ba}(\Delta\mathbf{F})_{a;i}\right) + R_{bb}\mathbf{S}_{b;i} + R_{bc}\mathbf{S}_{c;i} - R_{bb}(\Delta\mathbf{F})_{b;i} - R_{bc}(\Delta\mathbf{F})_{c;i} \quad \text{(C.35)}$$

The evolutionary equation at time level "$c$" at any nodal point "$i$" can be written as

$$\mathbf{U}_{c;i} = \left(\mathbf{U}_{a;i} + R_{ca}\mathbf{S}_{a;i} - R_{ca}(\Delta\mathbf{F})_{a;i}\right) + R_{cb}\mathbf{S}_{b;i} + R_{cc}\mathbf{S}_{c;i} - R_{cb}(\Delta\mathbf{F})_{b;i} - R_{cc}(\Delta\mathbf{F})_{c;i} \quad \text{(C.36)}$$

The terms inside the round brackets should be evaluated only once. The above two equations show how the update equations can be compactly and explicitly evaluated.

**Appendix D**

This Appendix provides an explicit catalogue of the ADER-CG update, i.e. eqn. (5.28), at fourth order on the reference equilateral prism. All edges of the prism have unit length. In this Appendix, we denote the spatial nodes with numbers, as shown in Fig. 7c. The time levels are subscripted with letters just to make it easy to distinguish between different time levels. The time levels in eqn. (5.17) are, therefore, labeled as

$$\tau_a = 0 \; ; \; \tau_b = \frac{1}{2} - \frac{1}{2}\sqrt{\frac{3}{5}} \; ; \; \tau_c = \frac{1}{2} \; ; \; \tau_d = \frac{1}{2} + \frac{1}{2}\sqrt{\frac{3}{5}} \tag{D.1}$$

The locations of the spatial nodes on the 3D reference triangular prism in Fig. 7c (with all edges having unit length, and centroid at the origin) are given by

$$(\xi_{01}, \eta_{01}, \zeta_{01}) = \left(-\frac{1}{2}, -\frac{1}{2\sqrt{3}}, \frac{1}{2}\right) \tag{D.2}$$

$$(\xi_{02}, \eta_{02}, \zeta_{02}) = \left(\frac{1}{2}, -\frac{1}{2\sqrt{3}}, \frac{1}{2}\right) \tag{D.3}$$

$$(\xi_{03}, \eta_{03}, \zeta_{03}) = \left(0, \frac{1}{\sqrt{3}}, \frac{1}{2}\right) \tag{D.4}$$

$$(\xi_{04}, \eta_{04}, \zeta_{04}) = \left(-\frac{1}{2}, -\frac{1}{2\sqrt{3}}, -\frac{1}{2}\right) \tag{D.5}$$

$$(\xi_{05}, \eta_{05}, \zeta_{05}) = \left(\frac{1}{2}, -\frac{1}{2\sqrt{3}}, -\frac{1}{2}\right) \tag{D.6}$$

$$(\xi_{06}, \eta_{06}, \zeta_{06}) = \left(0, \frac{1}{\sqrt{3}}, -\frac{1}{2}\right) \tag{D.7}$$

$$(\xi_{07}, \eta_{07}, \zeta_{07}) = \left(-\frac{1}{2}, -\frac{1}{2\sqrt{3}}, \frac{1}{6}\right) \tag{D.8}$$

$$(\xi_{08}, \eta_{08}, \zeta_{08}) = \left(\frac{1}{2}, -\frac{1}{2\sqrt{3}}, \frac{1}{6}\right) \tag{D.9}$$

$$(\xi_{09}, \eta_{09}, \zeta_{09}) = \left(0, \frac{1}{\sqrt{3}}, \frac{1}{6}\right) \tag{D.10}$$

$$(\xi_{10}, \eta_{10}, \zeta_{10}) = \left(-\frac{1}{2}, -\frac{1}{2\sqrt{3}}, -\frac{1}{6}\right) \tag{D.11}$$

$$(\xi_{11}, \eta_{11}, \zeta_{11}) = \left(\frac{1}{2}, -\frac{1}{2\sqrt{3}}, -\frac{1}{6}\right) \tag{D.12}$$

$$(\xi_{12}, \eta_{12}, \zeta_{12}) = \left(0, \frac{1}{\sqrt{3}}, -\frac{1}{6}\right) \tag{D.13}$$

$$(\xi_{13}, \eta_{13}, \zeta_{13}) = \left(-\frac{1}{6}, -\frac{1}{2\sqrt{3}}, \frac{1}{2}\right) \tag{D.14}$$

$$(\xi_{14}, \eta_{14}, \zeta_{14}) = \left(\frac{1}{6}, -\frac{1}{2\sqrt{3}}, \frac{1}{2}\right) \tag{D.15}$$

$$(\xi_{15}, \eta_{15}, \zeta_{15}) = \left(\frac{1}{3}, 0, \frac{1}{2}\right) \tag{D.16}$$

$$(\xi_{16}, \eta_{16}, \zeta_{16}) = \left(\frac{1}{6}, \frac{1}{2\sqrt{3}}, \frac{1}{2}\right) \tag{D.17}$$

$$(\xi_{17}, \eta_{17}, \zeta_{17}) = \left(-\frac{1}{6}, \frac{1}{2\sqrt{3}}, \frac{1}{2}\right) \tag{D.18}$$

$$(\xi_{18}, \eta_{18}, \zeta_{18}) = \left(-\frac{1}{3}, 0, \frac{1}{2}\right) \tag{D.19}$$

$$(\xi_{19}, \eta_{19}, \zeta_{19}) = \left(-\frac{1}{6}, -\frac{1}{2\sqrt{3}}, -\frac{1}{2}\right) \tag{D.20}$$

$$(\xi_{20}, \eta_{20}, \zeta_{20}) = \left(\frac{1}{6}, -\frac{1}{2\sqrt{3}}, -\frac{1}{2}\right) \tag{D.21}$$

$$(\xi_{21}, \eta_{21}, \zeta_{21}) = \left(\frac{1}{3}, 0, -\frac{1}{2}\right) \tag{D.22}$$

$$(\xi_{22}, \eta_{22}, \zeta_{22}) = \left(\frac{1}{6}, \frac{1}{2\sqrt{3}}, -\frac{1}{2}\right) \tag{D.23}$$

$$(\xi_{23}, \eta_{23}, \zeta_{23}) = \left(-\frac{1}{6}, \frac{1}{2\sqrt{3}}, -\frac{1}{2}\right) \tag{D.24}$$

$$(\xi_{24},\eta_{24},\zeta_{24}) = \left(-\frac{1}{3}, 0, -\frac{1}{2}\right) \tag{D.25}$$

$$(\xi_{25},\eta_{25},\zeta_{25}) = \left(0, 0, \frac{1}{2}\right) \tag{D.26}$$

$$(\xi_{26},\eta_{26},\zeta_{26}) = \left(0, 0, -\frac{1}{2}\right) \tag{D.27}$$

Once these nodes are specified, any computer algebra system can be used to obtain the nodal basis in the reference element.

We do not write down the solution vector at all space and time locations in the reference element because that would consume too much space and it would not be very illustrative. The twenty-six update equations for the solution at time level "b", and the twenty-six update equations for the solution at time level "c", as well as the twenty-six update equations for the solution at time level "d" are given by first specifying a matrix. This matrix gives us the contribution of the source terms from any one level to the state at any of the other three dynamically active time levels, i.e. "b", "c" and "d". That matrix is given by

$$\mathbf{R} = \begin{pmatrix} 0 & 0 & 0 & 0 \\ \dfrac{-4}{35} & \dfrac{295+24\sqrt{15}}{1260} & \dfrac{184-84\sqrt{15}}{1260} & \dfrac{295-66\sqrt{15}}{1260} \\ \dfrac{-5}{28} & \dfrac{145+36\sqrt{15}}{504} & \dfrac{13}{126} & \dfrac{145-36\sqrt{15}}{504} \\ \dfrac{-4}{35} & \dfrac{295+66\sqrt{15}}{1260} & \dfrac{184+84\sqrt{15}}{1260} & \dfrac{295-24\sqrt{15}}{1260} \end{pmatrix} \tag{D.28}$$

The construction of the update equations can be automated as follows. At time level "a" and at nodal point "i" we can write the gradient of the flux terms very compactly as

$$(\Delta \mathbf{F})_{a;i} \equiv \sum_{j=1}^{26}\left[\hat{\mathbf{F}}_{a;j}\frac{\partial \psi_j(\xi_i,\eta_i,\zeta_i)}{\partial \xi} + \hat{\mathbf{G}}_{a;j}\frac{\partial \psi_j(\xi_i,\eta_i,\zeta_i)}{\partial \eta} + \hat{\mathbf{H}}_{a;j}\frac{\partial \psi_j(\xi_i,\eta_i,\zeta_i)}{\partial \zeta}\right] \tag{D.29}$$

Exactly analogous equations to the one above can be written with $a \to b$, $b \to c$ and $c \to d$. This gives us the gradient of the flux terms at other time levels. The evolutionary equation at time level "b" at any nodal point "i" can be written as

$$\mathbf{U}_{b;i} = \left(\mathbf{U}_{a;i} + R_{ba}\mathbf{S}_{a;i} - R_{ba}(\Delta\mathbf{F})_{a;i}\right) + R_{bb}\mathbf{S}_{b;i} + R_{bc}\mathbf{S}_{c;i} + R_{bd}\mathbf{S}_{d;i} - R_{bb}(\Delta\mathbf{F})_{b;i} - R_{bc}(\Delta\mathbf{F})_{c;i} - R_{bd}(\Delta\mathbf{F})_{d;i}$$
(D.30)

The evolutionary equation at time level "c" at any nodal point "i" can be written as

$$\mathbf{U}_{c;i} = \left(\mathbf{U}_{a;i} + R_{ca}\mathbf{S}_{a;i} - R_{ca}(\Delta\mathbf{F})_{a;i}\right) + R_{cb}\mathbf{S}_{b;i} + R_{cc}\mathbf{S}_{c;i} + R_{cd}\mathbf{S}_{d;i} - R_{cb}(\Delta\mathbf{F})_{b;i} - R_{cc}(\Delta\mathbf{F})_{c;i} - R_{cd}(\Delta\mathbf{F})_{d;i}$$
(D.31)

The evolutionary equation at time level "d" at any nodal point "i" can be written as

$$\mathbf{U}_{d;i} = \left(\mathbf{U}_{a;i} + R_{da}\mathbf{S}_{a;i} - R_{da}(\Delta\mathbf{F})_{a;i}\right) + R_{db}\mathbf{S}_{b;i} + R_{dc}\mathbf{S}_{c;i} + R_{dd}\mathbf{S}_{d;i} - R_{db}(\Delta\mathbf{F})_{b;i} - R_{dc}(\Delta\mathbf{F})_{c;i} - R_{dd}(\Delta\mathbf{F})_{d;i}$$
(D.32)

The terms inside the round brackets should be evaluated only once. The above two equations show how the update equations can be compactly and explicitly evaluated.